\documentclass[aps,pra,twocolumn,amsmath,amssymb,superscriptaddress,floatfix]{revtex4-2}

\usepackage{comment}

\usepackage[utf8]{inputenc}
\usepackage[T1]{fontenc}
\usepackage{microtype}
\usepackage{bbm}
\usepackage{amsmath}
\usepackage{graphicx}
\usepackage{times}
\usepackage{amssymb}
\usepackage{mathtools}
\usepackage{bm}
\usepackage{txfonts}
\usepackage{physics}
\usepackage{float}
\usepackage[table]{xcolor}
\usepackage{adjustbox}
\usepackage{mathrsfs}
\usepackage{hhline}
\usepackage{todonotes}
\PassOptionsToPackage{hyphens}{url}
\usepackage[colorlinks = true, linkcolor = blue, urlcolor  = blue, citecolor = blue, anchorcolor = blue]{hyperref}
\usepackage{etoolbox}
\apptocmd{\sloppy}{\hbadness 10000\relax}{}{}
\usepackage{tikz}
\usepackage{tensor}
\usepackage{tabularx}
\usepackage{pgfplots}
\pgfplotsset{compat=newest}

\usepackage[caption = false, subrefformat=parens,labelformat=parens]{subfig}
\usepackage{ragged2e} 
\DeclareCaptionJustification{justified}{\justifying}

\newcolumntype{Y}{>{\centering\arraybackslash}X}
\newcommand{\One}{\mathbbm{1}}

\newcommand{\OldCoordsVec}{\bm{\mathsf{r}}}

\newcommand{\OldCoordsTime}{\textsf{t}}

\newcommand{\OldCoordsX}{\mathsf{x}}
\newcommand{\OldCoordsg}{\mathsf{g}}

\newcommand{\NewCoordsVec}{\bm{r}}

\newcommand{\NewCoordsTime}{t}
\newcommand{\NewCoordsBigVec}{\bm{R}}

\newcommand{\NewCoordsX}{x}
\newcommand{\NewCoordsg}{g}
\newcommand{\rNew}{\hat{\bm{r}}}
\newcommand{\pNew}{\hat{\bm{p}}}
\newcommand{\R}{\hat{\bm{R}}}
\newcommand{\PR}{\hat{\bm{P}}}
\newcommand{\ZZ}{\hat{Z}}
\newcommand{\PZ}{\hat{P}}

\newcommand{\DimTime}{\tau}

\newcolumntype{L}[1]{>{\raggedright\arraybackslash}p{#1}}
\newcolumntype{C}[1]{>{\centering\arraybackslash}p{#1}}
\newcolumntype{R}[1]{>{\raggedleft\arraybackslash}p{#1}}

\newcommand{\HamiltonianInternal}{\text{I}}
\newcommand{\HamiltonianMotional}{\text{M}}
\newcommand{\HamiltonianAtomLight}{\text{A-L}}
\newcommand{\HamiltonianLightField}{\text{L}}
\newcommand{\ii}{\mathrm{i}}

\newcommand{\GZero}{\mathcal{G}_0}
\newcommand{\GOneR}{\mathcal{G}_{1, \text{R}}}
\newcommand{\GOneB}{\mathcal{G}_{1, \text{B}}}
\newcommand{\GTwoR}{\mathcal{G}_{2, \text{R}}}
\newcommand{\GTwoB}{\mathcal{G}_{2, \text{B}}}
\newcommand{\GThreeR}{\mathcal{G}_{3, \text{R}}}

\newcommand{\ZZero}{\mathcal{Z}_0}
\newcommand{\VZero}{\mathcal{V}_0}
\newcommand{\RR}{\mathcal{R}_\text{R}}
\newcommand{\RB}{\mathcal{R}_\text{B}}
\newcommand{\Comp}{\omega_\text{C}}
\newcommand{\FreqR}{\mathcal{F}_\text{R}}

\begin{document}
\title{Atom interferometers in weakly curved spacetimes using Bragg diffraction and Bloch oscillations}

\author{Michael Werner}
\affiliation{Institut für Theoretische Physik, Leibniz Universität Hannover, Appelstraße 2, 30167 Hannover, Germany}

\author{Philip K. Schwartz}
\affiliation{Institut für Theoretische Physik, Leibniz Universität Hannover, Appelstraße 2, 30167 Hannover, Germany}

\author{Jan-Niclas Kirsten-Siemß}
\affiliation{Institut für Theoretische Physik, Leibniz Universität Hannover, Appelstraße 2, 30167 Hannover, Germany}
\affiliation{Leibniz Universität Hannover, Institut für Quantenoptik, Welfengarten 1, 30167 Hannover, Germany}

\author{Naceur Gaaloul}
\affiliation{Leibniz Universität Hannover, Institut für Quantenoptik, Welfengarten 1, 30167 Hannover, Germany}

\author{Domenico Giulini}
\affiliation{Institut für Theoretische Physik, Leibniz Universität Hannover, Appelstraße 2, 30167 Hannover, Germany}
\affiliation{Center of Applied Space Technology and Microgravity, Universität Bremen, Am Fallturm 1, 28359 Bremen, Germany}

\author{Klemens Hammerer}
\affiliation{Institut für Theoretische Physik, Leibniz Universität Hannover, Appelstraße 2, 30167 Hannover, Germany}

\date\today

\begin{abstract}
We present a systematic approach to determine all relativistic phases up to $\mathcal{O}(c^{-2})$ in light-pulse atom interferometers in weakly curved spacetime that are based on elastic scattering --- namely Bragg diffraction and Bloch oscillations. Our analysis is derived from first principles using the parameterized post-Newtonian formalism. In the treatment developed here, we derive algebraic expressions for relativistic phases for arbitrary interferometer geometries in an automated manner. As case studies, we consider symmetric and antisymmetric Ramsey-Bordé interferometers, as well as a symmetric double diffraction interferometer with baseline lengths of $10 \, \mathrm{m}$ and $100 \, \mathrm{m}$. We compare our results to previous calculations conducted for a Mach-Zehnder interferometer.
\end{abstract}

\maketitle



\section{Introduction}

Atom interferometers (IFs), at the forefront of quantum metrology, are highly precise instruments widely utilized in various research domains. They have been employed in diverse fields, including the determination of the fine-structure constant \cite{morel2020determination, parker2018measurement}, serving as quantum sensors for measuring the gravitational field of the Earth \cite{Wu_2019, Rosi_2015, del_Aguila_2018, Fang_2016}, proposed measurements of gravitational waves \cite{beaufils2022coldatom, chen2023enhancing, Bertoldi_2021, canuel2020elgar}, investigations of fundamental physics and alternative gravitational models \cite{ufrecht2020atom,schlippert2014quantum,damour2012theoretical,colladay1997cpt, coleman2018magis100}, as well as measurements of time dilation and gravitational redshift \cite{loriani2019interference,roura2021measuring,zych2011quantum, roura2020gravitational,di2021gravitational}.

The interpretation of measurements of the gravitational redshift have ignited extensive discussions regarding the influence of relativistic effects in atom IFs \cite{muller2010precision, wolf2010atom, wolf2011does, hohensee2011significance}. These discussions have underscored the need for interferometry with internal superposition states~\cite{roura2020gravitational, di2021gravitational} enabling the effective detection of gravitational redshift effects. As a result, there has been significant research focus on IFs employing inelastic scattering processes, such as single-photon or Raman transitions, commonly referred to as `clock interferometry'. However, inelastic scattering introduces additional systematic effects due to the different internal atomic states. In contrast, atom IFs based on elastic scattering processes, such as Bragg diffraction~\cite{Giltner1995Theoretical, Martin1988Bragg} and Bloch oscillations~\cite{Dahan1996Bloch, peik1997bloch}, currently exhibit the highest sensitivity. This advancement has facilitated groundbreaking measurements, such as the precise determination of atomic recoil --- and, consequently, the fine-structure constant --- with unprecedented accuracy~\cite{parker2018measurement}. The gravitational redshift cannot be directly measured with these IFs, it is worth noting that phases involving relativistic effects and even extensions to the standard model (SME) can still manifest in these atom IFs~\cite{diPumpo2021GravitationalRedshift, diPumpo2022LightPropagation}.

Dimopoulos et al.~\cite{dimopoulos2007PRL, dimopoulos2008general} presented the determination and detailed listing of phases induced by special- and general-relativistic effects specifically for the Mach-Zehnder IF. However, the laborious algebraic calculations involved make it difficult to reproduce and extend these results to more general IF geometries. Here, we propose a systematic framework for computing relativistic phases in arbitrary atom IFs realized by elastic scattering. Our approach employs rigorous expansions in relevant small parameters, implemented through computer algebra in Python~\cite{PythonCode}. This enables automated algebraic calculations of relativistic phases up to the desired order of accuracy. We compute and display the phases for three common IF geometries: symmetric Ramsey-Bordé interferometer (SRBI), antisymmetric Ramsey-Bordé interferometer (ARBI), and symmetric double diffraction interferometer (SDDI). The computer algorithm is, however, capable of calculating phases for more general IF geometries. For each geometry, we algebraically list and quantitatively illustrate the leading relativistic phases. Our analysis focuses on atom IFs with baseline lengths of $10\,\mathrm{m}$ and $100\,\mathrm{m}$, inspired by numerous operational or under-development setups \cite{schlippert2020matter, canuel2020elgar, Dimopoulos2008AGIS, dickerson2013multiaxis, zhan2020zaiga, badurina2020aion}. Furthermore we provide a detailed comparison to the results of Dimopoulos et al. \cite{dimopoulos2007PRL, dimopoulos2008general} for the Mach-Zehnder IF, where we find overall good agreement and comment on the remaining discrepancies.

The starting point and basis of our calculation is a quantum optical Hamiltonian, as developed in~\cite{schwartz2019post}. It accounts for both the internal and external degrees of freedom of an atom, as well as the electromagnetic (EM) field, within a weak post-Newtonian gravitational background field which is described by the Eddington-Robertson parametrized post-Newtonian metric~\cite{schwartz2020post}. Consequently, this Hamiltonian captures the phases resulting from extensions of GR and provides a comprehensive description of the leading-order effects for scalar particles. Importantly, this treatment eliminates the need for additional relativistic corrections and avoids the requirement of ad-hoc time reparametrization in the Schrödinger equation to account for time dilation, cf. the analysis of the mass-defect in~\cite{martinez2022abinitio}.

To achieve a self-consistent description of experiments that involve the measurement of time and length using local clocks and rulers --- i.e., laser fields --- we transform the Hamiltonian from the asymptotically flat coordinates, as formulated in~\cite{schwartz2019post}, to coordinates which reduce to the Minkowski metric at the point of the experiment --- e.g., that of the experiment’s reference atomic clock. This reformulation of the Hamiltonian holds significance not only for the description of atom IFs but also for any other local experiments.

We begin in Sec.~\ref{Section 2} by introducing the notation, explaining the gravitational model, and deriving the relativistically corrected Hamiltonian. We do this within the context of newly established coordinates, which are suitable for describing atom IFs. We continue in Sec.~\ref{Section 3} to solve the Schrödinger equation for a general class of atom IF geometries which rely on elastic scattering. We will then present the resulting phase shifts for the aforementioned IF geometries in Sec.~\ref{Section 4}. We end in Sec.~\ref{Section 5} with a summary and conclusion. For improved readability we supply intermediate calculation steps and supplemental information in Appendices~\ref{Appendix: New Coordinate System} - \ref{Appendix: Mach-Zehnder-Interferometer}.


\section{Relativistically corrected Hamiltonian of an atom in weakly curved spacetime}\label{Section 2}

We adopt the following convention: Greek indices range from 0 to 3, whereas Latin indices range from 1 to 3. The components of the Minkowski metric are $\eta_{\mu \nu}= \text{diag}(-1,1,1,1)$. Bold letters like $\bm{x}$ and $\bm{R}$ will always represent three dimensional quantities, like elements in $\mathbb{R}^3$ or 3-tuples of operators. Commas indicate partial differentiation, whereas semicolons abbreviate covariant derivatives --- that is, $\tensor{A}{_{\mu,\nu}} = \partial_\nu A_\mu $ and $\tensor{A}{_{\mu;\nu}} = \nabla_\nu A_\mu $. 

\subsection{Gravitational model}

We are going to model gravity by a \emph{metric theory} in which test bodies follow geodesics with respect to the corresponding Levi-Civita connection and to which matter couples by the standard minimal scheme. In this way Einstein’s equivalence principle is implemented  and spacetime is represented by the triple $(\mathcal{M}, g, \nabla)$, where $\mathcal{M}$ is a 4-dimensional differentiable manifold, $g$ a Lorentzian metric, and $\nabla$ the corresponding Levi-Civita connection. This scheme may be extended to scalar-tensor~\cite{brans1961mach} and vector-tensor theories~\cite{nordtvedt1972conservation}, which however we will not consider here. 

To perform a Newtonian \emph{weak-field} and \emph{slow-motion} expansion of the physics in a relativistic spacetime, one first has to introduce a background structure w.r.t.\ which these notions can be defined~\cite[sec.\ 16.2.1]{giulini2023coupling}. This background structure we take to consist of a background metric and the worldline of a preferred observer. Then one can implement the Newtonian expansion by a power series expansion in small parameters: The Newtonian gravitational potential $\phi/c^2 \ll 1$ and $v^2/c^2 \ll 1$, where $v$ is a typical velocity of the matter sourcing gravity. With respect to the background structure, the metric tensor $g$ then splits into the sum of i) the background Minkowski metric, ii) a first order part describing Newtonian mechanics, and iii) post-Newtonian corrections, special to each metric theory. This approach gives rise to a 10-parameter class of different metric theories and is described in the \emph{parametrized post-Newtonian} (PPN) formalism \cite{will1974gravitation}. The low-order post-Newtonian behaviour of each metric theory is then uniquely determined by those ten \emph{PPN parameters}.

In the following, we only consider two out of those ten possible parameters --- namely, the `Eddington-Robertson (ER) parameters' $\beta, \gamma \in \mathbb{R}$ corresponding to the biggest relativistic corrections. In a local coordinate system $\OldCoordsX^\mu = \{c \OldCoordsTime, \OldCoordsVec\}$ the covariant PPN metric tensor components can, in the case of a spherically symmetric and static spacetime, be written using the line element
\begin{multline}
	\dd s^2 
	= - \left( c^2 + 2 \phi(\OldCoordsVec) + 2 \beta \frac{\phi(\OldCoordsVec)^2}{c^2} \right) \dd \OldCoordsTime^2 \\
	+ \left( 1-2 \gamma \frac{\phi(\OldCoordsVec)}{c^2} \right) \dd \OldCoordsVec^2 + \mathcal{O}(c^{-4}) \label{eq: PPN_metric},
\end{multline}
where the scalar field $\phi: \mathbb{R}^3 \longrightarrow \mathbb{R}$ is the gravitational potential that arises from solving the Poisson equation~\footnote{See, e.g., the discussion in Chapter~6 of \cite{wald2010general} and Eq.~(6.2.14) therein.}. In standard GR, the ER parameters are unity. Upper bounds for the ER parameters are long known to be on the orders $\abs{\gamma-1} \approx \abs{\beta-1} \lesssim 10^{-5}$ obtained by tracking, e.g., the trajectory of the Cassini mission or measuring perihelion shifts of Mercury and Mars (cf. \cite{bertotti2003test, konopliv2011mars, fienga2011inpop10a, verma2014use}). We model Earth as a point source with mass $M_\oplus$ --- i.e., the Newtonian gravitational potential takes the form
\begin{align}
	\phi(\OldCoordsVec) = - \frac{G M_\oplus}{\abs{\OldCoordsVec}}. \label{eq: Potential_in_original_coords}
\end{align}
Notice that the metric tensor is asymptotically flat in these coordinates, since $\phi(\OldCoordsVec) \longrightarrow 0$ for $\abs{\OldCoordsVec} \longrightarrow \infty$, which means that metric tensor is not Minkowskian at the point of an earthbound experiment.

\subsection{Transformation to adapted coordinates} \label{Sec: Transformation_to_new_coords}

Using coordinates in which the metric tensor asymptotically approximates the Minkowskian values makes asymptotic flatness manifest, but also means that  these coordinates cannot be employed for a direct representation of local measurements of spacetime distances in terms of simple coordinate expressions. Such measurements refer to colocated clocks and spatial length-references set by light signals (e.g., by a laser). That itself would have to be described by a Hamiltonian operator, whereby the corresponding clock signal would contain the metric at the location of the experiment. In order to allow for a simpler and more direct interpretation of coordinates as quantities measurable with a local clock and length scale, it is convenient to transform to coordinates in which the metric components at the reference point of the experiment (set by clock and laser) reduce to the Minkowski metric. 

The transformation from asymptotically Minkowskian coordinates into such coordinates that we use is explained in App.~\ref{Appendix: New Coordinate System}, and is given by $(\OldCoordsX^\mu) = (c \OldCoordsTime, \OldCoordsVec ) \longmapsto (\NewCoordsX^\mu) = (c \NewCoordsTime, \NewCoordsVec )$, with
\begin{align}
	\NewCoordsTime &= \left( 1 + \frac{\phi_0}{c^2} + \frac{2 \beta - 1}{2} \frac{\phi_0^2}{c^4} \right) \OldCoordsTime , &
	\NewCoordsVec &= \left(1 - \gamma \frac{\phi_0}{c^2} \right) \OldCoordsVec. \label{eq: Definition_new_coordinates}
\end{align}
where we have defined $\phi_0 = \phi(\NewCoordsBigVec_\oplus)$. Here, we assume the point of reference is at rest at Earth's surface --- i.e., at a radius of $\NewCoordsBigVec_\oplus$ in the new coordinates. It will be convenient to define the shifted gravitational potential
\begin{align}
    \overline{\phi}(\NewCoordsVec) = \phi(\NewCoordsVec) - \phi_0
\end{align}
such that the line element in the new coordinates reads
\begin{multline}
	\dd s^2 = -\left( c^2 + 2\overline{\phi} + 2 \beta \frac{\overline{\phi}^2}{c^2} + 4 (\beta-1) \frac{\phi_0 \overline{\phi}}{c^2} \right) \dd \NewCoordsTime^2  \\
	+ \left( 1-2 \gamma \frac{\overline{\phi}}{c^2} \right) \dd \NewCoordsVec^2 + \mathcal{O} (c^{-4} ), \label{eq: PPN_metric_new_coords}
\end{multline}
which reduces to the Minkowski metric for $\NewCoordsVec = \NewCoordsBigVec_\oplus$.  This is the form of the metric that we shall use in the sequel. 

\subsection{Relativistically corrected Hamiltonian} \label{Sec: Relativistically_corrected_Hamiltonian}

The Hamiltonian for an atom, modelled as a system of two electromagnetically bound spinless point charges, coupled to an external EM field and the weakly curved PPN spacetime metric from Eq.~\eqref{eq: PPN_metric} has been derived in~\cite{schwartz2019post}. We perform the same derivation as in~\cite{schwartz2019post} but use the coordinates from Eq.~\eqref{eq: Definition_new_coordinates} --- i.e., the metric tensor in Eq.~\eqref{eq: PPN_metric_new_coords}. To zeroth order in $1/c^2$, this Hamiltonian corresponds to the standard description of an atom in non-relativistic Quantum Optics. Terms of order $1/c^2$ correspond to the leading relativistic corrections to the energies of the centre-of-mass (COM) and internal (electronic) degrees of freedom of the atom, as well as their mutual interactions and that with the external EM field. The Hamiltonian can be grouped as 
\begin{align}
    \hat{H} = \hat{H}_\HamiltonianMotional + \hat{H}_\HamiltonianInternal +
    \hat{H}_\text{M-I} +
    \hat{H}_\HamiltonianAtomLight  + \mathcal{O}(c^{-4}). \label{eq: Full_Hamiltonian}
\end{align}
It consists of the Hamiltonian for COM motion $\hat{H}_\HamiltonianMotional$, the Hamiltonian for the internal degrees of freedom $\hat{H}_\HamiltonianInternal$, their relativistic coupling $\hat{H}_\text{M-I}$, and the relativistically corrected dipole interaction of the atom with the external EM field $\hat{H}_\HamiltonianAtomLight$. The external EM field is a classical solution to the source-free Maxwell equations in our curved spacetime \footnote{Note that in order to keep the treatment of the external EM field consistent, differently to the original discussion in \cite{schwartz2019post} (and to the non-gravitational one in \cite{sonnleitner2018mass}, on which the derivation in \cite{schwartz2019post} is based), we need to treat the external EM field not as a dynamical variable, but as a background field in which the motion of the atom takes place.  A detailed explanation and discussion of this point may be found in \cite[chapter 4]{schwartz2020post}, and in \cite[sec.\ 16.2.1.2, footnote 15]{giulini2023coupling}}. In the following the relativistically corrected canonical position and momentum operators of COM and internal degrees of freedom will be denoted by $\R$, $\PR$ and $\rNew$, $\pNew$, respectively. The total (rest) mass of the atom will be denoted by $m$, and the reduced mass of the internal degree of freedom by $\mu$. The Hamiltonian for COM motion is
\begin{subequations} \label{eq: COM Hamiltonian} 
\begin{align}
	\hat{H}_\HamiltonianMotional &= m \overline{\phi}(\R)+\frac{\PR^2}{2m} \label{eq: Motional_Hamiltonain_Non_Rel} \\
	& + \frac{1}{m c^2}\Bigg[ \frac{2 \gamma +1}{2} \PR \overline{\phi}(\R) \PR - \frac{\PR^4}{8 m^2} + \frac{2 \beta - 1}{2} m^2 \overline{\phi}(\R)^2 \label{eq: Motional_Hamiltonain_Rel_1} \\
	& + 2(\beta-1) m^2 \phi_0 \overline{\phi}(\R) \Bigg] + \mathcal{O}(c^{-4}). \label{eq: Motional_Hamiltonain_Rel_2}
\end{align}
\end{subequations}
The terms in the square brackets --- i.e., Eqs.~\eqref{eq: Motional_Hamiltonain_Rel_1} and \eqref{eq: Motional_Hamiltonain_Rel_2} --- comprise the relativistic corrections of the COM energy, and will be the most relevant for our analysis. The first of those terms corresponds to the metric correction of the length of the vector $\PR$ determining the kinetic energy, written in symmetric ordering. The second term is the special relativistic correction to the kinetic energy, the third term describes the relativistic non-linear correction to the Newtonian potential, and the final term represents non-linear relativistic effects which may arise in theories of gravity deviating from GR.

The internal Hamiltonian,
\begin{align}
    &\hat{H}_\HamiltonianInternal = \frac{\pNew^2}{2 \mu} - \frac{e^2}{4 \pi \epsilon_0 \abs{\rNew}} +  H_\text{FS}, \label{eq: Atomic Hamiltonian}
\end{align}
consists of the non-relativistic kinetic energy and Coulomb interaction and relativistic corrections subsumed in the fine-structure Hamiltonian $H_\text{FS}$, which contains the special relativistic corrections of kinetic and Coulomb energy, as well as spin-orbit interaction etc., if spin was included. We refer to \cite{schwartz2019post} for the explicit form of $H_\text{FS}$. In the following, we take these corrections to be accounted for in the internal stationary states and energies --- e.g. the ground state $\hat{H}_\HamiltonianInternal\ket{g}=E_g\ket{g}$.

The relativistic coupling of COM and internal degrees of freedom has the form
\begin{multline}
	\hat{H}_\text{M-I} = \frac{1}{m c^2} \qty(m \phi(\R)- \frac{\PR^2}{2m})\otimes \hat{H}_\HamiltonianInternal \\
	+ \frac{\phi(\R)}{ c^2} \otimes \left( 2 \gamma \frac{\pNew^2}{2 \mu} - \gamma \frac{e^2}{4 \pi \epsilon_0 \abs{\rNew}} \right).\label{eq: Motional_Hamiltonain_Internal_COM_Mix} 
\end{multline}
The first line of this equation can be interpreted as arising from the correction of the mass of the atom due to internal binding energy, as can be seen by replacing $m \rightarrow m + \hat{H}_\HamiltonianInternal/c^2$ in Eq.~\eqref{eq: Motional_Hamiltonain_Non_Rel} and expanding in $1/c^2$. The second line describes the metric corrections of the internal kinetic and Coulomb energy, similar to Eqs.~\eqref{eq: Motional_Hamiltonain_Rel_1} and \eqref{eq: Motional_Hamiltonain_Rel_2}, cf.~\cite{giulini2023coupling}. For atom IFs involving elastic scattering processes only, the atom remains in its internal ground state $\ket{g}$ at all times. In this case the terms in Eq.~\eqref{eq: Motional_Hamiltonain_Internal_COM_Mix} contribute only trivially to the dynamics of the problem: As explained above, the effect of the first term can be absorbed in a rescaling of the atomic mass $m \rightarrow m + E_g/c^2$. The second term does not contribute at all since it has vanishing matrix elements for stationary states~\cite{martinez2022abinitio} as a consequence of the virial theorem~\footnote{This follows from $\frac{\mathrm{i}}{\hbar}\left[\rNew \cdot \pNew, H_\HamiltonianInternal\right]=2 \frac{\pNew^2}{2 \mu}-\frac{e^2}{4 \pi \varepsilon_0} \frac{1}{\abs{\rNew}} + \mathcal{O}(c^{-2})$, and taking the matrix element with respect to $\ket{g}$}. However, they can be the main contribution to the phase in quantum clock IFs with inelastic scattering processes~\cite{roura2020gravitational, di2021gravitational, chiba2022QuantumClocks, DiPumpo2023T3Scaling}.

Finally, the Hamiltonian for the interaction of the atom with the external light field is
\begin{align}
	\hat{H}_\HamiltonianAtomLight &= - \hat{\bm{d}} \cdot \bm{E}^\perp(\R)  + \frac{1}{2m} \left[ \PR \cdot \left( \hat{\bm{d}} \times \bm{B}(\R) \right) + \text{h.c.} \right]  \label{eq: Atom Light Hamiltonian},
\end{align}
where $\hat{\bm{d}}$, $\bm{E}$, and $\bm{B}$ denote the dipole moment operator and the electric and magnetic fields, respectively. The atom-light interaction in Eq.~\eqref{eq: Atom Light Hamiltonian} is written in the dipole approximation and includes the Röntgen term as the dominant relativistic correction~\footnote{We disregard further relativistic corrections relevant in strong magnetic fields, cf.~Eq.~(5.9) in Schwartz et al.~\cite{schwartz2019post}.}. In comparison to the Hamiltonian from~\cite{schwartz2019post}, only the last term in the motional Hamiltonian --- i.e., Eq.~\eqref{eq: Motional_Hamiltonain_Rel_2} is a new contribution. Its specific scaling with the PPN parameter $\beta$ is commented on in App.~\ref{Appendix: New Coordinate System}. Apart from this new term, the only difference is the dependence on $\overline{\phi}$ instead of $\phi$. 

We note that a physical misinterpretation of the coordinates used in Eq.~\eqref{eq: PPN_metric} easily results in erroneous terms scaling with $\phi_0/c^2$, as was the case, e.g., in the debate about such contributions in the measurements of the electronic gyromagnetic factor $g_e$~\cite{morishima2018general,venhoek2018analyzing,nikolic2018can}. Such pitfalls can be avoided by working in the coordinates in Eq.~\eqref{eq: Definition_new_coordinates}, where the Hamiltonian in Eq.~\eqref{eq: Full_Hamiltonian} has no dependence at all on $\phi_0$ for GR ($\beta=1$).


\section{Atom interferometers using elastic scattering}\label{Section 3}

\begin{figure*}
    \subfloat[Symmetric Ramsey-Bordé]{
		\centering
		\begin{tikzpicture}[scale=.6]
			\draw[->,thick] (-0.2,0) -- (8,0) node[right] {$\NewCoordsTime$}; 		
			\draw[->,thick] (0,-0.2) -- (0,6.7) node[above] {$z(\NewCoordsTime)$}; 	
			\draw[color=red,dashed] (1,-0.5) -- (1,6.5);					
			\draw[color=red,dashed] (3,-0.5) -- (3,6.5);
			\draw[color=red,dashed] (5,-0.5) -- (5,6.5);
			\draw[color=red,dashed] (7,-0.5) -- (7,6.5);
			\fill[blue,opacity=0.5] (3.9,0) rectangle ++(0.2,6.5);			
			\draw[color=green,thick] (0.5,1) -- (1,1) -- (3,3.5) -- (4,3.5) -- (5,4) -- (7,5) -- (7.5,5.25);		
			\draw[color=green,thick] (1,1) -- (3,1) -- (4,1) -- (5,1.5) -- (7,5) -- (7.5,5.75);						
			\draw[color=black,thick] (-0.075,1) -- (0.075,1);				
			\draw (-0.35,1) node {$z_0$};									
			\draw (2,-0.35) node {$T_R$};									
			\draw (4,-0.35) node {$T_B$};
			\draw (6,-0.35) node {$T_R$};
			\draw[color=green] (2, 3) node {$z_\text{up}$};
			\draw[color=green] (2, 0.7) node {$z_\text{low}$};
			\draw[->,thick] (1,-0.35) -- (1.5,-0.35);						
			\draw[<-,thick] (2.5,-0.35) -- (3,-0.35);
			\draw[->,thick] (3,-0.35) -- (3.5,-0.35);
			\draw[<-,thick] (4.5,-0.35) -- (5,-0.35);
			\draw[->,thick] (5,-0.35) -- (5.5,-0.35);
			\draw[<-,thick] (6.5,-0.35) -- (7,-0.35);
			\draw[color=black,thick] (1,-0.2) -- (1,-0.5);					
			\draw[color=black,thick] (3,-0.2) -- (3,-0.5);
			\draw[color=black,thick] (5,-0.2) -- (5,-0.5);
			\draw[color=black,thick] (7,-0.2) -- (7,-0.5);
			\draw[color=blue] (4,7) node {$+ \hbar k_B$};				    
			\draw[color=red] (2,7) node {$\pm \hbar k_R$};					
			\draw[color=red] (6,7) node {$\pm \hbar k_R$};
			\fill[black] (1,1) circle (0.08);								
			\fill[black] (3,1) circle (0.08);
			\fill[black] (3,3.5) circle (0.08);
			\fill[black] (4,1) circle (0.08);
			\fill[black] (4,3.5) circle (0.08);
			\fill[black] (5,1.5) circle (0.08);
			\fill[black] (5,4) circle (0.08);
			\fill[black] (7,5) circle (0.08);
		\end{tikzpicture}
		\label{fig: Subfigure_SRBI}
	}
	\subfloat[Symmetric Double Diffraction]{
		\centering
		\begin{tikzpicture}[scale=.6]
			\draw[->,thick] (-0.2,0) -- (8,0) node[right] {$\NewCoordsTime$}; 		
			\draw[->,thick] (0,-0.2) -- (0,6.7) node[above] {$z(\NewCoordsTime)$}; 	
			\draw[color=red,dashed] (1,-0.5) -- (1,6.5);					
			\draw[color=red,dashed] (3,-0.5) -- (3,6.5);
			\draw[color=red,dashed] (5,-0.5) -- (5,6.5);
			\draw[color=red,dashed] (7,-0.5) -- (7,6.5);
			\fill[blue,opacity=0.5] (3.9,0) rectangle ++(0.2,6.5);			
			\draw[color=green,thick] (0.5,3) -- (1,3) -- (3,4.5) -- (4,4.5) -- (5,5) -- (7,4.5) -- (7.5,4.375);		
			\draw[color=green,thick] (1,3) -- (3,1.5) -- (4,1.5) -- (5,2) -- (7,4.5) -- (7.5,5.125);				
			\draw[color=black,thick] (-0.075,3) -- (0.075,3);				
			\draw (-0.35,3) node {$z_0$};								    
			\draw (2,-0.35) node {$T_R$};									
			\draw (4,-0.35) node {$T_B$};
			\draw (6,-0.35) node {$T_R$};
			\draw[->,thick] (1,-0.35) -- (1.5,-0.35);						
			\draw[<-,thick] (2.5,-0.35) -- (3,-0.35);
			\draw[->,thick] (3,-0.35) -- (3.5,-0.35);
			\draw[<-,thick] (4.5,-0.35) -- (5,-0.35);
			\draw[->,thick] (5,-0.35) -- (5.5,-0.35);
			\draw[<-,thick] (6.5,-0.35) -- (7,-0.35);
			\draw[color=black,thick] (1,-0.2) -- (1,-0.5);					
			\draw[color=black,thick] (3,-0.2) -- (3,-0.5);
			\draw[color=black,thick] (5,-0.2) -- (5,-0.5);
			\draw[color=black,thick] (7,-0.2) -- (7,-0.5);
			\draw[color=blue] (4,7) node {$+ \hbar k_B$};					
			\draw[color=red] (2,7) node {$\pm \hbar k_R$};				    
			\draw[color=red] (6,7) node {$\pm \hbar k_R$};
			\fill[black] (1,3) circle (0.08);							    
			\fill[black] (3,1.5) circle (0.08);
			\fill[black] (3,4.5) circle (0.08);
			\fill[black] (4,1.5) circle (0.08);
			\fill[black] (4,4.5) circle (0.08);
			\fill[black] (5,2) circle (0.08);
			\fill[black] (5,5) circle (0.08);
			\fill[black] (7,4.5) circle (0.08);
		\end{tikzpicture}
		\label{fig: Subfigure_SDDI}
	}
	\subfloat[Asymmetric Ramsey-Bordé]{
		\centering
		\begin{tikzpicture}[scale=.6]
			\draw[->,thick] (-0.2,0) -- (8,0) node[right] {$\NewCoordsTime$}; 		
			\draw[->,thick] (0,-0.2) -- (0,6.7) node[above] {$z(\NewCoordsTime)$}; 	
			\draw[color=red,dashed] (1,-0.5) -- (1,6.5);				    
			\draw[color=red,dashed] (3,-0.5) -- (3,6.5);
			\draw[color=red,dashed] (5,-0.5) -- (5,6.5);
			\draw[color=red,dashed] (7,-0.5) -- (7,6.5);
			\fill[blue,opacity=0.5] (3.9,0) rectangle ++(0.2,6.5);	
			\draw[color=green,thick] (0.5,1) -- (1,1) -- (3,3.5) -- (4,3.5) -- (5,4.5) -- (7,4) -- (7.5,3.875);	
			\draw[color=green,thick] (1,1) -- (3,1) -- (4,1) -- (5,2) -- (7,4) -- (7.5,4.5);					
			\draw[color=black,thick] (-0.075,1) -- (0.075,1);	
			\draw (-0.35,1) node {$z_0$};						
			\draw (2,-0.35) node {$T_R$};						
			\draw (4,-0.35) node {$T_B$};
			\draw (6,-0.35) node {$T_R$};
			\draw[->,thick] (1,-0.35) -- (1.5,-0.35);		    
			\draw[<-,thick] (2.5,-0.35) -- (3,-0.35);
			\draw[->,thick] (3,-0.35) -- (3.5,-0.35);
			\draw[<-,thick] (4.5,-0.35) -- (5,-0.35);
			\draw[->,thick] (5,-0.35) -- (5.5,-0.35);
			\draw[<-,thick] (6.5,-0.35) -- (7,-0.35);
			\draw[color=black,thick] (1,-0.2) -- (1,-0.5);		
			\draw[color=black,thick] (3,-0.2) -- (3,-0.5);
			\draw[color=black,thick] (5,-0.2) -- (5,-0.5);
			\draw[color=black,thick] (7,-0.2) -- (7,-0.5);
			\draw[color=blue] (4,7) node {$+ \hbar k_B$};		
			\draw[color=red] (2,7) node {$\pm \hbar k_R$};		
			\draw[color=red] (6,7) node {$\pm \hbar k_R$};
			\fill[black] (1,1) circle (0.08);				    
			\fill[black] (3,1) circle (0.08);
			\fill[black] (3,3.5) circle (0.08);
			\fill[black] (4,1) circle (0.08);
			\fill[black] (4,3.5) circle (0.08);
			\fill[black] (5,2) circle (0.08);
			\fill[black] (5,4.5) circle (0.08);
			\fill[black] (7,4) circle (0.08);
		\end{tikzpicture}
		\label{fig: Subfigure_ARBI}
	}
	\caption{
	Schematic pictures of atomic trajectories (green lines) for three different IF geometries in the freely falling frame. Bragg pulses are depicted in red dashed with a momentum transfer of $\pm \hbar k_R$ and Bloch oscillations in blue with a momentum transfer of $\hbar k_B$. Finite speed of light effects are neglected in this picture. (a) Symmetric Ramsey-Bordé Interferometer (SRBI), (b) Symmetric Double Diffraction Interferometer (SDDI), (c) Asymmetric Ramsey-Bordé Interferometer (ARBI). Note that (a) and (c) can be realized using single Bragg diffraction, whereas (b) relies on double Bragg diffraction.}
	\label{Fig: Interferometer_Geometries}
\end{figure*}
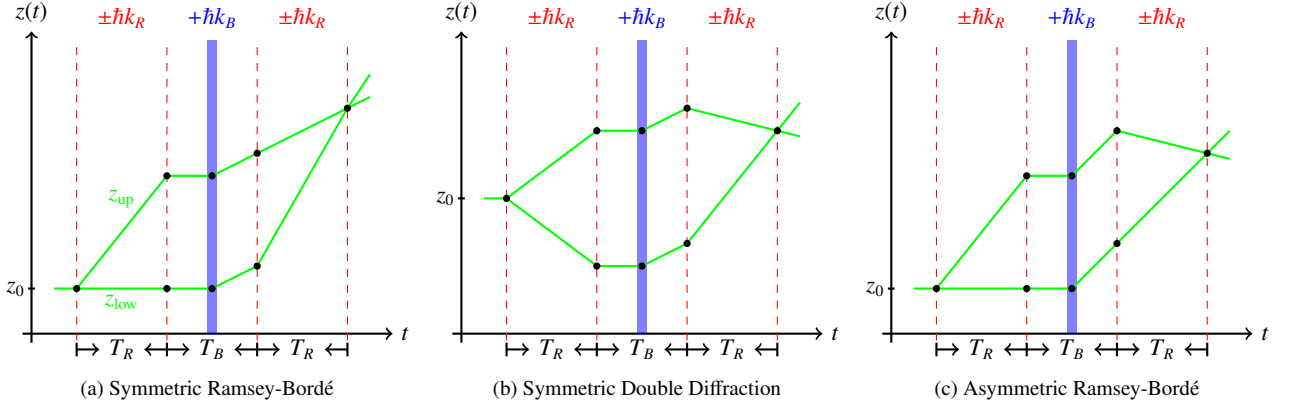

In this section, we solve the Schrödinger equation for a range of IF geometries, which involve elastic scattering processes. We illustrate our approach using three specific atom IF geometries, shown in Fig.~\ref{Fig: Interferometer_Geometries}. We will consider COM motion along the vertical direction. In the following we denote the vertical coordinate --- i.e., that pointing (radially) upwards in Earth's gravitational field --- by $Z$. For the earth-bound IF geometries considered in the following we assume height differences $\Delta Z$ on the order of $10\, \mathrm{m}$, as realized in~\cite{dickerson2013multiaxis, schilling2020vertical, zhou2011development}, up to several $100\, \mathrm{m}$ baselines, as currently under investigation~\cite{badurina2020aion, zhan2020zaiga}. Therefore, it will be appropriate to expand the gravitational potential at height $Z$ above ground as
\begin{align}
	\phi \Big\rvert_\text{height $Z$} 
	= \phi_0 + g \, Z - \frac{1}{2} \Gamma \, Z^2 + \frac{1}{3} \Lambda Z^3,
\end{align}
with $\phi_0$ as before, linear gravitational acceleration $g$, gravity gradient $\Gamma$ and the second-order gradient $\Lambda$. We choose to expand $\phi(\NewCoordsVec)$ to third order to facilitate a comparison of our results with Refs.~\cite{dimopoulos2008general, dimopoulos2007PRL}.It also allows us to include gravitational effects from other source masses, either due to test masses as in Refs.~\cite{overstreet2022observation, rosi2014precision} or due to mass inhomogeneities, such as studied for the \emph{Very Long Baseline Atom Interferometer} (VLBAI)~\cite{lezeik2022understanding}. 
Using this expansion, the Hamiltonian in Eq.~\eqref{eq: COM Hamiltonian} for COM motion along the vertical axis becomes
\begin{multline}
	\hat{H}_\HamiltonianMotional = m g \ZZ - \frac{m}{2} \Gamma \ZZ^2 + \frac{m}{3} \Lambda \ZZ^3 + \frac{\PZ^2}{2m} \\
	+ \frac{1}{m c^2} \Bigg[ \frac{2 \gamma+1}{2} g \PZ \ZZ \PZ - \frac{\PZ^4}{8 m^2} + \frac{2 \beta-1}{2} m^2 g^2 \ZZ^2 \\
	+ 2(\beta-1) m^2 \phi_0 g \ZZ \Bigg] + \mathcal{O}(\Gamma \, c^{-2}) \label{eq: New_Motional_Hamiltonian_1D}.
\end{multline}

In the IF geometries considered here, we assume atoms are initialized in a wave packet which is localized at coordinate height $z_0$ and has vertical coordinate velocity $v_0$. The IF sequences consist in each case of an initial and a final light pulse splitting and recombining the wave packet in momentum by Bragg diffraction. After the first beam splitter, a free propagation for a (Ramsey) time $T_R$ and another Bragg pulse, we consider a time $T_B$ in which (optional) Bloch oscillations accelerate both wave packet components. Subsequently, the atoms, again, interact with a Bragg pulse, propagate freely for a time $T_R$, after which the wave packet is recombined in a final Bragg pulse. The time decomposition ($T_R, T_B, T_R$) is chosen symmetrically to obtain compact final results, yet it is in principle straightforward to describe asymmetrical pulse sequences. This general class of IF geometries corresponds to the SRBI and ARBI like in~\cite{loriani2019interference} with intermediate Bloch oscillations as performed in the fine-structure measurements~\cite{parker2018measurement, morel2020determination} and a similar geometry using double Bragg diffraction~\cite{giese2013double, ahlers2016double}. These example IF geometries can be considered as representatives of different IF classes that exhibit different symmetry axes. Investigating how the different symmetries are reflected in the phase shift results might then be an interesting study in the future.

In solving the Schrödinger equation for geometries and pulse sequences as shown in Fig.~\ref{Fig: Interferometer_Geometries}, we approximate the light pulses as instantaneous on the scale of the Ramsey time $T_R$ \footnote{Note that the notion of simultaneity is well defined because of the metric being static.}. Relativistic contributions to phases imparted during the free propagation times $T_R$ will be evaluated in Subsec.~\ref{subsec:FreeProp}, those due to the Bragg pulses in Subsec.~\ref{subsec:BSPulses}. Finally, we assume that phases imprinted by the Bloch pulses, are common mode and therefore do not contribute to the IF signal. Accounting for differential phases (due to gravity gradients or relativistic corrections) accumulated during Bloch pulses is beyond the scope of this work.

\subsection{Free propagation}\label{subsec:FreeProp}

During free propagation the atoms remain in the ground state $\ket{g}$, and we need to solve the Schrödinger equation with the Hamiltonian in Eq.~\eqref{eq: New_Motional_Hamiltonian_1D} for the COM wave function $\psi(z,\NewCoordsTime)$. For wave packets which stay localized around the classical trajectories corresponding to the IF paths in Fig.~\ref{Fig: Interferometer_Geometries}, it is sufficient to determine the \emph{propagation phase} --- i.e., the relative phase accumulated along the two paths. Following the stationary phase approximation one can approximate the propagation phase by the difference of the action functional along the upper trajectory $z_\text{up}(\NewCoordsTime)$ and the lower trajectory $z_\text{low}(\NewCoordsTime)$~\cite{Kasevich_Chu, storey1994feynman}
\begin{align}
	\Delta \Phi_\text{Prop} = \frac{1}{\hbar} \int \left[ L_\HamiltonianMotional(z_\text{up}(\NewCoordsTime)) - L_\HamiltonianMotional(z_\text{low}(\NewCoordsTime)) \right] \dd \NewCoordsTime.
	\label{eq: Propagation phase}
\end{align}
$L_\HamiltonianMotional$ is the Lagrangian corresponding to Eq.~\eqref{eq: New_Motional_Hamiltonian_1D} whose explicit form can be found in App.~\ref{Appendix: Euler-Lagrange equations}. We point out that Eq.~\eqref{eq: Propagation phase} for the propagation phase only applies exactly for Lagrange functions, which depend quadratically on position and velocity. For a more general Lagrangian (cf. Eq.~\eqref{eq:RelLagrangian}) however, Eq.~\eqref{eq: Propagation phase} only applies under a stationary phase approximation. In the following we will perform a systematic expansion of the propagation phase with respect to the small, non-quadratic terms of the Lagrangian. Corrections beyond the stationary phase approximation would only contribute in a higher and negligibly small order.

In order to find the classical trajectories of the upper and lower arm of the IF we need to solve the geodesic equation, or equivalently, the Euler-Lagrange equation (ELE) corresponding to Eq.~\eqref{eq: New_Motional_Hamiltonian_1D} which is given explicitly in App.~\ref{Appendix: Euler-Lagrange equations}. Rather than seeking the exact trajectories, we construct approximate solutions of the ELE which are accurate in leading orders of $\Gamma$, $\Lambda$ and $c^{-1}$. These approximations can be performed systematically by rewriting the ELE in a dimensionless form in terms of scaled time and position variables
\begin{align}
    \DimTime &= \frac{\NewCoordsTime}{T_R}, & 
    \xi(\DimTime) &= \frac{z(\DimTime)}{c T_R}.
\end{align}
Here, we choose $T_R$ as the natural time scale of the IF sequences under consideration. In terms of these new variables, the ELE depends on a number of dimensionless parameters, summarized in Table~\ref{fig: Definition_of_dimensionless_parameters}. They are connected to the gravitational potential, initial conditions for the atomic wave packet and momentum recoils from light pulses. In terms of the dimensionless variables $\DimTime$ and $\xi$ and the dimensionless parameters of Table~\ref{fig: Definition_of_dimensionless_parameters}, the ELE reads 
\begin{align}
    \ddot{\xi}(\DimTime) &= - \GOneR + \GTwoR \xi(\DimTime) - \GThreeR \xi(\DimTime)^2 \nonumber \\
    & \quad - 2 (\beta + \gamma) \GOneR^2 \xi(\DimTime) + (\gamma + 2) \GOneR \dot{\xi}(\DimTime)^2 \nonumber \\
    & \quad+ 2 (\beta - 1) \GZero \qty[ - \GOneR + \GTwoR \xi(\DimTime)] + \mathcal{O}(4),
\end{align}
as shown in App.~\ref{Appendix: Euler-Lagrange equations}. Throughout this manuscript we use the notation $\mathcal{O}(n)$ to collect terms which are at least of $n$-th order in the small dimensionless parameters in Table~\ref{fig: Definition_of_dimensionless_parameters}. 
If we denote each time instance of light-matter interactions along the paths by $\DimTime_i$ with $i \in \mathbb{N}$, as indicated by the black dots in Fig.~\ref{Fig: Interferometer_Geometries}, we can define $\xi(\DimTime)$ piecewise by a set of functions $\xi_i(\DimTime)$ on the time intervals $\left[ \DimTime_i, \DimTime_{i+1} \right)$. A unique solution therefore requires the knowledge of two initial conditions for each segment of propagation. They can be constructed iteratively for $i > 0$ via
\begin{align}
    \xi_{i}(\DimTime_{i}) &= \xi_{i-1}(\DimTime_{i}), & 
    \dot{\xi}_{i}(\DimTime_{i}) &= \dot{\xi}_{i-1}(\DimTime_{i}) + N^{(i)}_R \RR + N^{(i)}_B \RB, \label{eq: Atomic boundary conditions}
\end{align}
whereas the first two initial conditions are
\begin{align}
    \xi_0(\DimTime_0) &= \ZZero, & \dot{\xi}_0(\DimTime_0) &= \VZero + N^{(0)}_R \RR + N^{(0)}_B \RB.
\end{align}
$N^{(i)}_R$ and $N^{(i)}_B$ denote the number of photon recoils imprinted at time $\DimTime_i$ via the Bragg or Bloch interaction respectively and will be discussed later. The dot in Eq.~\eqref{eq: Atomic boundary conditions} indicates differentiation w.r.t.~$\DimTime$. Note that $\ZZero$ denotes the initial height of the atoms compared to the reference position of the laser/clock, see Sec.~\ref{Sec: Transformation_to_new_coords}. This parameter can be significant in differential measurements between different IFs --- as for example in UFF tests, where an uncertainty of the relative difference in initial height will lead to relevant phase contributions~\cite{loriani2020colocation}. Despite its implementation in our algorithm~\cite{PythonCode}, we set $\ZZero = 0$ in the following for simplicity.
\begin{table}
\centering
\begin{tabularx}{0.95\linewidth}{|c|c|Y|Y|}
	\hline
	Parameter & Definition & \multicolumn{2}{|c|}{Magnitude for 10\,m baseline IF} \\
	\cline{3-4}
	& & $i = R$ (Bragg) & $i = B$ (Bloch) \\
	\hhline{|=|=|=|=|}
    $\ZZero$ & $\frac{z_0}{c T_R}$ & \multicolumn{2}{|c|}{0} \\
    \hline
    $\VZero$ & $\frac{v_0}{c}$ & \multicolumn{2}{|c|}{$4.3 \times 10^{-8}$} \\
    \hline
    $\FreqR$ & $\frac{\hbar \omega_R}{m c^2}$ & \multicolumn{2}{|c|}{$8.1 \times 10^{-20}$} \\
    \hline
    $\GZero$ & $\frac{\phi_0}{c^2}$ & \multicolumn{2}{|c|}{$7 \times 10^{-10}$} \\
    \hline
    $\mathcal{G}_{1,i}$ & $\frac{g T_i}{c}$ & $4.2 \times 10^{-8}$ & $1.3 \times 10^{-8}$ \\
    \hline
    $\mathcal{G}_{2,i}$ & $\Gamma T_i^2$ & $5.2 \times 10^{-6}$ & $4.9 \times 10^{-7}$ \\
    \hline
    $\mathcal{G}_{3,i}$ & $\Lambda c T_i^3$ & $4.8 \times 10^{-4}$ & $1.4 \times 10^{-5}$ \\
    \hline
    $\mathcal{R}_i$ & $ \frac{\hbar k_i}{m c}$ & $3.9 \times 10^{-11}$ & $1.2 \times 10^{-9}$ \\
    \hline
\end{tabularx}
\caption{Definitions of dimensionless parameters for an atom IF based on elastic scattering processes. As a case study, we give the magnitudes for a $10 \, \mathrm{m}$ baseline Rb IF assuming the following values: $T_R = 1.3 \, \mathrm{s}, \, T_B = 0.4 \, \mathrm{s}, \, z_0 = 0, \, v_0 = 13 \, \mathrm{m}/\mathrm{s}, \, m = 87 \, \mathrm{u}, \, \omega_R = 10^7 \, \mathrm{Hz}, \, k_R = 16 \times 10^6 \, \mathrm{m}^{-1}$ and $k_B = 5 \times 10^8 \, \mathrm{m}^{-1}$.}
\label{fig: Definition_of_dimensionless_parameters}
\end{table}

Using the example of a drop tower with $10 \, \mathrm{m}$ height, we show in Table~\ref{fig: Definition_of_dimensionless_parameters} that the dimensionless parameters are small in case of the IFs in Fig.~\ref{Fig: Interferometer_Geometries}. We exploit this fact to consistently construct an approximate solution of the ELE. For example, the trajectory of the path segment starting at $\DimTime_0$ is, up to second order in these small parameters, given by
\begin{align}
	\xi(\DimTime) &= \xi(\DimTime_0) + \dot{\xi}(\DimTime_0) \DimTime - \frac{1}{2} \qty(1 + 2(\beta - 1) \GZero) \GOneR \DimTime^2 \nonumber \\
    & \quad + \frac{\GTwoR}{2} \qty(\xi(\DimTime_0) \DimTime^2 + \frac{1}{3} \dot{\xi}(\DimTime_0) \DimTime^3 - \frac{1}{12} \GOneR  \DimTime^4)  + \mathcal{O}(3).
	\label{eq: Trajectory}
\end{align}
Note that corrections induced by the third-order expansion of the gravitational potential --- i.e. $\GThreeR$ --- would only manifest at $\mathcal{O}(3)$. With this solution strategy, we calculate each of the IF paths --- i.e., $\xi_\text{up}(\DimTime)$ and $\xi_\text{low}(\DimTime)$ for all three IF geometries. The propagation phase in Eq.~\eqref{eq: Propagation phase} can now be calculated in dimensionless form using those trajectories 
\begin{align}
	\Delta \Phi_\text{Prop} = \Comp T_R \int \left[ \frac{L_\HamiltonianMotional(\xi_\text{up}(\DimTime))}{m c^2} - \frac{L_\HamiltonianMotional(\xi_\text{low}(\DimTime))}{m c^2} \right] \dd \DimTime,
	\label{eq: Propagation phase dimensionless}
\end{align}
where we introduced the atomic Compton frequency 
\begin{align}
    \Comp = \frac{m c^2}{\hbar} . \label{eq:Compton}
\end{align}

The propagation phase has to be consistently expanded in terms of the small dimensionless parameters up to the desired order. If the trajectory is known to $\mathcal{O}(2)$ one can calculate the propagation phase to $\mathcal{O}(3)$, since the Lagrangian depends on the velocity to second order and on the trajectory to first order after multiplication with $\GOneR$ and additional higher-order terms, see App.~\ref{Appendix: Euler-Lagrange equations} for details. Similarly, the propagation phase can be evaluated to $\mathcal{O}(4)$ by first determining the trajectory to order $\mathcal{O}(3)$. The required integrals over the segments of the full IF sequence are in principle straightforward, but result tedious and error-prone calculations. We perform these calculations to $\mathcal{O}(4)$ using computer algebra based on \texttt{sympy} in Python~\cite{PythonCode}. The results for trajectories are too lengthy to be reproduced here, but are given in~\cite{PythonCode}.

So far, we have only considered the propagation phase in Eq.~\eqref{eq: Propagation phase dimensionless}. However, taking into account the gravity gradient $\Gamma$ and its spatial derivative $\Lambda$, one encounters the problem that the IF will not exactly close at the output port without special mitigation schemes \cite{Roura2017Circumventing, Overstreet2018Effective, DAmico2017Canceling}. In this case one also has to take into account the \emph{separation phase}, which is usually obtained by taking the spatial separation at the output port multiplied by the average momentum of the two atomic ports~\cite{hogan2008light, dimopoulos2008general}. Written directly in terms of the dimensionless trajectory from before we can denote the output port separation as $\Delta \xi = \xi_\text{l}(\DimTime_\text{f}) - \xi_\text{u}(\DimTime_\text{f})$ and the average output velocity as $\dot{\xi}_\text{aver.} = \frac{1}{2} \left( \dot{\xi}_\text{l}(\DimTime_\text{f}) + \dot{\xi}_\text{u}(\DimTime_\text{f}) \right)$, such that the separation phase can be written as
\begin{align}
	\Delta \Phi_\text{Sep} &= \Comp T_R \Delta \xi \cdot \dot{\xi}_\text{aver.}. \label{eq: Separation_Phase_and_Compton_Frequency}
\end{align}
The quantities $\Delta \xi$ and $\dot{\xi}_\text{aver.}$ need to be evaluated to the desired order to obtain the separation phase. Since trajectory separations only appear due to non-linear gravitational effects, which manifests at second order in the trajectory, one should note that $\Delta \xi$ is also at least $\mathcal{O}(2)$ in the absence of any mitigation schemes~\cite{Roura2017Circumventing}. Since $\dot{\xi}_\text{aver.}$ is an $\mathcal{O}(1)$-term, the separation phase overall is of $\mathcal{O}(3)$. Additional relativistic $\mathcal{O}(c^{-2})$-corrections to $\Delta \Phi_\text{Sep}$ will be at least of order $\mathcal{O}(5)$, see Table~\ref{fig: Definition_of_dimensionless_parameters}.

Note that due to the finite speed of light (FSL) the interaction times between the upper and lower interferometry path will differ slightly and therefore alter the integration times in the propagation phase. FSL effects like those arise naturally at $\mathcal{O}(c^{-1})$ --- i.e. any additional gravitational effects would manifest at $\mathcal{O}(c^{-3})$. Furthermore, FSL related effects are highly dependent on the actual experimental realization. In particular, they are determined by the initial laser positions, additional mirrors or other optical elements. The calculation of FSL effects thus requires various experiment-specific parameters, each of which would significantly complicate the notation presented here. Conversely, assuming constraints on FSL effects and light paths requires selecting specific experimental configurations, contradicting our goal to provide a general framework. This is why, for now, we have chosen to omit FSL terms. However, we will compare our results to the Mach-Zehnder IF from~\cite{dimopoulos2008general} in App.~\ref{Appendix: Mach-Zehnder-Interferometer} and demonstrate how to add the dominant FSL phases, in the case of an explicitly given experimental setup, by hand. We refer the interested reader to the paper of Tan et al.~\cite{tan2016finite} analyzing contributions of FSL effects in a very general context as well as to~\cite{peters2001high, DiPumpo2023T3Scaling} for direct applications in atom IFs. An implementation of the full FSL effect into our algorithm for given experimental setups is going to be addressed in future work.

\subsection{Maxwell equations in gravity}

Following Refs.~\cite{dimopoulos2008general, wajima1997post}, so far we have considered only relativistic corrections of atomic degrees of freedom. Now, we proceed to analyze the EM field and its interaction with atoms in the PPN curved spacetime. The logic here will be to first identify the eigenmodes of the relativistically corrected wave equation for the light field, and then in the next section to use these when describing the interaction of atoms with a coherent laser field in a semiclassical approximation --- namely considering a coherent laser field in (some of) these eigenfunctions coupled to an atom via $H_\HamiltonianAtomLight$ in Eq.~\eqref{eq: Atom Light Hamiltonian}.  An approach that is similar to ours has been taken in~\cite{diPumpo2022LightPropagation}.

The most straightforward way to obtain the Maxwell equations in general coordinates is to start from the Lagrangian and use the variational principle. The Lagrangian for an EM field in vacuum with EM field strength tensor $F_{\mu \nu}$ is given by
\begin{align}
	L_\HamiltonianLightField = \int \frac{\sqrt{-\NewCoordsg}}{4 \mu_0} F_{\mu \nu} F^{\mu \nu} \, \dd^3 \bm{x},
\end{align}
where $g$ is the determinant of the matrix of spacetime metric components, and the field strength tensor is taken as derived from a vector potential $A_\mu$ via $\tensor{F}{_{\mu\nu}}= \tensor{A}{_{\nu ; \mu}} - \tensor{A}{_{\mu ; \nu}} = \tensor{A}{_{\nu ,\mu}} - \tensor{A}{_{\mu , \nu}}$ (thus, the homogeneous Maxwell equations $\dd F = 0$, or in components $F_{[\mu\nu;\rho]} = F_{[\mu\nu,\rho]} = 0$, are automatically satisfied). The variational principle, varying with respect to $A_\mu$, then directly leads to the inhomogeneous Maxwell equations (in our case with vanishing source)
\begin{align}\label{eq:MEQ}
	\tensor{F}{^\alpha ^\beta_{;\beta}} = 0.
\end{align}

We solve Eqs.~\eqref{eq:MEQ} in Lorenz gauge $\tensor{A}{_\mu^{;\mu}} = 0$, see App.~\ref{Appendix: Electromagnetism} for details.  Note that for the solutions of interest to us and to our order of approximation, the Lorenz gauge agrees with the curved-spacetime generalization $\tensor{A}{_i^{;i}} = 0$ of the Coulomb gauge.  The solution for a light field propagating in the $z$-direction results in a geometric optics approximation of the potential $A_\mu$ with $A_0 =0$ and a covariant spatial part $\bm{A}= (A_i)_i$ of the form
\begin{align}
	\bm{A}= \bm{\mathcal{A}} \, e^{\ii \Phi(z, \NewCoordsTime)} + \mathcal{O} ( \Gamma \, c^{-2} ). \label{eq: Solution to Maxwell}
\end{align}
Here, $\bm{\mathcal{A}} = (\mathcal{A}_x, \mathcal{A}_y, 0)$ are the amplitudes and 
\begin{align}
	\Phi(z, \NewCoordsTime) = k_0 c \NewCoordsTime \pm \left( 1 - \frac{\gamma + 1}{2} \frac{g z}{c^2} \right) k_0 z + \mathcal{O}(\Gamma \, c^{-2}) \label{eq: Phase_general_expression}
\end{align}
is the phase of the EM field, where the sign depends on the direction of propagation. $k_0$ is the zeroth component of the covariant 4-wave vector and we have omitted dependencies on the transverse coordinates $x,y$ in a plane wave approximation. The derivation of Eqs.~\eqref{eq: Solution to Maxwell} and \eqref{eq: Phase_general_expression} can be found in App.~\ref{Appendix: Electromagnetism}.  

The effective wave vector of the EM field in the relevant (1+1)-dimensional $(c \NewCoordsTime, z)$-spacetime is
\begin{align}
	(k_\mu (z)) = (\partial_\mu \Phi(z, \NewCoordsTime)) = 
	\begin{pmatrix} 
		1 \\ \pm \left( 1 - (\gamma + 1) \frac{g z}{c^2} \right)
	\end{pmatrix} k_0
	+ \mathcal{O} ( \Gamma \, c^{-2}). \label{eq: general 4-wave vector with k_0}
\end{align}
We note that the component $k_0$ is a coordinate dependent quantity and has, a priori, no direct physical meaning. To interpret $k_0$ one needs to convert it into a measurable frequency, which requires the notion of a local observer. For this, we consider an observer resting at height $z$ above Earth's surface, defined by its 4-velocity $\qty(u^\mu_\text{obs}(z))$. Since 4-velocities are normalized to $u^\mu u_\mu = -c^2$ it follows that
\begin{align}
	\qty(u^\mu_\text{obs}(z)) = \left( 1 - \frac{g z}{c^2} + \mathcal{O} (\Gamma \, c^{-2} ) \right) 
	\begin{pmatrix} 
	    c \\ 0 
	\end{pmatrix}.
\end{align}
The frequency of the light field as measured by this observer at rest at height $z$ is then
\begin{align}
	\omega (z) &= - k_\mu (z) u^\mu_\text{obs}(z) = - \left(1 - \frac{g z}{c^2} + \mathcal{O} (\Gamma \, c^{-2} ) \right) c k_0,
\end{align}
which may be interpreted as the gravitational redshift as witnessed by observers at rest at different heights. In the following, we will assume that this observer is located at $z = 0$ and equipped with a clock and a laser of frequency $\omega_0 = \omega(0)$, such that $k_0 = -\frac{\omega_0}{c}$. One may think here of an atomic clock and a laser which is stabilized to its reference frequency. 

\subsection{Light-matter interactions and the laser phase}\label{subsec:BSPulses}

We are now in a position to analyze the atom-light interaction in order to describe beam splitter and mirror pulses including relativistic corrections. Coming back to the interaction Hamiltonian in Eq.~\eqref{eq: Atom Light Hamiltonian}, we see that there are no direct gravitational effects apparent on the Hamiltonian level. However, since the EM fields themselves are gravitationally altered, we need to analyze how those corrections might affect beam splitter and mirror pulses.

\subsubsection{Bragg interactions: Two counter-propagating light fields}

We start by considering relativistically altered Bragg transitions which realize the mirror and beam splitter operations in Fig.~\ref{Fig: Interferometer_Geometries}. For this we will describe the two counter-propagating light beams aligned in the $z$-direction. The light fields are assumed to have frequencies $\omega_{R, 1}$ and $\omega_{R, 2}$ and wave vectors $k_{R, i} = \frac{\omega_{R,i}}{c} > 0$. We denote the `effective' frequency and wave vector as $\omega_R = \omega_{R,1} - \omega_{R,2}$ and $k_R = k_{R,1} + k_{R,2}$ and assume that those parameters induce a resonant two-photon process. The index $R$ refers to the Ramsey sequence opened and closed by Bragg pulses, cf. Fig.~\ref{Fig: Interferometer_Geometries}, and follows the notation used, e.g., by Morel et al. in~\cite{morel2020determination}. Later on, in Sec.~\ref{Bloch_Oscillations_description}, we will consider also light pulses transferring the momentum $k_B$ via Bloch oscillations.

In principle, the dynamics during a light pulse requires to solve the Schrödinger equation with the Hamiltonian in Eq.~\eqref{eq: Atom Light Hamiltonian}, which reads in one-dimensional form
\begin{align}
	\hat{H}_\HamiltonianAtomLight = - \hat{\bm{d}} \cdot \hat{\bm{E}}(\ZZ) + \frac{1}{2m} \left[ \PZ \left( \hat{\bm{d}} \times \hat{\bm{B}}(\ZZ) \right)_z + \text{h.c.} \right]
\end{align}
with position and momentum operators $\ZZ, \PZ$. This calculation is deferred to App.~\ref{Appendix: Atom Light Hamiltonian}, and we only summarize its result in the following.  We first discuss the case of atoms at rest, and include effects due to Doppler shifts in a second step. The $\ZZ$-dependence of the EM fields, which follow from the vector potential in Eq.~\eqref{eq: Solution to Maxwell}, result in $\ZZ$-dependent corrections the EM phases. This height dependence of the phase will be of relevance for the net interferometric phase, and will be the focus in the following.

The wave vector of each laser will be gravitationally altered according to
\begin{align}
    \kappa_i(z) = \pm \qty(1 - (\gamma + 1) \frac{g z}{c^2} ) k_{R, i} + \mathcal{O}(\Gamma \, c^{-2}). \label{eq: Effective_k_vector_Not_Doppler_Shifted}
\end{align}
Hence, in order to achieve a resonant transition we aim to transfer only the desired momentum of $k_{R,i}$ upon interaction. One therefore needs to detune the each laser s.t. $\kappa(z_\text{int}) = \pm k_{R, i}$, where $z_\text{int}$ is the interaction height between the light pulse and (a component of) the atomic wave packet on the respective path --- i.e., $k_{R, i} \longmapsto \qty(1 + (\gamma + 1) \frac{g z_\text{int}}{c^2} )k_{R, i}$. 

The effective laser phase  imprinted on the atoms in a two-photon process at the interaction height $z_\text{int}$ is given by 
\begin{align}
	\Phi_L(z_\text{int}) = \pm \left(1 + \frac{\gamma + 1}{2} \frac{g z_\text{int}}{c^2} \right) k_R z_\text{int} + \Phi_\text{FSL} + \mathcal{O}(\Gamma \, c^{-2}), \label{eq: Effective_Laser_Phase_Not_Doppler_Shifted}
\end{align}
where the sign corresponds to a net gain or loss in momentum, respectively. We absorb FSL effects in $\Phi_\text{FSL}$, and refer to App.~\ref{Appendix: Atom Light Hamiltonian} for details. 

As shown in App.~\ref{Appendix: Atom Light Hamiltonian}, the scattering matrix for a Bragg diffraction transferring a momentum $N \hbar k_R$ reads
\begin{align}
	U^{(\theta)}(z_\text{int}) = \frac{1}{\sqrt{2}}
	\begin{pmatrix}
		\cos(\theta) & \ii \sin(\theta) e^{\ii N \Phi_L(z_\text{int})} \\
		\ii \sin(\theta) e^{-\ii N \Phi_L(z_\text{int})} & \cos(\theta)
	\end{pmatrix}, \label{eq: Scattering_Matrix_Bragg}
\end{align}
in the basis of the momentum eigenstates $\ket{0 \hbar k_R}$ and $\ket{N \hbar k_R}$. In the experiment, the angle $\theta$ is controlled via pulse intensities and durations~\cite{mueller2008Atom-wave, siemss2020analytic}, and is tuned to $\pi/2 + n \pi$ for a beam splitter and $\pi + 2 n \pi$ for a mirror pulse with $n \in \mathbb{Z}$. Note that due to the various relativistic effects in the Rabi frequency and the detuning, this angle $\theta$ will, in principle, also depend on position and momentum, but at an insignificant level as argued above.

\subsubsection{Doppler effect and laser phase}

To address the Doppler effect, we must examine the frequencies of the two light fields in the atoms' rest frame. Assuming that the atoms have a velocity of $v_\text{int}$ upon interacting with these light fields, we can deduce that the atoms will undergo both first and second-order Doppler shifts, as detailed in Appendix~\ref{Appendix: Atom Light Hamiltonian}. In order to compensate this shift, one needs to, similarly to the gravitational detuning before, rescale the laser frequencies beforehand, according to
\begin{subequations}
\begin{align}
    \omega_{R, 1} &\longmapsto \qty(1 + \frac{v_\text{int}}{c} - \frac{v_\text{int}^2}{2c^2}) \, \omega_{R, 1}, \\ 
    \omega_{R, 2} &\longmapsto \qty(1 - \frac{v_\text{int}}{c} - \frac{v_\text{int}^2}{2c^2}) \, \omega_{R, 2}.
\end{align}
\label{eq: Detuned_frequencies}
\end{subequations}
The atoms then interact with light fields of appropriate frequencies --- i.e. $\omega_{R, i}$, as measured in their own frame of reference --- in order to resonantly stimulate the desired two-photon transition with a momentum kick of $\hbar k_R$. From this point onward, consider the frequencies $\omega_{R, i}$ as the values already detuned as described in Eqs.~\eqref{eq: Detuned_frequencies}. The imprinted phase Eq.~\eqref{eq: Effective_Laser_Phase_Not_Doppler_Shifted} will then be additionally Doppler shifted, and is given by
\begin{multline}
    \Phi_L(z_\text{int}, v_\text{int}) = \pm \bigg(k_R + \frac{v_\text{int}}{c} \frac{\omega_R}{c} - \frac{v_\text{int}^2}{2c^2} k_R \\
    + \frac{\gamma + 1}{2} \frac{g z_\text{int}}{c^2} k_R \bigg) z_\text{int} + \mathcal{O}(\Gamma \, c^{-2}), \label{eq: Effective_Laser_Phase_Atomic_Rest_Frame}
\end{multline}
where we suppressed the FSL correction. For later reference, we rewrite each laser phase contribution in Eq.~\eqref{eq: Effective_Laser_Phase_Atomic_Rest_Frame} in a dimensionless form, similar to the propagation phase Eq.~\eqref{eq: Propagation phase dimensionless} and the separation phase Eq.~\eqref{eq: Separation_Phase_and_Compton_Frequency} 
\begin{multline}
    \Phi_L(\xi_\text{int}, \dot{\xi}_\text{int}) = \pm \Comp T_R \bigg( \bigg(1 - \frac{\dot{\xi}_\text{int}^2}{2} 
    + \frac{\gamma + 1}{2} \GOneR \xi_\text{int} \bigg) \RR 
    \\ + \dot{\xi}_\text{int} \FreqR \bigg) \xi_\text{int} + \mathcal{O}(5), \label{eq: Effective_Laser_Phase_Atomic_Rest_Frame_Dimensionless}
\end{multline}
where we, again, used the Compton frequency from Eq.~\eqref{eq:Compton}. The Doppler term proportional to $\FreqR$ and the recoil term proportional to $\RR$ make a contribution to the laser phase of $\mathcal{O}(3)$ and $\mathcal{O}(4)$, respectively. The terms of $\mathcal{O}(\Gamma \, c^{-2})$ in Eq.~\eqref{eq: Effective_Laser_Phase_Atomic_Rest_Frame} translate to order $\mathcal{O}(5)$. We therefore aim to correctly determine each phase shift contribution consistently to $\mathcal{O}(4)$.

In summary, the overall \emph{relative} laser phase accumulated along the upper and lower IF paths $\xi_\text{up}(\DimTime)$ and $\xi_\text{low}(\DimTime)$, also referred to as kick phase~\cite{loriani2019interference}, is
\begin{align}
	\Delta \Phi_{L,R} &= \sum_i \qty(\Phi_L(\xi_\text{up}(\DimTime_i), \dot{\xi}_\text{up}(\DimTime_i)) - \Phi_L(\xi_\text{low}(\DimTime_i), \dot{\xi}_\text{low}(\DimTime_i)) ). \label{eq: Bragg Kick Phase Formula Abstract}
\end{align}
Here, the sum extends over the time instances $\DimTime_i$ of all Bragg pulses transferring momenta along the two paths, see Fig.~\ref{Fig: Interferometer_Geometries}.

\subsubsection{Atomic velocity after a photon interaction}

Understanding how the Doppler effect and spacetime curvature affect the photon momentum transferred to the atoms undergoing Bragg transitions is necessary to calculate the boundary conditions of the atomic trajectories in Eq.~\eqref{eq: Atomic boundary conditions}. Note that calculating the momentum kicks bears a subtlety: The light field's momentum $\hbar k_\mu$ is a covector, whereas the atomic 4-velocity $u^\mu$ is a (contravariant) vector. Therefore, in order to compute the atomic velocity after the momentum kick, we need to raise the index of $\hbar k_\mu$ using the metric~\footnote{This is also evident from the upper index in Eq. (63) in Dimopoulos et al.~\cite{dimopoulos2008general}.}.

As an example, consider a Bragg pulse interacting with a wavepacket at a height $z_\text{int}$ as before. The additional velocity after the kick will then be given by
\begin{align}
	v_\text{Kick} = \frac{\hbar}{m} \qty( 1 + 2 \gamma \frac{g z_\text{int}}{c^2} ) k_R  + \mathcal{O}(\Gamma c^{-2}).
\end{align}

\subsubsection{Bloch oscillations: Accelerated optical lattices}\label{Bloch_Oscillations_description}

For completeness, we also allow for accelerations of the atomic ensemble common to both IF arms using Bloch oscillations. In the experiment, the atoms are initially loaded into an optical lattice which is then accelerated. After unloading the atoms they have gained an effective momentum, which we will denote by $\pm \hbar k_B$ the sign of the momentum transfer depends on whether momentum was gained in the positive or negative z-direction. We adopt here a highly simplified description by assuming that Bloch oscillations only impart the desired momentum of $\pm \hbar k_B$, that the interaction is infinitely short --- i.e., negligibly short compared to the time scale of the IF --- and that the whole process is lossless. A microscopic description of the underlying physics~\cite{peik1997bloch, fitzek2020universal} and its relativistic corrections are beyond the scope of the current article. Indeed, for the regime of large-momentum transfer~\cite{PhysRevLett.107.130403, PhysRevLett.121.133201}, the theoretical description of Bloch oscillations is the subject of current investigations~\cite{fitzek2023accurate}. In analogy to the case of Bragg pulses treated before, we will denote the imprinted laser phase during one Bloch interaction as $\Phi_{L,B}(z) = \pm k_B z$,  or written dimensionless, as $\Phi_{L,B}(\xi) = \pm \Comp T_R \RB \xi$. Hence, the relative Bloch laser phase can be written in terms of dimensionless quantities as
\begin{align}
	\Delta \Phi_{L,B} &=\sum_i \Phi_{L, B}(\xi_\text{up}(\DimTime_i)) - \Phi_{L, B}(\xi_\text{low}(\DimTime_i)), \label{eq: Bloch Kick Phase Formula Abstract}
\end{align}
where the summation is taken over all interaction times $\DimTime_i$ that imprint a Bloch momentum.

	
\section{Results}\label{Section 4}

So far, we have calculated the three different kinds of phase shift contributions of an atom IF --- namely the propagation phase [Eq.~\eqref{eq: Propagation phase dimensionless}], the separation phase [Eq.~\eqref{eq: Separation_Phase_and_Compton_Frequency}] and the laser phases [Eqs.~\eqref{eq: Bragg Kick Phase Formula Abstract} and~\eqref{eq: Bloch Kick Phase Formula Abstract} for, respectively, Bragg and Bloch pulses] -- all of which include relativistic corrections up to the $\mathcal{O}(c^{-2})$ level. We are now in a position to determine the phase for a given IF geometry, pursuing a number of goals: (i) The phase is to be determined algebraically, similar to the results for the Mach-Zehnder IF in Dimopoulos et al.~\cite{dimopoulos2008general}. (ii) The algebraic expressions shall achieve correctness in $\mathcal{O}(c^{-2})$ and $\mathcal{O}(4)$ in the small parameters from Table~\ref{fig: Definition_of_dimensionless_parameters}. (iii) The routine for calculating the phase should be applicable to a general class of IFs consisting of arbitrary sequences of Bragg and Bloch pulses. 

We consider IFs characterized by (a) a list of $n$ time intervals of free propagation between light pulses and (b) two lists of $n+1$ momenta transferred by Bragg or Bloch pulses in the upper and lower IF paths, see Eq.~\eqref{eq: Atomic boundary conditions}. For an IF geometry defined by these lists, first the classical trajectories --- and from these the propagation, separation and laser phases --- must be calculated to the desired accuracy $\mathcal{O}(c^{-2})$ and $\mathcal{O}(4)$. The required algebra is rather tedious and error prone and therefore relegated entirely to Computer algebra based on Python, see~\cite{PythonCode}. The code exploits that the required arithmetics and integrals can be mapped on list manipulations for keeping track of the small dimensionless parameters in Table~\ref{fig: Definition_of_dimensionless_parameters}. 

In the following, we will discuss the results for the three specific IF geometries shown in Fig.~\ref{Fig: Interferometer_Geometries}, which we refer to as symmetric Ramsey-Bordé interferometer (SRBI), symmetric double diffraction interferometer (SDDI), and asymmetric Ramsey-Bordé interferometer (ARBI). In Table~\ref{Table: All Phases Dimensionless Parameters} we present the main result of this analysis --- i.e. a list of phase shifts of the SRBI, ARBI and SDDI geometries from Figs.~\subref*{fig: Subfigure_SRBI} ${-}$ \subref*{fig: Subfigure_ARBI} where we display all terms of orders $\mathcal{O}(2)$ and $\mathcal{O}(3)$. The higher-order terms of $\mathcal{O}(4)$ are too numerous to be listed, but can be explicitly obtained from~\cite{PythonCode}.

The three $\mathcal{O}(2)$-terms are the well known non-relativistic phases due to linearized gravity and Bragg as well as Bloch recoils, see rows $\# 1 \, {-} \, \# 3$ in Table~\ref{Table: All Phases Dimensionless Parameters}. The 21 $\mathcal{O}(3)$-terms can be grouped as arising from thee main sources: i) The first is caused from PPN terms which result from non-linear gravitational effects as indicated by their dependence on $\GZero$. ii) The second arises by the gravity gradient as one can infer from their dependence on $\GTwoR$ or $\GTwoB$, while iii) the third stems from the Doppler effect, as is evident form the dependence on $\FreqR$. Other $\mathcal{O}(c^{-2})$ or $\Lambda$-dependent phase shifts naturally appear at the $\mathcal{O}(4)$ level. Comparing the phases for the three IF geometries, one can see that most of the terms are identical in the SRBI and ARBI, and differ from the SDDI by a factor of two. This is due to the fact that the enclosed spacetime area in the SDDI is twice as big as in the other two IFs. Connections between the enclosed spacetime area and the IF phases were analyzed in detail by~\cite{mcdonald2013space}. Terms like $\# 2$ and $\# 9$, however, differ quite significantly between the different IF geometries. The first of those terms was described in~\cite{loriani2019interference} via a special relativistic proper time difference, whereas the latter was phrased as a `1st gradient recoil' effect in~\cite{dimopoulos2008general} and was explained in the Appendix F of~\cite{roura2020gravitational}. Phases $\# 19 \, {-} \, \# 24$, which relate to the Doppler effect due to the transferred Bragg momentum $\RR$ also differ between the IF geometries in a non-trivial manner. The phases linear in $\RR$ cancel in the SDDI due to its symmetry, but are non-zero in the ARBI and the SRBI.

\begin{figure*}
    \centering
	\begin{adjustbox}{minipage=0.95\textwidth,frame}
		\centering
        \subfloat{\includegraphics[height=0.6cm]{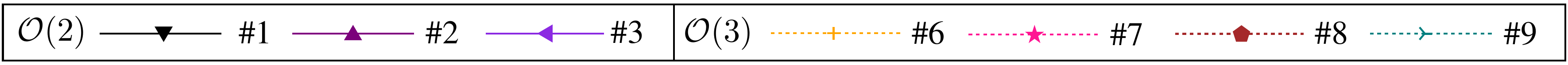}}
        \newline
        \setcounter{subfigure}{0}
		\subfloat[10 meter SRBI]{\includegraphics[width=0.315\textwidth]{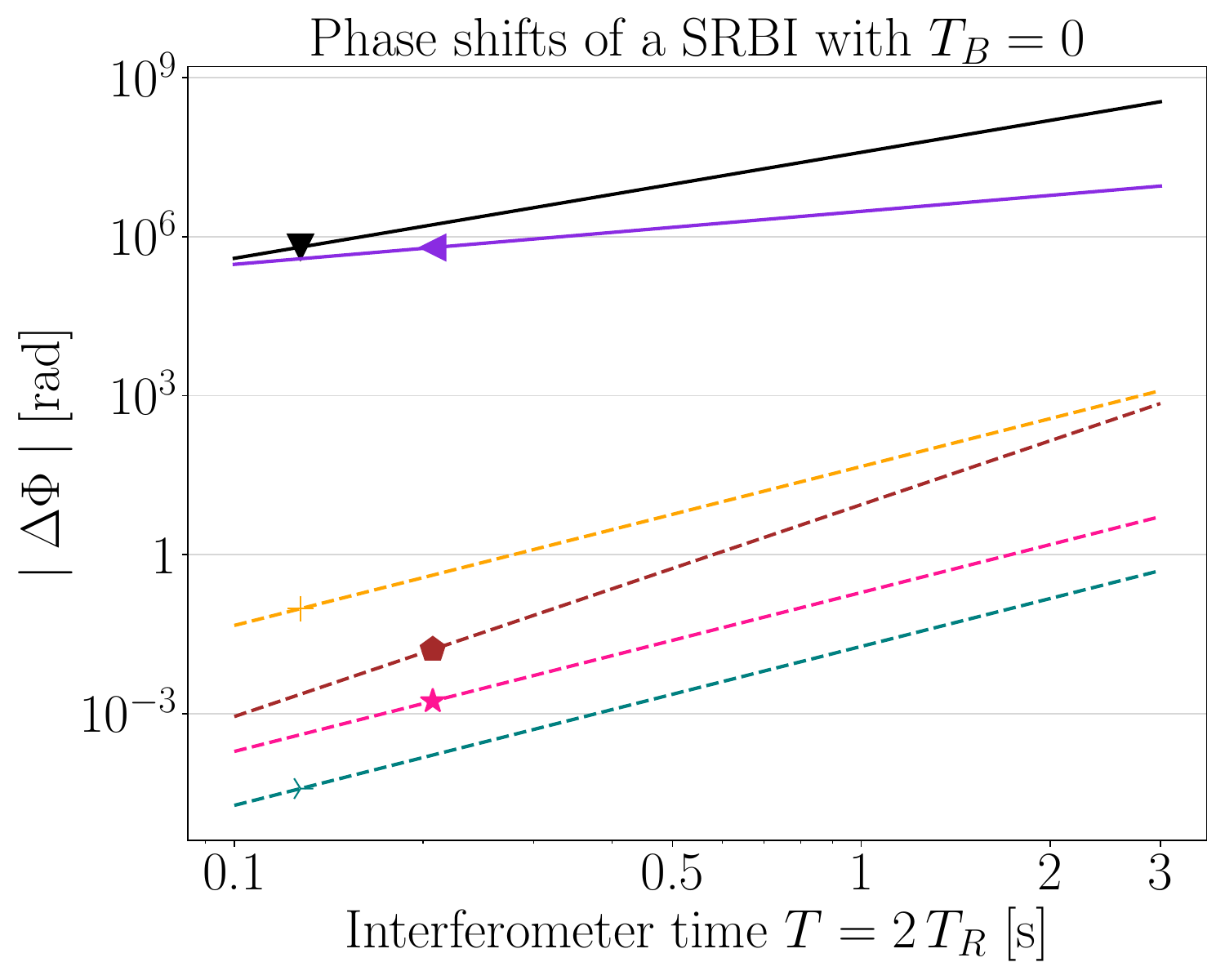} 
		\label{Fig: 10m_Numerical_Plot_SRBI_Without_Bloch}}
		\hfil
		\subfloat[10 meter SDDI]{\includegraphics[width=0.315\textwidth]{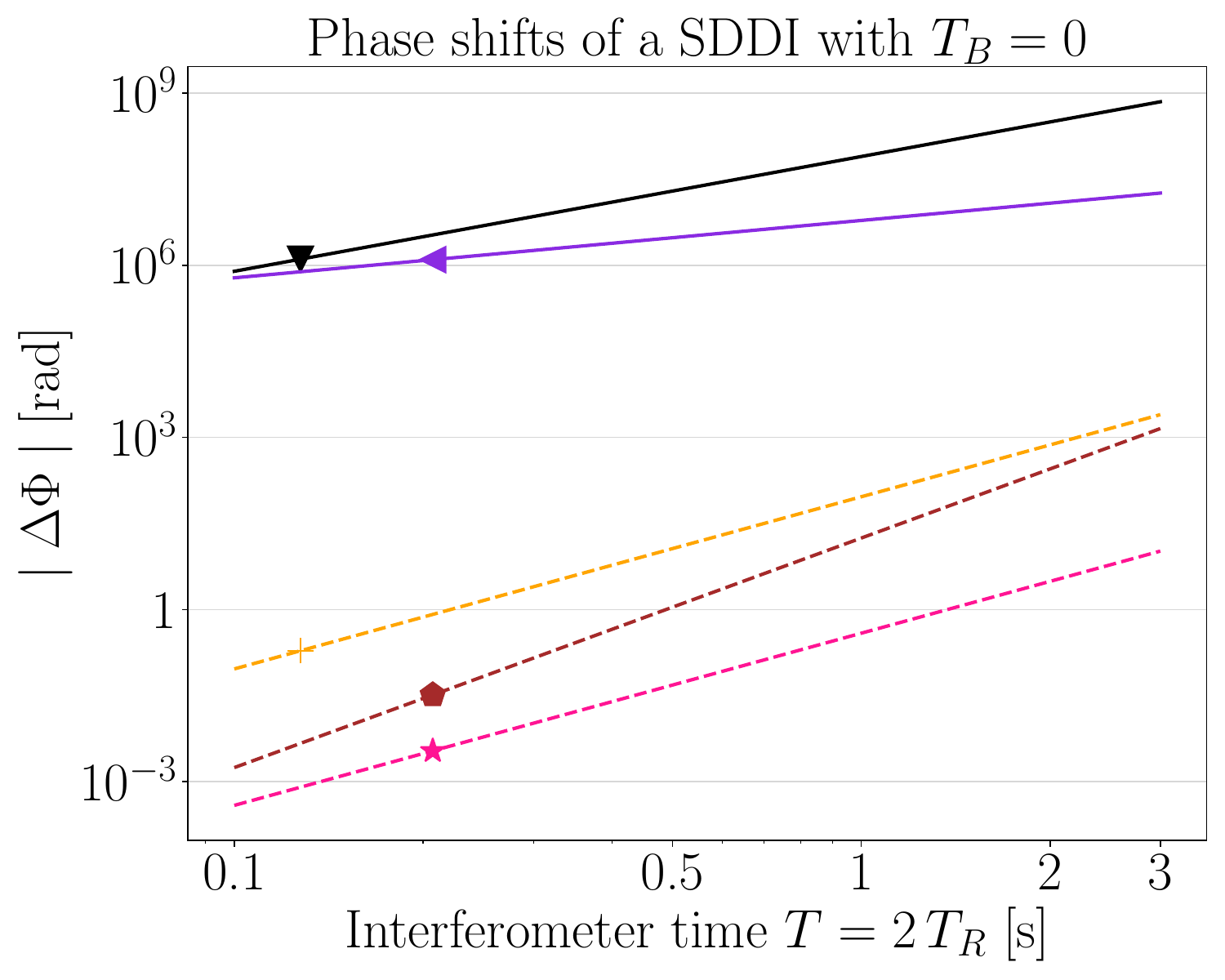} 
		\label{Fig: 10m_Numerical_Plot_SDI_Without_Bloch}}
        \hfil
        \subfloat[10 meter ARBI]{\includegraphics[width=0.315\textwidth]{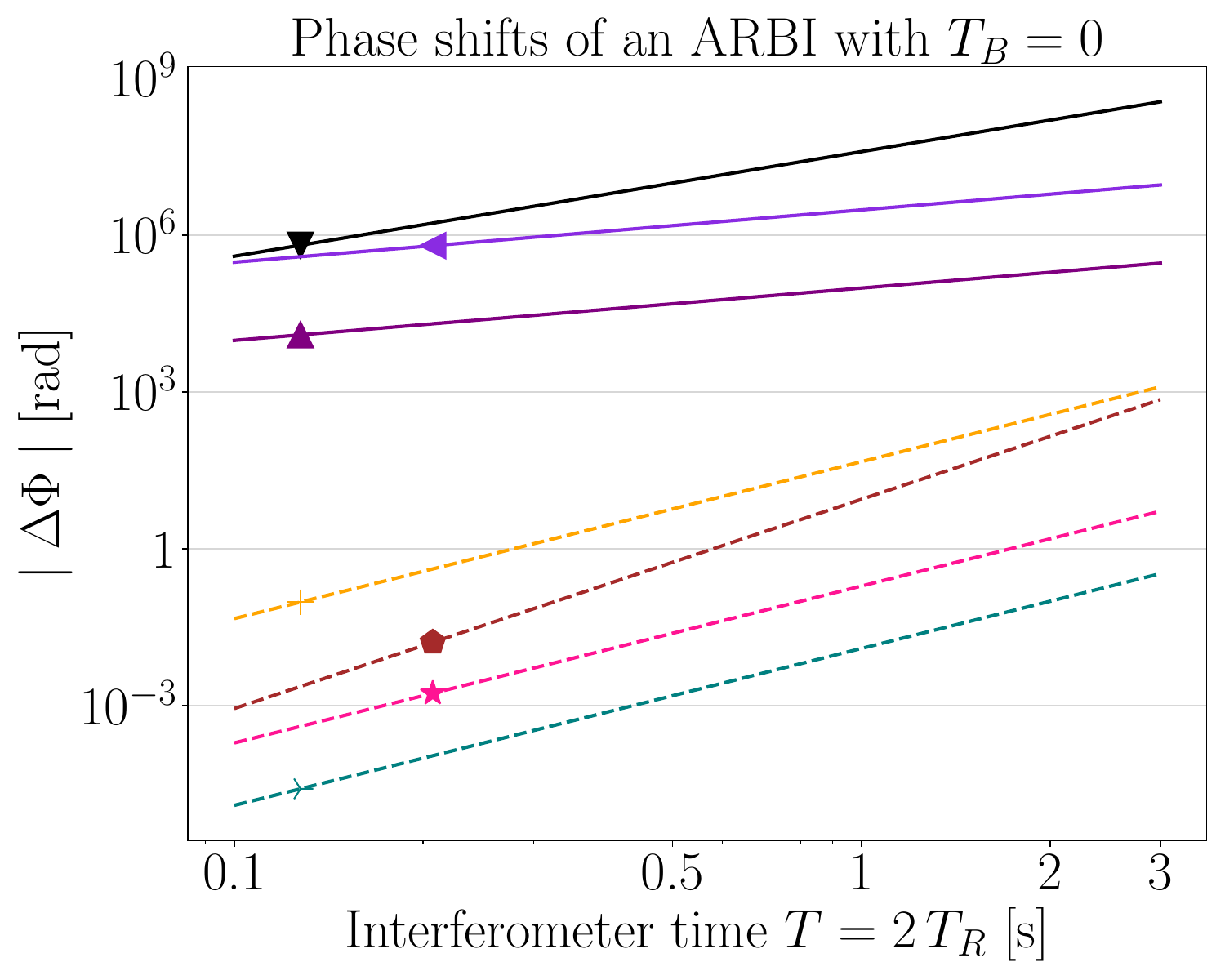}
		\label{Fig: 10m_Numerical_Plot_ARBI_Without_Bloch}}
        \newline
        \subfloat[100 meter SRBI]{\includegraphics[width=0.315\textwidth]{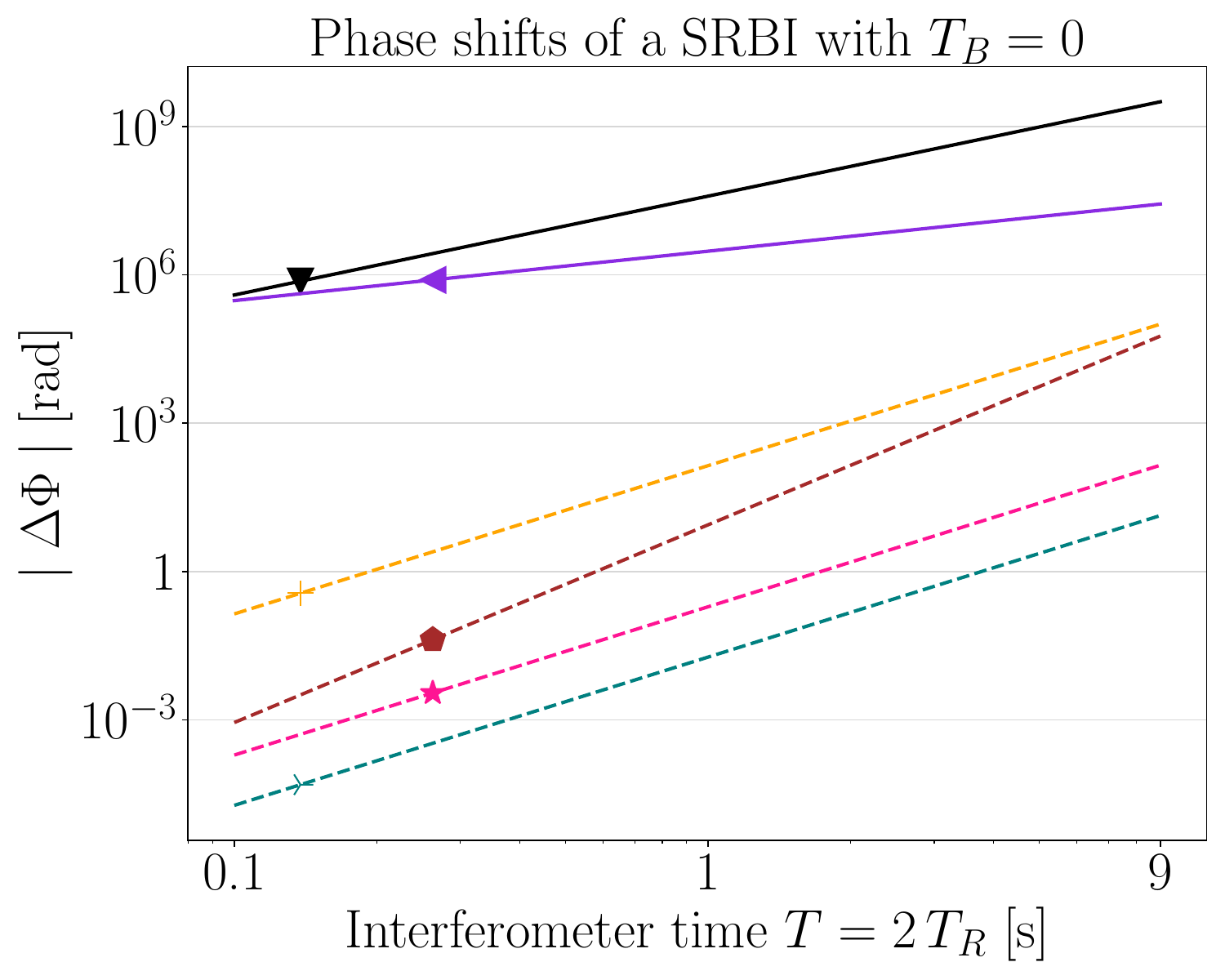} 
		\label{Fig: 100m_Numerical_Plot_SRBI_Without_Bloch}}
        \hfil
        \subfloat[100 meter SDDI]{\includegraphics[width=0.315\textwidth]{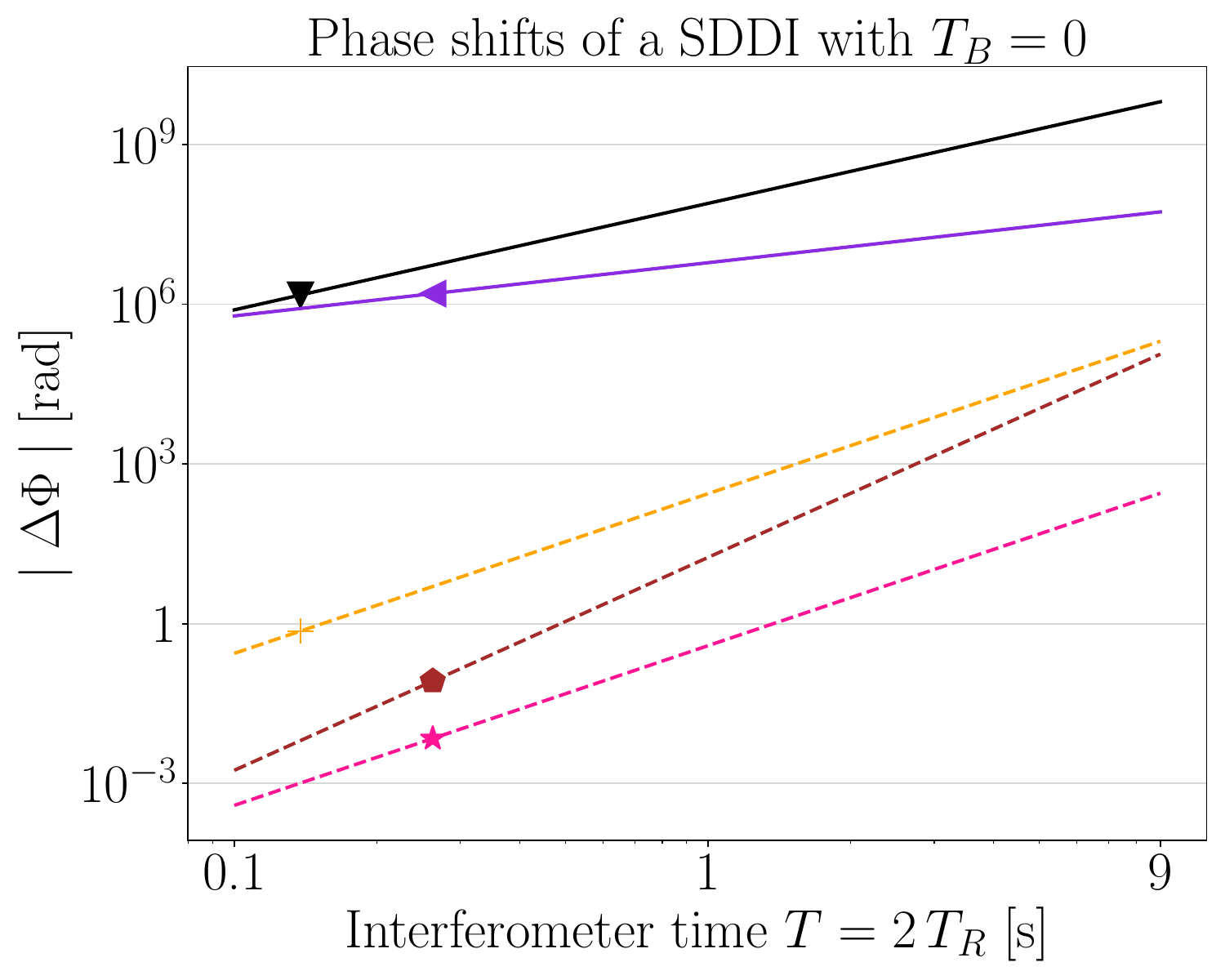} 
		\label{Fig: 100m_Numerical_Plot_SDDI_Without_Bloch}}
        \hfil
        \subfloat[100 meter ARBI]{\includegraphics[width=0.315\textwidth]{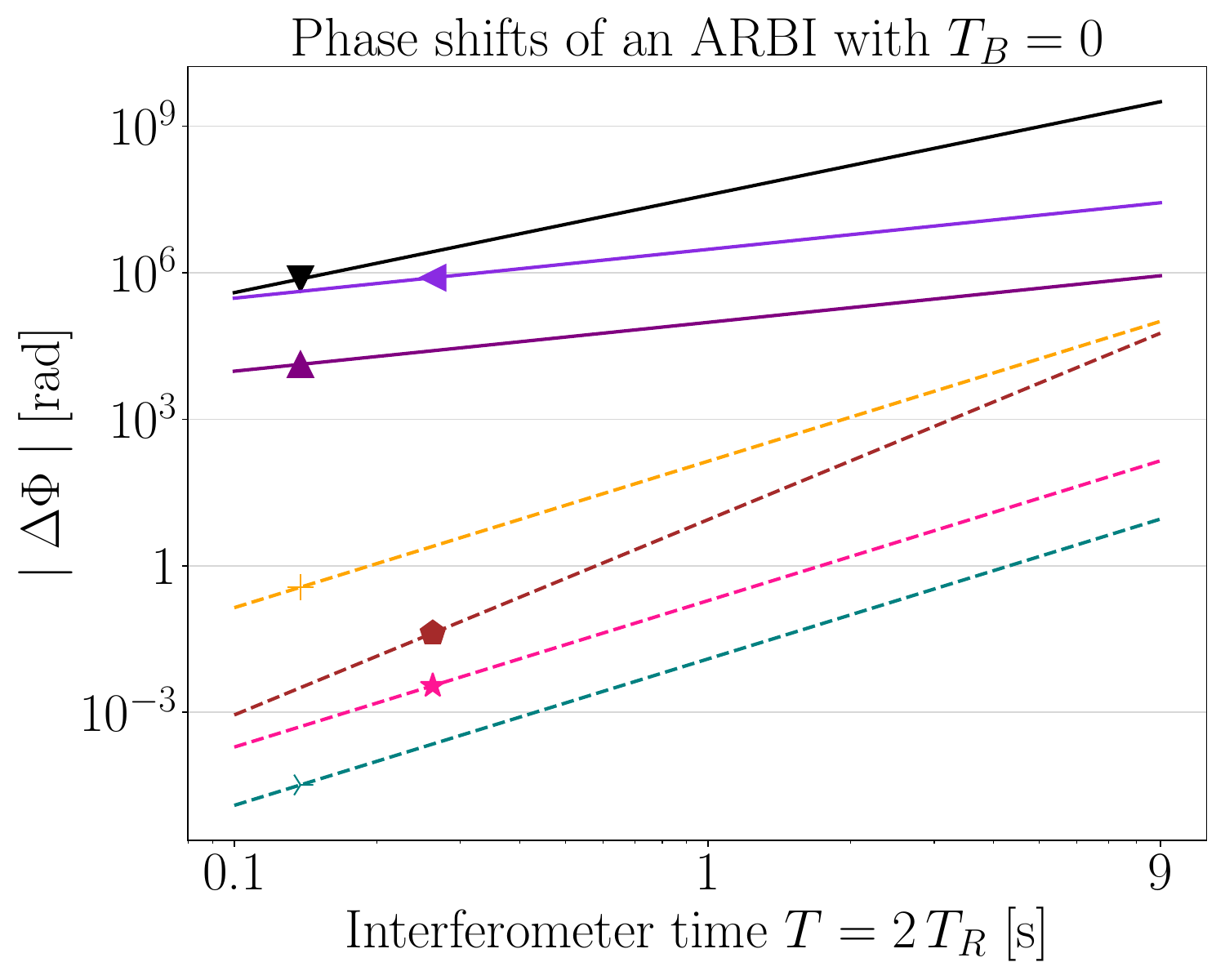}
		\label{Fig: 100m_Numerical_Plot_ARBI_Without_Bloch}}
	\end{adjustbox}
	
	\begin{adjustbox}{minipage=0.95\linewidth,frame}
		\centering
		\subfloat[10 meter SRBI]{\includegraphics[width=0.315\textwidth]{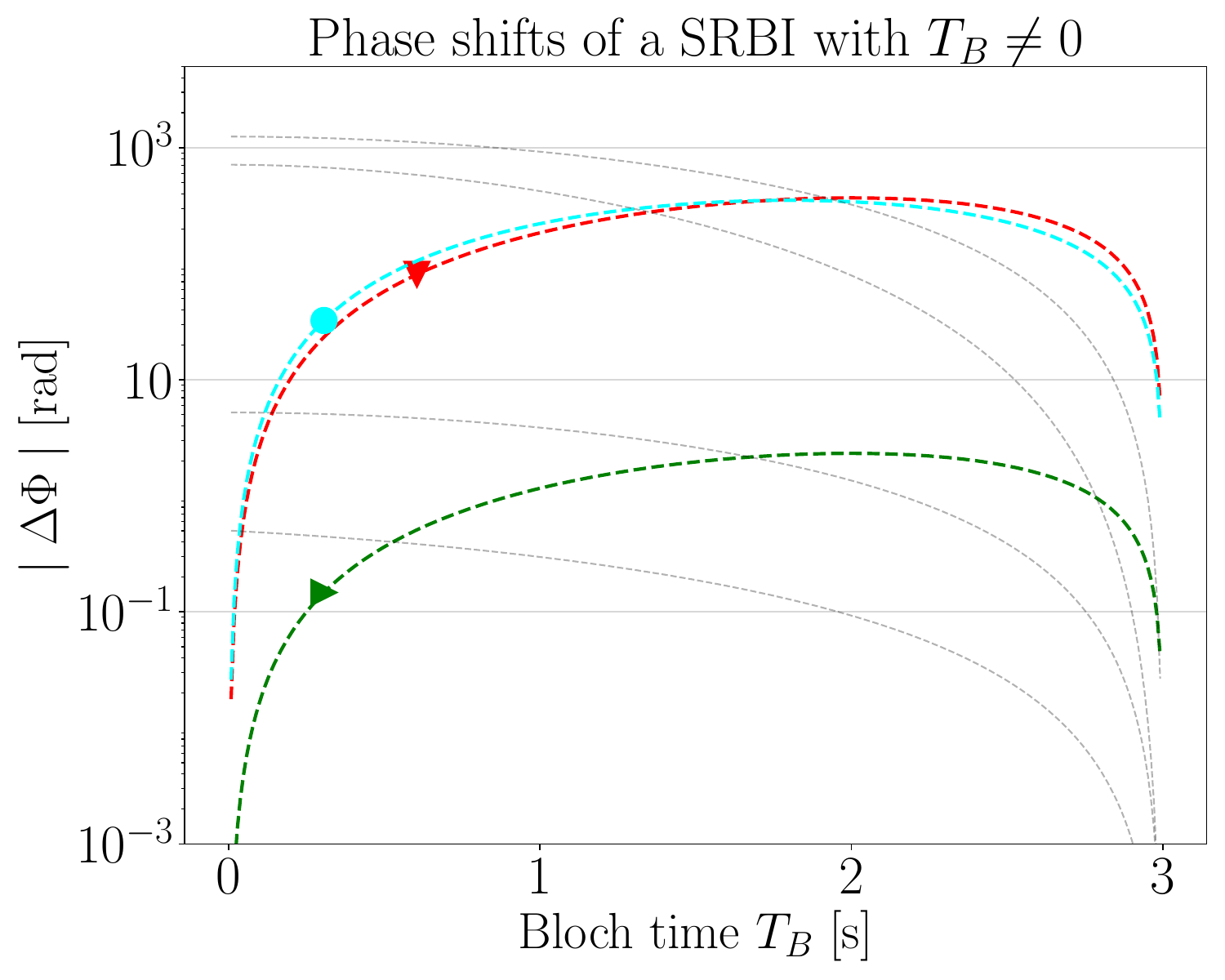} 
		\label{Fig: 10m_Numerical_Plot_SRBI_With_Bloch}}
		\hfil
		\subfloat[10 meter SDDI]{\includegraphics[width=0.315\textwidth]{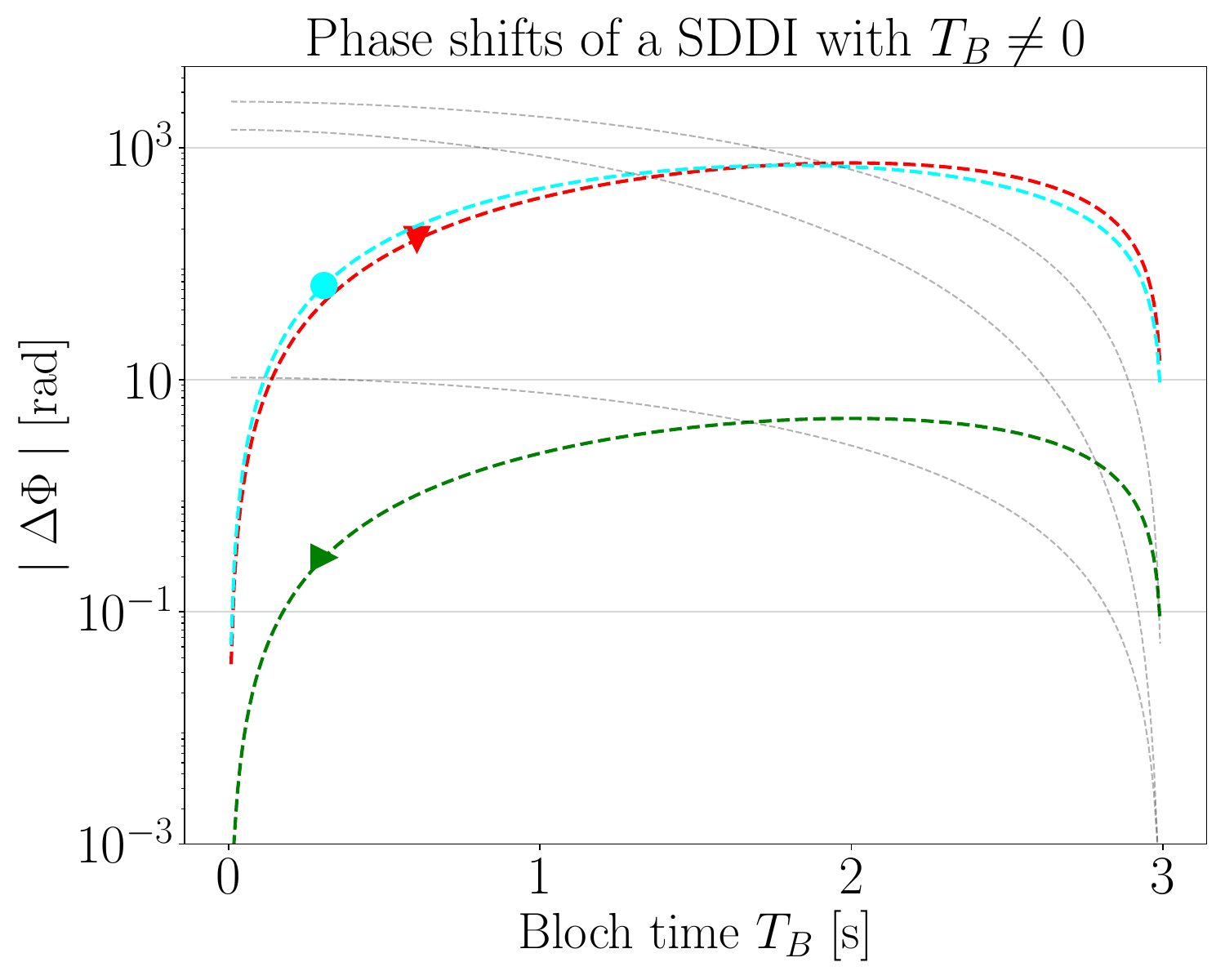} 
		\label{Fig: 10m_Numerical_Plot_SDDI_With_Bloch}}
        \hfil
        \subfloat[10 meter ARBI]{\includegraphics[width=0.315\textwidth]{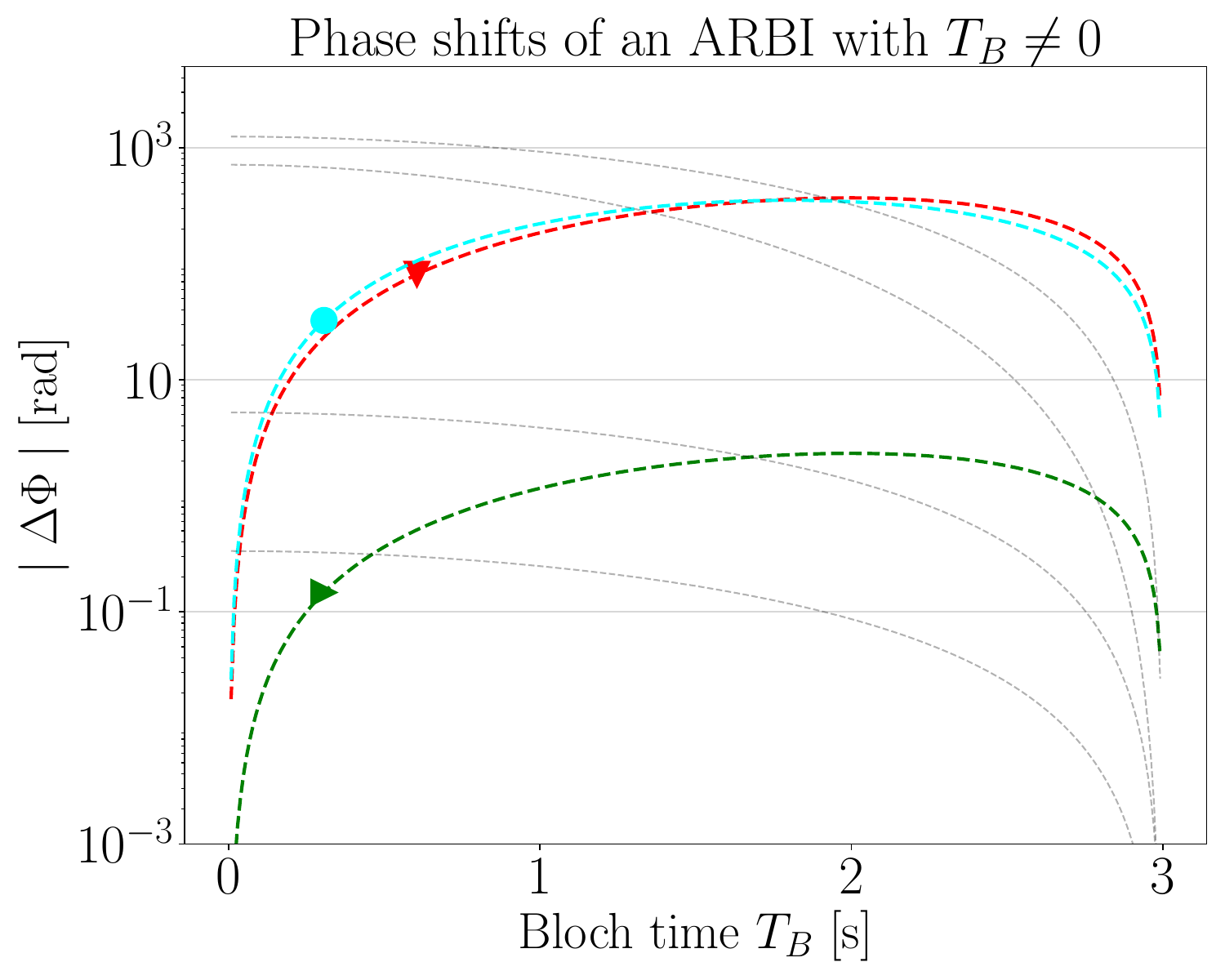} 
		\label{Fig: 10m_Numerical_Plot_ARBI_With_Bloch}}
        \newline
        \subfloat[100 meter SRBI]{\includegraphics[width=0.315\textwidth]{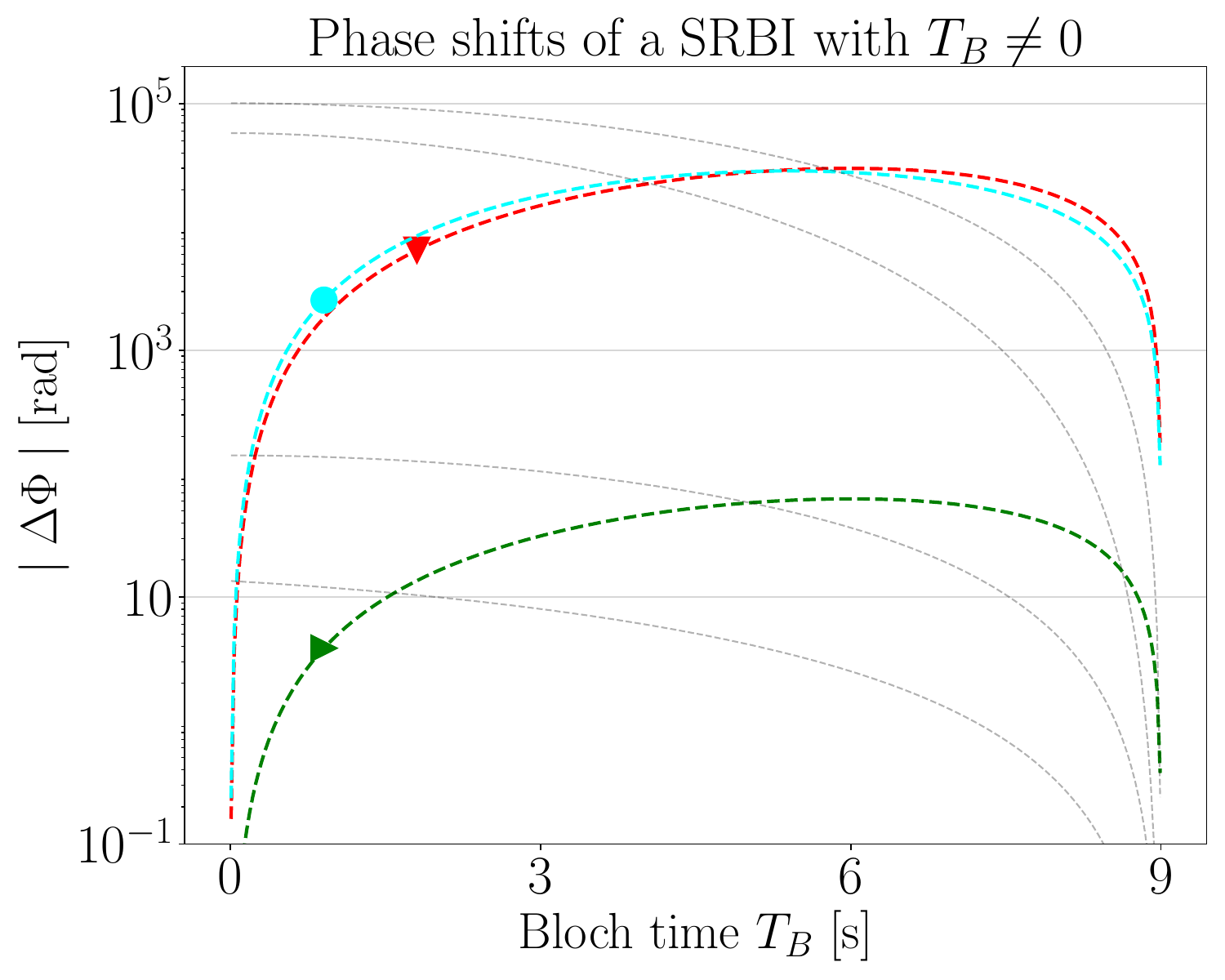} 
		\label{Fig: 100m_Numerical_Plot_SRBI_With_Bloch}}
        \hfil
        \subfloat[100 meter SDDI]{\includegraphics[width=0.315\textwidth]{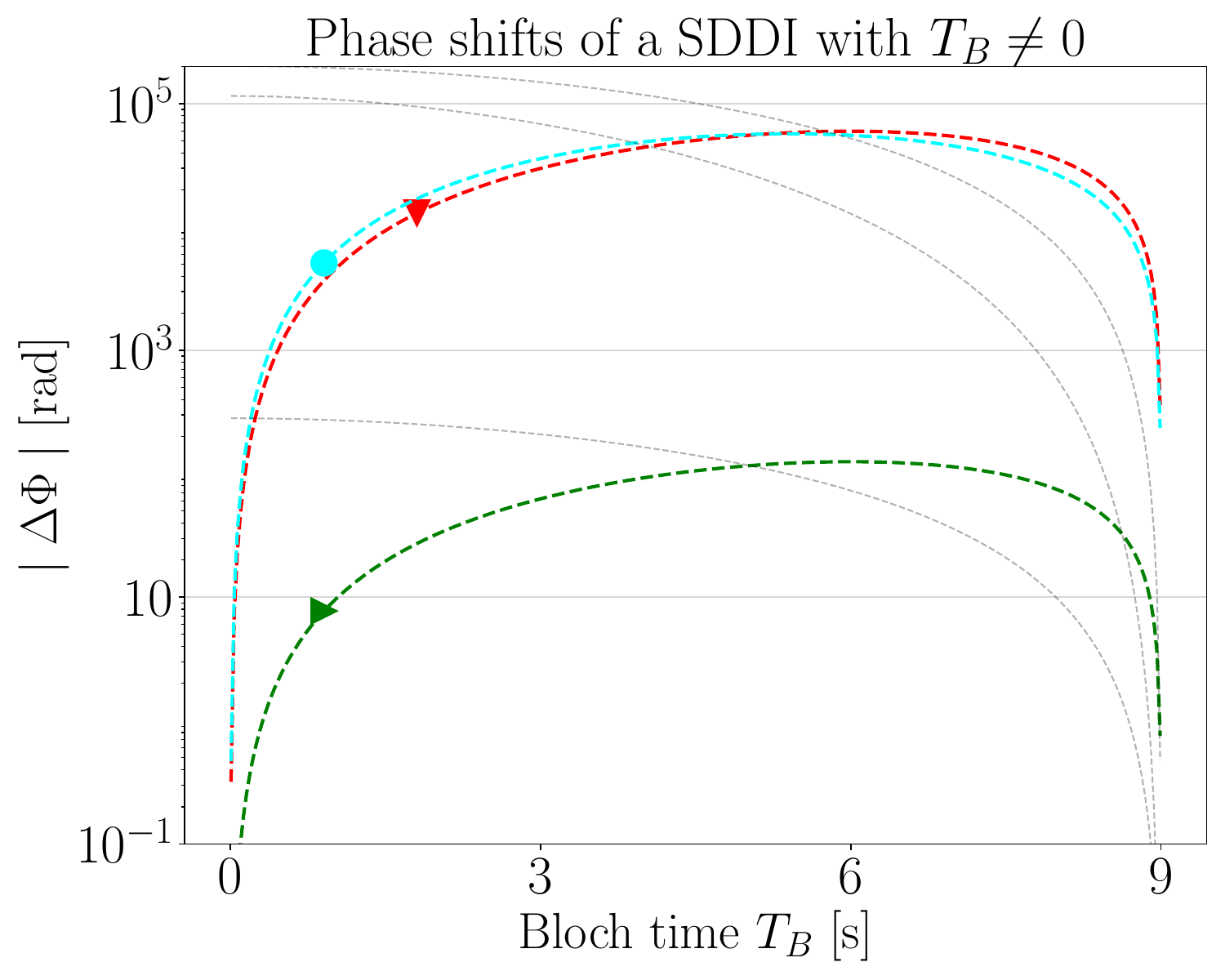}
		\label{Fig: 100m_Numerical_Plot_SDDI_With_Bloch}}
        \hfil
        \subfloat[100 meter ARBI]{\includegraphics[width=0.315\textwidth]{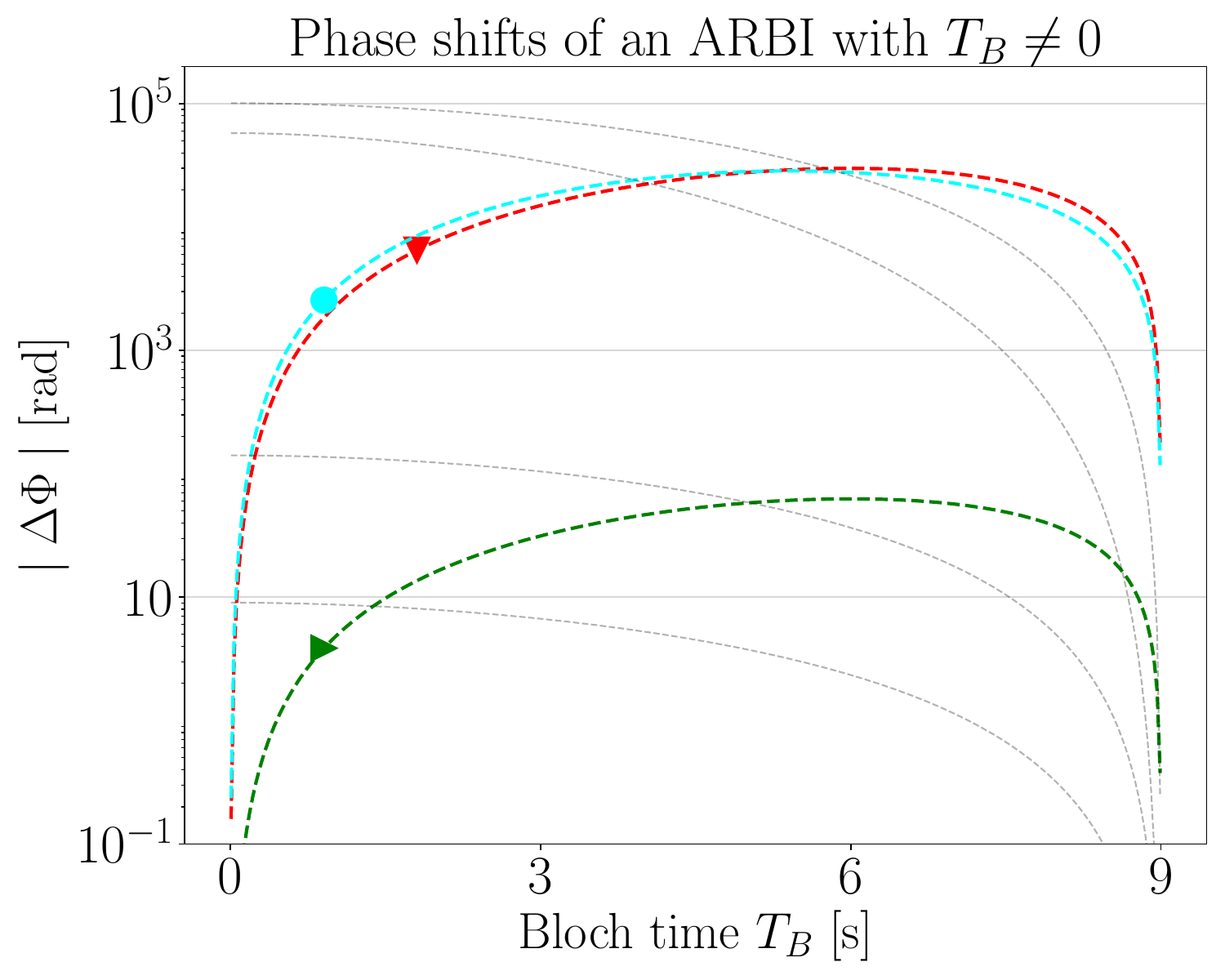}
        \label{Fig: 100m_Numerical_Plot_ARBI_With_Bloch}} 
		\newline	
		\subfloat{\includegraphics[height=0.7cm]{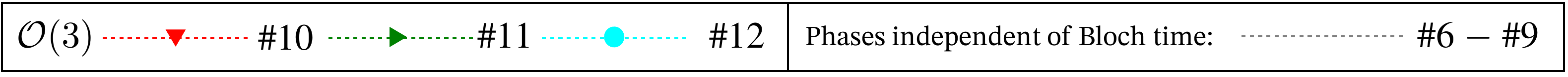}}
	\end{adjustbox}
	\caption{Phase shift contributions in the three IF geometries SRBI, SDDI, ARBI for $10 \, \mathrm{m}$ and $100 \, \mathrm{m}$ baselines. Solid curves correspond to phase shifts of order $\mathcal{O}(2)$ and dashed curves to $\mathcal{O}(3)$. (a)--(f): The Bloch time $T_B$ is set to zero and all non-zero phase shift contributions are plotted w.r.t. time $T = 2 T_R$. (f) ${-}$ (l): Colored phase shift contributions depend non-trivially on $T_B$ and are plotted against $T_B$ for fixed time $T = 2 T_R + T_B$ of 3 seconds (9 seconds) for the $10 \, \mathrm{m}$ ($100 \, \mathrm{m}$) baseline; the gray curves correspond to the Bloch-time independent phase contributions from (a)--(f). In addition to the numerical values in Table~\ref{fig: Definition_of_dimensionless_parameters}, the plots assumed numerical values of $g = 9.81 \, \mathrm{m} \mathrm{s}^{-2}$ and $\Gamma = 1.54 \times 10^{-6} \, \mathrm{s}^{-2}$.}
	\label{Fig: Phase_Comparison}
\end{figure*}

In Fig.~\ref{Fig: Phase_Comparison}, we plot the phase shifts of $\mathcal{O}(2)$ and the leading contributions of $\mathcal{O}(3)$ evaluated for two case studies corresponding to a $10\, \mathrm{m}$ and a $100\, \mathrm{m}$ baseline IF. The resulting list of phase shifts can be grouped into two: Figs.~\subref*{Fig: 10m_Numerical_Plot_SRBI_Without_Bloch} ${-}$ \subref*{Fig: 100m_Numerical_Plot_ARBI_Without_Bloch} show phases that are maximal for $T_B = 0$ and therefore would preferably be analyzed in an IF without Bloch pulses (vanishing Bloch time $T_B$). Figs.~\subref*{Fig: 10m_Numerical_Plot_SRBI_With_Bloch} ${-}$ \subref*{Fig: 100m_Numerical_Plot_ARBI_With_Bloch} display phases that are functions of the Bloch recoil and therefore are maximal for a non-trivial combination of $T_R$ and $T_B$, since the corresponding phase shift will vanish in both limiting cases --- i.e., $T_B=0$ and $T_R =0$. It can be observed that the curves of $\mathcal{O}(2)$ and $\mathcal{O}(3)$, respectively, cluster with a gap of several orders of magnitude between them. The magnitude of the $\mathcal{O}(4)$ terms will be smaller by about the same factor --- i.e., those terms contribute around $10^{-10} \, \mathrm{rad}$ for short IF times of around $0.1$ seconds. The Doppler related $\mathcal{O}(3)$-terms will, however, also be present at $\mathcal{O}(4)$, because of the comparably small value of $\omega_R$ and therefore $\FreqR$, see Table~\ref{fig: Definition_of_dimensionless_parameters}.

For the specific case of a Mach-Zehnder IF (equivalent to the SRBI with $k_B=0$, $T_B=0$) we can compare the results of our treatment to the one of Dimopoulos et al.~\cite{dimopoulos2007PRL, dimopoulos2008general}, and find good, although not exact, agreement. A detailed comparison of the terms up to $\mathcal{O}(4)$ can be found in App.~\ref{Appendix: Mach-Zehnder-Interferometer}, in which we also summarize where our approach differs from that of~\cite{dimopoulos2007PRL, dimopoulos2008general} in methodology and notation and discuss how these differences affect the final results.

\begin{turnpage}

\begin{table*}
	\footnotesize
	\centering
	\begin{tabularx}{0.95\linewidth}{|c|c|c|l|l|l|c|X|}
		\hline
		\multicolumn{8}{|c|}{\textbf{Phases in units of $\Comp$}} \\
		\hline
		$\#$ & Order & 
		Proportionality & Symmetric Ramsey-Bordé IF (SRBI) \hspace{0.4cm} & 
		Symmetric double diffraction IF (SDDI) \hspace{0.4cm} & 
		Asymmetric Ramsey-Bordé IF (ARBI) \hspace{0.4cm} & 
		\hspace{0.1cm} $\alpha$ \hspace{0.1cm} & 
		Origin  \\ 
		\hline
		1 & $\mathcal{O}\left( 2 \right)$ & 
		$\GOneR \RR$ &
		$T_B + T_R$ & 
		$2 T_B + 2 T_R$ & 
		$T_B + T_R$ & 
		2 & 
		Non-relativistic  \\ 
		\cline{3-7}
		2 & & 
		$\RR^2$ & 
		0 & 
		0 & 
		$T_R$ & 
		1 & 
		\\
		\cline{3-7}
		3 & & 
		$\RR \RB$ &
		$-T_R$ & 
		$-2 T_R$ & 
		$-T_R$ & 
		1 & 
		\\
		\cline{2-8}
		4 & $\mathcal{O}\left(3 \right)$ & 
		$\RR \GZero \GOneR$ &
		$2 (\beta - 1) T_B + 2 (\beta - 1) T_R$ \hspace{1cm} & 
		$4 (\beta - 1) T_B + 4 (\beta - 1)T_R$ \hspace{1cm} & 
		$2 (\beta - 1) T_B + 2 (\beta - 1) T_R$ \hspace{1cm} & 
		2 & 
		PPN \\  
		\cline{3-8}
		5 & & 
		$\RR \ZZero \GTwoR$ &
		$-T_B - T_R$ & 
		$-2 T_B - 2 T_R$ & 
		$-T_B - T_R $ & 
		2 & 
		Gravity gradient \hspace{0.3cm} \\ 
		\cline{3-7}
		6 & & 
		$\RR \VZero \GTwoR$ & 
		$-\frac{3}{2} T_B - T_R$ & 
		$-3 T_B - 2 T_R$ & 
		$-\frac{3}{2} T_B - T_R$ & 
		3 &
		\\
		\cline{3-7}
		7 & & 
		$\RR \RB \GTwoR$ & 
		$-\frac{1}{4} T_B - \frac{1}{6} T_R$ & 
		$-\frac{1}{2} T_B - \frac{1}{3} T_R$ & 
		$-\frac{1}{4} T_B - \frac{1}{6} T_R$ & 
		4 &
		\\
		\cline{3-7}	
		8 & & 
		$\RR \GOneR \GTwoR$ & 
		$\frac{7}{6} T_B + \frac{7}{12} T_R$ & 
		$\frac{7}{3} T_B +\frac{7}{6} T_R$ & 
		$\frac{7}{6} T_B + \frac{7}{12} T_R$ & 
		4 & 
		\\
		\cline{3-7}
		9 & & 
		$\RR^2 \GTwoR$ & 
		$-\frac{1}{2} T_B - \frac{1}{2} T_R$ & 
		0 & 
		$-\frac{1}{2} T_B - \frac{1}{3} T_R$ & 
		3 & 
		\\
		\cline{3-7}
		10 & & 
		$\RR \VZero \GTwoB$ & 
		$- \frac{1}{2} T_R$ & 
		$- T_R$ & 
		$- \frac{1}{2} T_R$ & 
		3 & 
		\\
		\cline{3-7}
		11 & & 
		$\RR \RB \GTwoB$ & 
		$-\frac{1}{8} T_R$ & 
		$-\frac{1}{4} T_R$ & 
		$-\frac{1}{8} T_R$ & 
		3 & 
		\\
		\cline{3-7}
		12 & & 
		$\RR \GOneR \GTwoB$ & 
		$\frac{1}{6} T_B + \frac{3}{4} T_R$ & 
		$\frac{1}{3} T_B + \frac{3}{2} T_R$ & 
		$\frac{1}{6} T_B + \frac{3}{4} T_R$ & 
		4 & 
		\\
		\cline{3-8}
		13 & & 
		$\FreqR \GOneR^2$ & 
		$- \frac{9}{2} T_B - 3 T_R$ & 
		$- 9 T_B - 6 T_R$ & 
		$- \frac{9}{2} T_B - 3 T_R$ & 
		3 & 
		Doppler effect \\
		\cline{3-7}
		14 & & 
		$\FreqR \GOneR \VZero$ & 
		$3 T_B + 3 T_R$ & 
		$6 T_B + 6 T_R$ & 
		$3 T_B + 3 T_R$ & 
		2 & 
		\\
		\cline{3-7}
		15 & & 
		$\FreqR \GOneR \GOneB$ & 
		$- \frac{3}{2} T_B$ & 
		$- 3 T_B$ & 
		$- \frac{3}{2} T_B$ &
		3 & 
		\\
		\cline{3-7}
		16 & & 
		$\FreqR \RB \GOneR$ & 
		$\frac{5}{2} T_B + \frac{7}{2} T_R$ & 
		$5 T_B + 7 T_R$ & 
		$\frac{5}{2} T_B + \frac{7}{2} T_R$ &
		2 & 
		\\
		\cline{3-7}
		17 & & 
		$\FreqR \RB^2$ & 
		$- T_R$ & 
		$- 2 T_R$ & 
		$- T_R$ &
		1 & 
		\\ 
		\cline{3-7}
		18 & & 
		$\FreqR \RB \VZero$ & 
		$- 2 T_R$ & 
		$- 4 T_R$ & 
		$- 2 T_R$ &
		1 & 
		\\
		\cline{3-7}
		19 & & 
		$\FreqR \RR^2$ & 
		0 & 
		$2 T_R$ & 
		$T_R$ &
		1 & 
		\\
		\cline{3-7}
		20 & & 
		$\FreqR \RR \GOneB$ & 
		$\frac{1}{2} T_B$ & 
		0 & 
		$- \frac{1}{2} T_B$ &
		2 & 
		\\
		\cline{3-7}
		21 & & 
		$\FreqR \RR \GOneR$ & 
		$3 T_B - \frac{5}{2} T_R$ & 
		0 & 
		$- 3 T_B + \frac{9}{2} T_R$ &
		2 & 
		\\
		\cline{3-7}
		22 & & 
		$\FreqR \RR \RB$ &
		$- \frac{1}{2} T_B + 2 T_R$ & 
		0 & 
		$\frac{1}{2} T_B - 2 T_R$ &
		1 & 
		\\
		\cline{3-7}
		23 & & 
		$\FreqR \RR \VZero$ & 
		$- T_B - T_R$ & 
		0 & 
		$T_B + 5 T_R$ &
		1 & 
		\\
		\cline{3-7}
		24 & & 
		$\FreqR \RR \ZZero$ & 
		0 & 
		0 & 
		$2 T_R$ &
		1 & 
		\\
		\hline
	\end{tabularx}
	\caption{Complete list of phases of the SRBI, SDDI, ARBI geometries in PPN spacetime written in terms of dimensionless parameters of order $\mathcal{O}(2)$ ($\# 1 \, {-} \, \# 3$) and $\mathcal{O}(3)$ ($\# 4 \, {-} \, \# 24$). To extract this phase for one of the IF geometries, multiply the factor in the column `proportionality' by the time given in the column of the respective IF and the atomic Compton frequency $\Comp$ in Eq.~\eqref{eq:Compton}. For example, phase shift $\#1$ for the SRBI $\RR \GOneR \Comp (T_R + T_B)$ which translates into $g k_R (T_R^2 + T_R T_B)$, see Table~\ref{fig: Definition_of_dimensionless_parameters}. Written out in terms of dimensionful quantities, each contribution is a polynomial in $T_R$ and $T_B$ --- i.e. is proportional to $T_R^{\alpha_R}T_B^{\alpha_B}$. The overall exponent $\alpha = \alpha_R + \alpha_B$ determines the scaling of each phase with IF time.}
	\label{Table: All Phases Dimensionless Parameters}
\end{table*}

\end{turnpage}

\section{Discussion and Summary}\label{Section 5}

Starting from a post-Newtonian approximation of the Schrödinger equation in a curved spacetime, described by a metric theory of gravity, we have presented a calculation of the phase contributions of a whole class of light-pulse atom IFs. The phases are derived from a relativistically-corrected, quantum-optical Hamiltonian  for atoms and light in a PPN spacetime, written in a locally orthonormal coordinate system, and applied to the specific case of atom IFs that use elastic scattering processes. After following the standard procedure for calculating phases in atom IFs in the presence of relativistic corrections we have expressed all the resulting phase contributions as functions of dimensionless parameters that arise naturally from this description. The computation of all phase contributions up to a desired order in those dimensionless parameters for any IF geometry consisting of Bragg and Bloch pulses is automated in Python. We have illustrated the results of this algorithm using three exemplary IF geometries and compared their individual phase contributions algebraically and, for the dominant contributions, quantitatively for long-baseline IFs. With a suitable choice of IF geometries, it may then be possible to find new measurement strategies for differential IF setups, which particularly enhance individual phase contributions and suppress terms which are of minor interest. 

This analysis can be extended to IFs that use inelastic scattering processes --- e.g., Raman transitions~\cite{Kasevich_Chu, borde1989atomic, roura2020gravitational, diPumpo2021GravitationalRedshift} --- and would also be sufficient to describe stationary spacetimes, which could include effects of Earth's rotation and describe gravito-magnetic phenomena such as the Lense-Thirring effect~\cite{wajima1997post}. One would then, however, need to include all FSL effects, since instantaneous laser pulses, would require the notion of  simultaneity, which is not necessarily well-defined in non-static spacetimes. The description of spaceborne experiments could also be a natural extension of this formalism, in which one would need to go beyond our approximation of the gravitational potential, due to possibly elliptical orbits with a considerable height variation. Another interesting approach one could pursue is to start from a Hamiltonian that describes fermionic particles in curved spacetime, cf.~\cite{alibabaei2023geometric}, and investigate whether spin-related interactions with gravity might give rise to interesting tests of GR in light-pulse atom IFs.


\section*{Acknowledgments}

Funded by the Deutsche Forschungsgemeinschaft (DFG) – SFB 1227 – Project-ID 274200144 project A05. We thank Ernst Rasel, Dennis Schlippert, Christian Schubert, and Enno Giese for insightful discussions.

\appendix


\section{Transformation into new coordinates}\label{Appendix: New Coordinate System}

We want to construct a coordinate transformation $(\OldCoordsX^{\mu}) = (c \OldCoordsTime, \OldCoordsVec ) \longmapsto (\NewCoordsX^\mu) = (c \NewCoordsTime, \NewCoordsVec )$ that brings the metric tensor to Minkowskian form at the point of the experiment, which in this appendix we will generically denote by $p_0$ (in the main text, this is taken as a point on the surface of the Earth).  Since the asymptotic coordinates in which the metric was originally expressed in Eq.~\eqref{eq: PPN_metric} are already orthogonal to our order of approximation, in order to bring the metric to Minkowskian form at $p_0$, we only have to perform a coordinate transformation such as to normalize the new coordinate basis vectors at this point.  This can most easily be realized by globally rescaling the coordinates by the corresponding (constant) normalization factor: defining the new coordinates according to
\begin{equation}
    \NewCoordsX^\mu = \sqrt{|\OldCoordsg_{\mu\mu}(p_0))|} \, \OldCoordsX^\mu
\end{equation}
(no summation over $\mu$), the new coordinate basis vectors are given by
\begin{equation}
    \frac{\partial}{\partial\NewCoordsX^\mu}
    = \frac{1}{\sqrt{|\OldCoordsg_{\mu\mu}(p_0))|}} \, \frac{\partial}{\partial\OldCoordsX^\mu} \; ,
\end{equation}
which are normalized at $p_0$.

Explicitly, defining $\phi_0 = \phi(p_0)$, the metric length of the timelike coordinate basis vector $\left.\frac{\partial}{\partial\OldCoordsX^0}\right|_{p_0} = \left.\frac{1}{c} \frac{\partial}{\partial\OldCoordsTime}\right|_{p_0}$ of the asymptotic (i.e.\ old) coordinates is given by
\begin{subequations}
\begin{align}
    \sqrt{-\OldCoordsg_{00}(p_0)} 
    &= \sqrt{1 + 2 \frac{\phi_0}{c^2} + 2 \beta \frac{\phi_0^2}{c^4} + \mathcal{O}(c^{-6})} \nonumber\\
    &= 1 + \frac{\phi_0}{c^2} + \frac{2\beta - 1}{2} \frac{\phi_0^2}{c^4} + \mathcal{O}(c^{-6}), \\
\intertext{while the metric length of the spacelike coordinate basis vector $\left.\frac{\partial}{\partial\OldCoordsX^i}\right|_{p_0}$ of the asymptotic coordinates is (without summation over $i$)}
    \sqrt{\OldCoordsg_{ii}(p_0)} 
    &= \sqrt{1 - 2 \gamma \frac{\phi_0}{c^2} + \mathcal{O}(c^{-4})} \nonumber\\
    &= 1 - \gamma \frac{\phi_0}{c^2} + \mathcal{O}(c^{-4}).
\end{align}
\end{subequations}
Therefore, we may take the constantly rescaled new coordinates expressed in terms of the old coordinates as
\begin{subequations}
\begin{align}
    \NewCoordsTime &= \left( 1 + \frac{\phi_0}{c^2} + \frac{2 \beta - 1}{2} \frac{\phi_0^2}{c^4} \right) \OldCoordsTime, \\
	\NewCoordsVec &= \left(1 - \gamma \frac{\phi_0}{c^2} \right) \OldCoordsVec.
\end{align}
\end{subequations}
The inverse transformation is then given by
\begin{subequations}
\begin{align}
    \OldCoordsTime &= \left( 1 - \frac{\phi_0}{c^2} - \frac{2 \beta - 3}{2} \frac{\phi_0^2}{c^4} + \mathcal{O}(c^{-6}) \right) \NewCoordsTime, \\
	\OldCoordsVec &= \left(1 + \gamma \frac{\phi_0}{c^2} + \mathcal{O}(c^{-4})\right) \NewCoordsVec.
\end{align}   
\end{subequations}

Now computing the components of the metric in the new coordinates according to
\begin{align}
    \NewCoordsg_{\mu\nu} = \frac{\partial\OldCoordsX^\lambda}{\partial\NewCoordsX^\mu} \frac{\partial\OldCoordsX^\sigma}{\partial\NewCoordsX^\nu} \OldCoordsg_{\lambda\sigma} \; ,
\end{align}
we obtain the new spatial components
\begin{align}
    \tensor{\NewCoordsg}{_{i j}}(\NewCoordsVec) 
    = \left(1 - 2 \gamma \frac{\phi(\NewCoordsVec) - \phi_0}{c^2} \right) \tensor{\delta}{_{ij}} + \mathcal{O}(c^{-4})
\end{align}
\begin{widetext}
\noindent
and the new temporal component
\begin{align}
    \tensor{\NewCoordsg}{_{00}}(\NewCoordsVec)
    = -1 - 2 \frac{\phi(\NewCoordsVec) - \phi_0}{c^2} - 2 \beta \frac{ (\phi(\NewCoordsVec) - \phi_0)^2}{c^4}  - 4 (\beta - 1) \phi_0 \frac{ \phi(\NewCoordsVec) - \phi_0}{c^4} + \mathcal{O}(c^{-6}).
\end{align}
Using the definition of the shifted gravitational potential, i.e., $\overline{\phi}(\NewCoordsVec) = \phi(\NewCoordsVec) - \phi_0$, the line element in the new coordinates reads
\begin{align}
	\dd s^2 = -\left( c^2 + 2\overline{\phi} + 2 \beta \frac{\overline{\phi}^2}{c^2} + 4 (\beta-1) \frac{\phi_0 \overline{\phi}}{c^2} \right) \dd \NewCoordsTime^2 
	+ \left( 1-2 \gamma \frac{\overline{\phi}}{c^2} \right) \dd \NewCoordsVec^2 + \mathcal{O} (c^{-4} ),
\end{align}
which is Minkowskian at the reference point. The components of the inverse metric are then
\begin{align}
	\NewCoordsg^{\mu \nu}(\NewCoordsVec) =
	\begin{pmatrix}
		-1 + 2 \frac{\overline{\phi}(\NewCoordsVec)}{c^2}  + (2 \beta -4) \frac{\overline{\phi}(\NewCoordsVec)^2}{c^4} + 4 (\beta-1) \phi_0 \frac{\overline{\phi}(\NewCoordsVec)}{c^4} + \mathcal{O}(c^{-6}) & \mathcal{O}(c^{-5})	\\[4pt]
		\mathcal{O}(c^{-5}) & \left( 1+ 2 \gamma \frac{\overline{\phi}(\NewCoordsVec)}{c^2} \right) \One_3 + \mathcal{O}(c^{-4})
	\end{pmatrix}. \label{eq: NEW_PPN_metric_Contravariant} 
\end{align}
\end{widetext}
Using these expressions, we can calculate the Christoffel symbols
\begin{align}
	\tensor{\Gamma}{^\mu_{\nu \sigma}} = \frac{1}{2} \tensor{\NewCoordsg}{^{\mu \lambda}} \left( \tensor{\NewCoordsg}{_{\lambda \nu , \sigma}} + \tensor{\NewCoordsg}{_{\lambda \sigma , \nu}} - \tensor{\NewCoordsg}{_{\nu \sigma , \lambda}}\right).
\end{align}
We assume that the Newtonian gravitational potential be time independent, i.e., $\partial_\NewCoordsTime \phi = 0$. The trivial Christoffel symbols are
\begin{align}
	\tensor{\Gamma}{^0_{00}} = \mathcal{O}(c^{-7}), \quad \tensor{\Gamma}{^0_{ij}} = \mathcal{O}(c^{-5}), \quad \tensor{\Gamma}{^i_{j0}} = \mathcal{O}(c^{-5}),
\end{align}
whereas the only non-vanishing Christoffel symbols are
\begin{subequations}
	\begin{align}
		\tensor{\Gamma}{^0_{i0}} &= \tensor{\Gamma}{^0_{0i}} = \left( 1 + 2( \beta-1) \frac{\phi}{c^2} \right) \frac{\overline{\phi}_{,i} }{c^2} + \mathcal{O}(c^{-6}) \\[10pt]
		\tensor{\Gamma}{^i_{00}} &=  \left( 1+ 2 (\beta + \gamma) \frac{\phi}{c^2} - 2 (\gamma + 1) \frac{\phi_0}{c^2} \right) \tensor{\delta}{^{ij}} \frac{\overline{\phi}_{,j}}{c^2} + \mathcal{O}(c^{-6})\\[10pt]
		\tensor{\Gamma}{^i_{jk}} &= -\gamma \frac{\tensor{\delta}{^i_j} \overline{\phi}_{,k} + \tensor{\delta}{^i_k} \overline{\phi}_{,j} - \tensor{\delta}{_{jk}}\tensor{\delta}{^{il}} \overline{\phi}_{,l}}{c^2} + \mathcal{O}(c^{-4}).
	\end{align}
\end{subequations}

\begin{widetext}
	
\section{Euler-Lagrange equation and propagation phase}\label{Appendix: Euler-Lagrange equations}

The motional relativistic Lagrangian corresponding to a point particle with mass m in a spacetime with a metric tensor $\NewCoordsg_{\mu \nu}$ can be obtained via $L_\HamiltonianMotional = -m c \sqrt{-\NewCoordsg_{\mu \nu}\, \frac{\dd \NewCoordsX^\mu}{\dd \NewCoordsTime} \, \frac{\dd \NewCoordsX^\nu}{\dd \NewCoordsTime}}$. This evaluates for a trajectory in z-direction to
\begin{align}
	L_\HamiltonianMotional = -m c^2 + \frac{m \dot{z}(\NewCoordsTime)^2}{2} - m \overline{\phi}(z(\NewCoordsTime)) + \frac{m \dot{z}(\NewCoordsTime)^4}{8 c^2} - m\frac{2 \beta-1}{2 } \frac{\overline{\phi}(z(\NewCoordsTime))^2}{c^2} 
	- m\frac{2 \gamma +1}{2}  \frac{ \overline{\phi}(z(\NewCoordsTime)) \dot{z}(\NewCoordsTime)^2 }{c^2} -2(\beta - 1) m \phi_0 \frac{ \overline{\phi}(z(\NewCoordsTime)) }{c^2} + \mathcal{O}(c^{-4}).~\label{eq:RelLagrangian}
\end{align}
From this we can deduce the Euler-Lagrange equation. We first calculate the derivatives as
\begin{subequations}
\begin{align}
	\frac{\partial L_\HamiltonianMotional }{\partial z} &= - m \partial_z \overline{\phi}(z) + \frac{m}{c^2} \left[ -(2 \beta-1) \overline{\phi}(z) \partial_z \overline{\phi}(z) - \frac{2 \gamma+1}{2} \dot{z}^2 \partial_z \overline{\phi}(z) - 2(\beta-1) \phi_0 \partial_z \overline{\phi}(z)  \right] + \mathcal{O}(c^{-4}) \\
	\frac{\partial L_\HamiltonianMotional }{\partial \dot{z}} &= m \dot{z} + \frac{m}{c^2} \left[ \frac{1}{2} \dot{z}^3 - (2\gamma+1) \overline{\phi}(z) \dot{z} \right] + \mathcal{O}(c^{-4}) \\
	\frac{\dd }{\dd \NewCoordsTime}\frac{\partial L_\HamiltonianMotional }{\partial \dot{z}} &= m \ddot{z} + \frac{m}{c^2} \left[ \frac{3}{2} \dot{z}^2 \ddot{z} - (2\gamma+1)  \dot{z}^2 \partial_z \overline{\phi}(z) - (2\gamma+1) \overline{\phi}(z) \ddot{z} \right] + \mathcal{O}(c^{-4}),
\end{align}
\end{subequations}
where we have used that $\dot{\overline{\phi}}(z) \dot{z} = \dot{z}^2 \partial_z \overline{\phi}(z)$, which holds because $\frac{\dd}{\dd \NewCoordsTime} \overline{\phi} (z(\NewCoordsTime))= \frac{\partial \overline{\phi}}{\partial z} \frac{\dd z}{\dd \NewCoordsTime}$. The Euler-Lagrange equation $\frac{\dd }{\dd \NewCoordsTime}\frac{\partial L_\HamiltonianMotional }{\partial \dot{z}} - \frac{\partial L_\HamiltonianMotional }{\partial z} =0$ can then be written recursively in terms of $\ddot{z}$ as
\begin{align*}
	\ddot{z} = - \partial_z \overline{\phi}(z) + 
	\frac{1}{c^2} \left[(2\gamma+1) \dot{z}^2 \partial_z \overline{\phi}(z) - \frac{3}{2} \dot{z}^2 \ddot{z} + (2\gamma+1) \overline{\phi}(z) \ddot{z} - (2 \beta-1)  \overline{\phi}(z) \partial_z \overline{\phi}(z) - \frac{2 \gamma+1}{2} \dot{z}^2 \partial_z \overline{\phi}(z) - 2(\beta-1) \phi_0 \partial_z \overline{\phi}(z) \right] + \mathcal{O}(c^{-4}).
\end{align*}
Inserting the $c^0$-contribution into the $\ddot{z}$-term on the right-hand side results in
\begin{align}
	\ddot{z} = - \partial_z \left[\left( 1 + 2 (\beta-1) \frac{\phi_0 }{c^2}\right) \overline{\phi}(z) + \frac{\beta + \gamma}{c^2} \overline{\phi}(z)^2   \right] + \frac{\gamma+2}{c^2} \dot{z}^2 \partial_z \overline{\phi}(z) + \mathcal{O}(c^{-4}) .
\end{align}
We now use the approximation of the gravitational potential to relevant order, i.e., $\overline{\phi}(z)= g z - \frac{1}{2} \Gamma z^2 + \frac{1}{3} \Lambda z^3 + \mathcal{O}\left(\partial^4_r \phi \right)$, which gives the Euler-Lagrange equation 
\begin{align}
	\ddot{z}(\NewCoordsTime) &= - \qty( 1 + 2 (\beta-1)\frac{\phi_0}{c^2} ) \, g + \qty(1 + 2 (\beta - 1) \frac{\phi_0}{c^2}) \, \Gamma \, z(\NewCoordsTime) - \Lambda \, z(\NewCoordsTime)^2 - 2 g^2 \frac{ \beta+\gamma }{c^2} z(\NewCoordsTime) + (\gamma+2) \frac{g}{c^2} \dot{z}(\NewCoordsTime)^2 + \mathcal{O}\left(\partial^4_r \phi , \Lambda c^{-2}, c^{-4} \right),
\end{align}
with initial conditions $z(0) = z_0, ~ \dot{z}(0)= v_0 + N_R \frac{\hbar k_R}{m} + N_B \frac{\hbar k_B}{m}$, which in the main text was written using the dimensionless trajectory $\xi(\DimTime)$ as
\begin{align}
    \ddot{\xi}(\DimTime) = - \GOneR + \GTwoR \xi(\DimTime) - \GThreeR \xi(\DimTime)^2 
    - 2 (\beta + \gamma) \GOneR^2 \xi(\DimTime) + (\gamma + 2) \GOneR \dot{\xi}(\DimTime)^2  
    + 2 (\beta - 1) \GZero \qty[ - \GOneR + \GTwoR \xi(\DimTime)] + \mathcal{O}(4),
\end{align}
with initial conditions $\xi(0) = \ZZero, ~ \dot{\xi}(0) = \VZero + N_R \RR + N_B \RB$. The perturbative solution of this equation to second order reads
\begin{align}
	\xi(\DimTime) = \xi(0) + \dot{\xi}(0) \DimTime - \frac{1}{2} \qty(1 + 2(\beta - 1) \GZero) \GOneR \DimTime^2 + \GTwoR \qty(\frac{1}{2} \xi(0) \DimTime^2 + \frac{1}{6} \dot{\xi}(0) \DimTime^3 - \frac{1}{24} \GOneR \GTwoR \DimTime^4) 
    + \mathcal{O}(3).
\end{align}
The propagation phase is given by the time integral over $L_\HamiltonianMotional/\hbar$, which we can now translate into dimensionless quantities as explained in the main text. We obtain
\begin{multline}
    \frac{1}{\hbar}\int L_\HamiltonianMotional \dd \NewCoordsTime 
    = \Comp T_R \int \bigg( - \GOneR \xi(\DimTime) + \frac{1}{2} \GTwoR \xi(\DimTime)^2 
    - \frac{1}{3} \GThreeR \xi(\DimTime)^3 + \frac{1}{2} \dot{\xi}(\DimTime)^2 + \frac{1}{8} \dot{\xi}(\DimTime)^4 - \frac{2 \beta - 1}{2} \GOneR^2 \xi(\DimTime) \\
    - \frac{2 \gamma + 1}{2} \GOneR \xi(\DimTime) \dot{\xi}(\DimTime)^2 
    - 2 (\beta - 1) \GZero \GOneR \xi(\DimTime)
    + 2 (\beta - 1) \GZero \GTwoR \xi(\DimTime)^2 \bigg) \dd \DimTime+ \mathcal{O}(5) \label{eq: Lagrangian_Dimensionless_written_out}.
\end{multline}
\end{widetext}

\section{Maxwell's equation in curved spacetime}\label{Appendix: Electromagnetism}

Let us start again with Maxwell's equations in vacuum, i.e.,
\begin{align}
	\tensor{F}{^\alpha ^\beta_{;\beta}} = \nabla_\beta \tensor{F}{^\alpha ^\beta} = \nabla_\beta \left( \nabla^\alpha \tensor{A}{^\beta} - \nabla^\beta \tensor{A}{^\alpha} \right) = 0, \label{eq: Maxwell equation Appendix}
\end{align}
where we have expressed the field strength tensor in terms of the (4-)vector potential $A^\alpha$. Commuting the covariant derivatives in the first term at the expense of introducing a curvature term, this becomes
\begin{align}
	\tensor{A}{^\beta_{;\beta}^{; \alpha}} + \tensor{R}{^\alpha_\beta} \tensor{A}{^\beta} - \tensor{A}{^{\alpha ;\beta}_{;\beta}} = 0.
\end{align}
From now on working in Lorenz gauge, with gauge condition
\begin{subequations}
\begin{align}
    \tensor{A}{^\beta_{;\beta}} &= 0, \label{eq: Curvy_Lorenz_gauge} \\
\intertext{this simplifies to}
	-\tensor{A}{^{\alpha ;\beta}_{;\beta}} + \tensor{R}{^\alpha_\beta} \tensor{A}{^\beta} &= 0. \label{eq: Curvy-Maxwell_Lorenz_gauge}
\end{align}
\end{subequations}
(We will later comment on how the Lorenz gauge is in our case related to the Coulomb gauge, which is the one that is commonly employed in quantum optics.)

\subsection*{Geometric optics approximation}

We now want to solve these equations in the approximation of geometric optics, which following~\cite[§22.5.]{misner1973gravitation}\ we implement by a series expansion. This approach was also recently used in~\cite{di2022light} to describe the light propagation in atom IFs with a non-vanishing Dilaton field. Denoting the wavelength of the light field by $\lambda$, we can deduce the two important length scales for the approximation to be
\begin{align*}
	\mathscr{R} &= \abs{\substack{ \text{typical component of } \tensor{R}{^{\mu}_{\nu \sigma \tau}} \text{ as}\\ \text{measured in a local Lorentz frame} } }^{-\frac{1}{2}} \approx \abs{\partial_r \frac{\overline{\phi}}{c^2}}^{-\frac{1}{2}} \approx 10^8 \mathrm{m} \\
	\mathscr{L} &= \abs{\substack{\text{radius of curvature} \\ \text{of a wave front}}}  \approx 1 \mathrm{mm}.
\end{align*}
We then define the small expansion parameter
\begin{align*}
    \epsilon = \lambda / (2\pi \min ( \mathscr{R},\mathscr{L} )) = \lambda/(2\pi \mathscr{L}),
\end{align*}
which can be thought of as the ratio of the wavelength of the light field to the length scale $\mathscr{L}$ of variation of a slowly varying envelope of the field, divided by $2\pi$.  We now assume the 4-potential to be given by
\begin{align}
	A_\mu = \left( a_\mu + \mathcal{O}(\epsilon) \right) e^{\ii \Phi/\epsilon}, \label{eq: Geo_Optics_Ansatz}
\end{align}
where the components $a_\mu \colon \mathcal{M} \rightarrow \mathbb{C}$ of the leading-order amplitude are complex-valued functions on spacetime and $\Phi \colon \mathcal{M} \rightarrow \mathbb{R}$ is the real-valued phase. The relativistic wave vector (or 4-wave vector) is then defined as $k = \mathrm{d} \Phi$, i.e.\ in components $k_\mu = \partial_\mu \Phi$.  Following~\cite[§22.5.]{misner1973gravitation}, we will refer to everything of order up to $\epsilon^{-1}$ as `geometric optics', whereas we ignore `post-geometric optics' orders, i.e., orders $\epsilon^{0}$ and higher.

Plugging the ansatz Eq.~\eqref{eq: Geo_Optics_Ansatz} into Maxwell's equations Eq.~\eqref{eq: Curvy-Maxwell_Lorenz_gauge} and the Lorenz gauge condition Eq.~\eqref{eq: Curvy_Lorenz_gauge} and grouping terms by their orders in $\epsilon$, one obtains the following (for derivations, see~\cite[§22.5.]{misner1973gravitation}):
\begin{itemize}
    \item Eq.~\eqref{eq: Curvy-Maxwell_Lorenz_gauge} at leading order $\epsilon^{-2}$ is equivalent to the wave vector $k_\mu$ being lightlike, i.e., satisfying $k_\mu k^\mu = g^{\mu \lambda} k_\mu k_\lambda = 0$.  Taking the covariant derivative of this condition and using that $k = \mathrm{d} \Phi$, one further obtains that $k_\mu$ must be geodesic, i.e., satisfy $k^\mu \nabla_\mu k^\nu = 0$.
    \item Eq.~\eqref{eq: Curvy_Lorenz_gauge} at leading order $\epsilon^{-1}$ is equivalent to $k^\mu a_\mu = 0$, i.e., the leading-order amplitude being orthogonal to the wave vector.
    \item Eq.~\eqref{eq: Curvy-Maxwell_Lorenz_gauge} at subleading-order $\epsilon^{-1}$ determines the propagation law for $a_\mu$ along the `light rays' (i.e.\ null geodesics integrating the vector field $k^\mu$): it must satisfy $k^\mu \nabla_\mu a_\nu = -\frac{1}{2} a_\nu \nabla_\mu k^\mu$.  Note that combining this propagation law with the geodesic equation for $k^\mu$ implies $k^\mu \nabla_\mu (a_\nu k^\nu) = -\frac{1}{2} (\nabla_\mu k^\mu) (a_\nu k^\nu)$, such that if we start integrating the propagation law along a geodesic with an initial $a_\mu$ that is orthogonal to $k^\mu$, this will automatically continue to hold along the geodesic.
\end{itemize}

\subsubsection*{Calculating the wave vector}

Because we have assumed our spacetime to be stationary with time-like Killing vector field $\partial_0 = c^{-1} \partial_\NewCoordsTime$, we know that, to the level of approximation we are working with, the wave vector's corresponding timelike component $k_0 = k_\mu (\partial_0)^\mu = c^{-1} \partial_t\Phi$ is constant along each geodesic / light ray that arises as an integral curve of the vector field $k^\mu$.  Note that due to the normalization of the phase in the ansatz Eq.~\eqref{eq: Geo_Optics_Ansatz}, differently to the main text, now the light's angular frequency as measured by a stationary observer at rest at the experiment's reference point is given by $\omega_0 = - c k_0/\epsilon$.  We will keep this normalization for the remainder of this appendix.  Also, since all the remaining equations that we will be considering are linear in $k_\mu$, the overall normalization of $k_\mu$ does not matter, and for notational convenience we will treat it to be of order $c^0$ for the remainder of this appendix.

We will only consider EM waves traveling in one spatial dimension, namely the $z$ (i.e.\ vertical) direction.  Therefore, we will take $k_x = k_y = 0$ and neglect $x$ and $y$ dependence of $k_\mu$.  The condition of $k^\mu$ being lightlike then takes the form $0 = g^{\mu\lambda} k_\mu k_\lambda = \left(-1 + 2 \frac{\overline{\phi}(z)}{c^2}\right) k_0^2 + \left(1 + 2 \gamma \frac{\overline{\phi}(z)}{c^2}\right) k_z^2 + \mathcal{O}(c^{-4})$, or equivalently $k_z^2 = \left(1 - 2(\gamma +1) \frac{\overline{\phi}(z)}{c^2}\right) k_0^2 + \mathcal{O}(c^{-4})$.  We now also restrict our consideration to the case of neighboring `light rays' (i.e.\ null geodesics integrating the vector field $k^\mu$) having the same timelike wave vector component $k_0$, such that for our purposes $k_0$ is constant.  Therefore, to our order of approximation, the wave vector's $z$ component is determined as
\begin{align}
	k_z = k_z(z) = \pm \left( 1 - (\gamma+1) \frac{\overline{\phi}(z)}{c^2} \right) k_0 + \mathcal{O}(c^{-4}), \label{eq: Solution_k_z_in_terms_of_k_0}
\end{align}
with the sign depending on the direction of propagation.  Hence, restricting to the $(c \NewCoordsTime, z)$ components, we obtain the relativistic wave vector
\begin{align}
	(k_\mu (z)) = 
	\begin{pmatrix} 
		1 \\ 
		\pm \left( 1 - (\gamma + 1) \frac{\overline{\phi}(z)}{c^2} \right)
	\end{pmatrix} 
	k_0 + \mathcal{O}(c^{-4}).
\end{align}
Since $k_\mu = \Phi_{,\mu}$ we therefore know that the phase of the EM field is to our order of approximation given by
\begin{align}
	\Phi (z, \NewCoordsTime) 
	= \Phi_0 + k_0 c \NewCoordsTime \pm \left( 1 - \frac{\gamma + 1}{2} \frac{g z}{c^2} \right) k_0 z + \mathcal{O}(\Gamma \, c^{-2}), \label{eq: Solution_Phase}
\end{align}
where $\Phi_0$ is an arbitrary offset, which we will set to zero.  The contravariant components $k^\mu = g^{\mu\nu} k_\nu$ of the wave vector, which we will need below, are given by
\begin{align}
	(k^\mu (z)) = 
	\begin{pmatrix} 
	    -1 + 2 \frac{\overline{\phi}(z)}{c^2} \\ 
		\pm \left( 1 + (\gamma - 1) \frac{\overline{\phi}(z)}{c^2} \right)
	\end{pmatrix} 
	k_0 + \mathcal{O}(c^{-4}). \label{eq: Solution_k_contravar}
\end{align}

\subsubsection*{Calculating the amplitude}

We are now going to determine the leading-order amplitude $a_\mu$ by integrating its propagation law
\begin{equation}
    k^\mu \nabla_\mu a_\nu = -\frac{1}{2} a_\nu \nabla_\mu k^\mu
\end{equation}
along the light rays / lightlike geodesics integrating $k^\mu$.  Considering the amplitude $a_\nu(s)$ along a single such geodesic, with $s$ denoting the affine parameter measured from $z=0$, the propagation law takes the form
\begin{equation}
    {a_\nu}'(s) = k^\mu \Gamma^\lambda_{\mu\nu} a_\lambda(s) - \frac{1}{2} a_\nu(s) (\nabla_\mu k^\mu), \label{eq: Propagation_amplitude}
\end{equation}
where the prime denotes the derivative with respect to $s$, and $k^\mu$, the Christoffel symbols, and $\nabla_\mu k^\mu$ are of course also to be evaluated along the geodesic.

The divergence of $k^\mu$ evaluates to
\begin{align}
    \nabla_\mu k^\mu &= \partial_\mu k^\mu + k^\mu \Gamma^{\rho}_{\mu\rho} \nonumber\\
    &= \partial_\mu k^\mu + k^\mu \partial_\mu \ln\sqrt{-g} \nonumber\\
    &= \mp 2 \gamma \frac{g}{c^2} k_0 + \mathcal{O}(\Gamma c^{-2}),
\end{align}
which follows from the explicit form of $k^\mu$ in Eq.~\eqref{eq: Solution_k_contravar} and $\ln\sqrt{-g} = \ln\sqrt{1 + (2 - 6 \gamma) \frac{\overline\phi}{c^2} + \mathcal{O}(c^{-4})} = (1 - 3 \gamma) \frac{\overline\phi}{c^2} + \mathcal{O}(c^{-4})$.  Inserting this and the explicit form of the Christoffel symbols, the amplitude propagation law Eq.~\eqref{eq: Propagation_amplitude} takes the explicit form
\begin{subequations} \label{eq: Propagation_amplitude_explicit}
\begin{align}
    {a_0}'(s) &= \left(\pm(\gamma + 1) \frac{g}{c^2} a_0(s) - \frac{g}{c^2} a_z(s)\right) k_0 + \mathcal{O}(\Gamma c^{-2}), \\
    {a_x}'(s) &= \mathcal{O}(\Gamma c^{-2}), \\
    {a_y}'(s) &= \mathcal{O}(\Gamma c^{-2}), \\
    {a_z}'(s) &= - \frac{g}{c^2} a_0(s) k_0 + \mathcal{O}(\Gamma c^{-2}).
\end{align}
\end{subequations}
To solve this first-order system of ODEs, we need an initial condition for the amplitude $a_\mu$.  We want this initial value $a_\mu(s=0)$ (i.e.\ the value at the point of the light ray at height $z = 0$) to be purely transversal, i.e.\ given by $(a_0(0), a_x(0), a_y(0), a_z(0)) = (0, \mathcal{A}_x, \mathcal{A}_y, 0)$.  Note that this initial value is consistent with the condition $a_\mu k^\mu = 0$.  We directly see that up to our level of approximation, i.e., order $\Gamma/c^2$, the solution of Eq.~\eqref{eq: Propagation_amplitude_explicit} with this initial condition is constant, i.e., we have
\begin{equation}
    (a_0, a_x, a_y, a_z) = (0, \mathcal{A}_x, \mathcal{A}_y, 0) + \mathcal{O}(\Gamma c^{-2})
\end{equation}
along the whole geodesic.

Combining this with the phase determined above in Eq.~\eqref{eq: Solution_Phase}, we end up with an EM vector potential in geometric optics approximation given by
\begin{align}
	(A_i) = \bm{A} = \bm{\mathcal{A}} \, e^{\ii \left( k_0 c \NewCoordsTime \pm \left( 1 - \frac{\gamma + 1}{2} \frac{g z}{c^2} \right) k_0 z \right)} + \mathcal{O} (\Gamma \, c^{-2}),
\end{align}
where we introduced $\bm{\mathcal{A}} = (\mathcal{A}_x, \mathcal{A}_y, 0)$, and the electric scalar potential $-c A_0$ vanishing to our approximation. Note that we have now absorbed the factor $\epsilon^{-1}$ from the exponent of the ansatz Eq.~\eqref{eq: Geo_Optics_Ansatz} into the normalization of $k_0$, s.t. the light's angular frequency as measured by a stationary observer at rest at the origin is again given by $\omega_0 = - c k_0$ as in the main text.  Note also that this EM 4-potential satisfies (to our approximation) not only the Lorenz gauge condition using which it was derived, but also the `geometric Coulomb' gauge $\nabla_i A^i = 0$ (which is a somewhat sensible notion, given the $3+1$-split form of the metric): to our approximation, the difference term between the two gauge conditions is $\nabla_0 A^0 = \partial_0 A^0 + \Gamma^0_{0i} A^i + \dots = \partial_0 A^0 + \frac{\overline{\phi}_{,i}}{c^2} A^i + \dots$, and due to $A^0$, $A^z$, $\overline{\phi}_{,x}$ and $\overline{\phi}_{,y}$ vanishing in our approximation, this vanishes as well.

\section{Atom-Light Hamiltonian}\label{Appendix: Atom Light Hamiltonian}

We now consider counter-propagating light fields that describe interaction processes in atom IFs, i.e., we define two light fields with temporal wave vector coefficients $k_{0, i}$, for $i = 1, 2$, corresponding to frequencies $\omega_{R,i} = -k_{0, i} c$, as measured by a resting observer at the origin. The coefficient appearing in the spatial component of the wave four-vectors of the two light fields will be given by the $k_{0, i}$ with opposite signs, i.e., for the first laser it will be positive and denoted by $k_{R, 1} = |k_{0, 1}| = \omega_{R,1}/c$, for the second one it will be negative and given by $-k_{R, 2} = -|k_{0, 2}| = -\omega_{R,2}/c$. We will add a time dependence to the amplitudes, as this will be used to create certain pulse shapes in the experiment, and write each of the corresponding vector potentials as
\begin{align}
	\bm{A}_i(z, \NewCoordsTime) = \bm{\mathcal{A}}_i(\NewCoordsTime) 
	e^{\ii \Phi_i(z, \NewCoordsTime) } + \mathcal{O}(\Gamma \, c^{-2}), \label{eq: Vectorpotential Counterpropagating light fields}
\end{align}
where the phase expression is given by
\begin{align}
	\Phi_{i}(z, \NewCoordsTime) = - \omega_{R,i} \NewCoordsTime \pm \left(1 - \frac{\gamma + 1}{2} \frac{g z}{c^2}  \right) k_{R,i} z + \mathcal{O}(\Gamma \, c^{-2}), \label{eq: Phase of counterpropagating lasers}
\end{align} 
with $i=1,2$ respectively. Using canonical quantization we write every motional variable w.r.t. the position and momentum operators $\ZZ$ and $\PZ$. Using the vector potential we get expressions for the electric and magnetic fields via $\bm{E}_i(\ZZ, \NewCoordsTime) = - \partial_\NewCoordsTime \bm{A}_i(\ZZ, \NewCoordsTime)$ and $\bm{B}_i(\ZZ, \NewCoordsTime) = \nabla \times \bm{A}_i(\ZZ, \NewCoordsTime)$ as
\begin{subequations}
\begin{align}
    \bm{E}_i(\ZZ, \NewCoordsTime) &= \bm{\mathcal{E}}_i(\NewCoordsTime) 
     e^{\ii \Phi_i(\ZZ, \NewCoordsTime)} + \mathcal{O}(\Gamma \, c^{-2}), \\[6pt]
    \bm{B}_i(\ZZ, \NewCoordsTime) &= \pm \bm{\mathcal{B}}_i(\NewCoordsTime) 
     e^{\ii \Phi_i(\ZZ, \NewCoordsTime)} + \mathcal{O}(\Gamma \, c^{-2}),
\end{align}
\end{subequations}
where the amplitudes are given by
\begin{align*}
    \bm{\mathcal{E}}_i(\NewCoordsTime) &= - \ii \omega_{R,i} \bm{\mathcal{A}}_i(\NewCoordsTime) - \dot{\bm{\mathcal{A}}}_i(\NewCoordsTime), \\
    \bm{\mathcal{B}}_i(\NewCoordsTime) &= \ii k_{R, i} \, \bm{e}_z \times \bm{\mathcal{A}}_i(\NewCoordsTime).
\end{align*}
The interaction Hamiltonian Eq.~\eqref{eq: Atom Light Hamiltonian} then takes the form
\begin{align}
    \hat{H}_\HamiltonianAtomLight &= - \hat{\bm{d}} \cdot \bm{E}_i(\ZZ, \NewCoordsTime)  + \frac{1}{2m} \left[ \PZ \cdot \left( \hat{\bm{d}} \times \bm{B}_i(\ZZ, \NewCoordsTime) \right) + \text{h.c.} \right] . \label{eq: Interaction_Hamiltonian_1D_Appendix}
\end{align}
We now transform into the interaction picture corresponding to the internal Hamiltonian $H_\HamiltonianInternal$ in Eq.~\eqref{eq: Atomic Hamiltonian}, which only alters the dipole operator
\begin{align}
    \hat{\bm{d}}(\NewCoordsTime) = e^{\ii H_\HamiltonianInternal \NewCoordsTime / \hbar}\, \hat{\bm{d}} \, e^{-\ii H_\HamiltonianInternal \NewCoordsTime / \hbar} = \bm{d}_\text{eg} \ketbra{e}{g} e^{\ii \omega_\text{eg}\NewCoordsTime}
\end{align}
with $\bm{d}_\text{eg} = \bra{e} \hat{\bm{d}} \ket{g}$. Note that in order to commute the $\ZZ$-dependent phase of the magnetic field in the Röntgen term past the $\PZ$ we will use the relation $e^{\pm \ii k \ZZ} \PZ = \left( \PZ \mp \hbar k \right)e^{\pm \ii k \ZZ}$ (cf.~\cite{sonnleitner2017rontgen}). We also use the Graßmann identity to simplify $\left( \bm{d}_\text{eg} \times \left( \bm{e}_z \times \bm{\mathcal{A}}_j \right) \right)_z = \bm{d}_\text{eg} \cdot \bm{\mathcal{A}}_j$.

The full Hamiltonian, written in the rotating wave approximation and in the interaction picture w.r.t. $\hat{H}_\HamiltonianInternal$, takes the form
\begin{align}
    \hat{H} = \hat{H}_\HamiltonianMotional + \sum\limits_i \frac{\hbar \Omega_i(\ZZ, \PZ, \NewCoordsTime)}{2} \ketbra{\text{e}}{\text{g}} e^{\ii \left( \pm k_{i} (\ZZ) \ZZ - \left( \omega_{R, i} - \omega_\text{eg} \right) \NewCoordsTime \right)} \label{eq: Interaction_Hamiltonian_Appendix},
\end{align}
where the Rabi frequency has an additional dependency on the momentum
\begin{align*}
    \frac{\hbar \Omega_i (\PZ, \NewCoordsTime)}{2} =  - \bm{d}_\text{eg} \cdot \bm{\varepsilon}_i \, \mathcal{E}_i (\NewCoordsTime )
    \pm \qty(\PZ \mp \hbar k_{R,i}/2 ) \frac{ \bm{d}_\text{eg} \cdot \bm{\varepsilon}_i}{m} \mathcal{B}_i(\NewCoordsTime),
\end{align*}
and the coordinate wave vector is given by
\begin{align}
	 k_i (\ZZ) &= \qty(1 - \qty(\gamma + 1) \frac{g}{c^2} \ZZ ) k_{R,i} + \mathcal{O}(c^{-4}).
\end{align}

The non-relativistic part of the $\ZZ$-dependent exponential of Hamiltonian Eq.~\eqref{eq: Interaction_Hamiltonian_Appendix} is just the momentum translation
\begin{align}
    e^{\pm \ii k_{R, i} \ZZ} = \int \dd p \ketbra{p \pm \hbar k_{R, i}}{p},
\end{align}
whereas we now also have an additional contribution $\exp(\pm \ii \alpha k_{R,i} \ZZ^2)$ with $\alpha = -(\gamma+1) \frac{g}{c^2}$. Since this operator will act on spatially well defined Gaussian wave packets it is clear that the full operator $\exp(\pm \ii (1 + \alpha \ZZ ) k_{R,i} \ZZ)$ will map a momentum eigenstate $\ket{p}$ to $\ket{p \pm \hbar k_{R,i} (1 + \alpha z)}$, where $z$ is the expectation value of $\ZZ$ w.r.t. the initial wave packet.

\subsection*{Doppler effect}

Since we want to include all relativistic effects in atom IFs, including terms of the order $c^{-2}$, we need to not only consider the first-order Doppler effect, but also the second-order. Consider a atom-light interaction and the atoms have a velocity of $v$ as measured in the lab-frame. Note that a distinction between coordinate and proper velocity is not needed here, since both notion will differ by $\mathcal{O}(c^{-2})$-terms, which will manifest at the $\mathcal{O}(c^{-3})$ level for the first-order Doppler effect.

The light fields' frequencies will subsequently be perceived by the atoms as Doppler-shifted, characterized by $\left(1 \mp \frac{v}{c} + \frac{v^2}{2c^2}\right) \omega_{R,i}$, where the '+' corresponds to the first laser, moving upward, and vice versa. The wave vectors will, analogously, be shifted as $\pm \left( 1 \mp \frac{v}{c} + \frac{v^2}{2c^2} - (\gamma + 1)\frac{g z}{c^2} \right) k_{R,i}$. In order to compensate this Doppler shifts, one has to rescale the laser frequencies inversely, i.e.,
\begin{align*}
    \omega_{R, 1} &\longmapsto \tilde{\omega}_{R, 1}(v) = \qty(1 + \frac{v}{c} - \frac{v^2}{2c^2} + (\gamma + 1)\frac{g z}{c^2}) \omega_{R, 1}, \\ 
    \omega_{R, 2} &\longmapsto \tilde{\omega}_{R, 2}(v) = \qty(1 - \frac{v}{c} - \frac{v^2}{2c^2} + (\gamma + 1)\frac{g z}{c^2}) \omega_{R, 2},
\end{align*}
where the gravitational contribution was discussed previously and needs to be adjusted in order to resonantly induce a momentum kick of $\hbar(k_{R, 1} + k_{R, 2}) = \hbar k_R$.

\subsection*{Bragg scattering matrix and effective laser phase}

We can see that relativistic effects enter the atom-light interaction as corrections to the imprinted phase and momentum to the atoms. Those corrections consist (apart from the well-known first-order Doppler shift) of the second-order Doppler shift and a gravitational contribution. Similar to~\cite{siemss2020analytic, giese2015mechanisms} one can now understand Bragg transitions in terms of scattering matrices between atomic momentum eigenstates as in Eq.~\eqref{eq: Scattering_Matrix_Bragg} but now using the Doppler corrected phase from Eq.~\eqref{eq: Effective_Laser_Phase_Atomic_Rest_Frame_Dimensionless}. We are now able to write down an expression for the imprinted laser phase of the Bragg transition at each interaction event ($z_\text{int} = z, v_\text{int} = v)$ as
\begin{multline}
	\Phi_L(z, v) = \pm \qty(\Phi_{1}(z, v, \Delta \NewCoordsTime_1) - \Phi_{2}(z, v, \Delta \NewCoordsTime_2)) \\
	= \pm \qty[ \left( 1 + \frac{\gamma + 1}{2} \frac{g z}{c^2} - \frac{v^2}{2c^2} \right) k_R + \frac{v}{c} \frac{\omega_R}{c} ] \, z \pm \Phi_\text{FSL}(v, \Delta \NewCoordsTime_1, \Delta \NewCoordsTime_2), \label{eq: Effective_Laser_Phase_Rest_Frame_Appendix}
\end{multline} 
where $\Delta \NewCoordsTime_i$ is the photon flight time from emission to interaction of light field $i$, the sign corresponds to net momentum gain or loss and we subsumed the temporal parts of Eq.~\eqref{eq: Phase of counterpropagating lasers} into
\begin{align}
    \Phi_\text{FSL}(v, \Delta\NewCoordsTime_1, \Delta\NewCoordsTime_2) = - \tilde{\omega}_{R, 1}(v) \Delta \NewCoordsTime_1 + \tilde{\omega}_{R, 2}(v) \Delta \NewCoordsTime_2. \label{eq: FSL_Phase_1_Order}
\end{align}

\subsection*{Finite speed of light corrections}

Let us summarize which FSL effects the Python code in~\cite{PythonCode} will neglect. Consider for this a light field which is emitted at a time $\NewCoordsTime_\text{emit}$ and interacts with the atom at $\NewCoordsTime_\text{int} = \NewCoordsTime_\text{emit} + \Delta \NewCoordsTime$. 

\begin{enumerate}
	\item The temporal part of the phase for each single photon interaction, or analogously the contribution $\Phi_\text{FSL}$ in Eq.~\eqref{eq: FSL_Phase_1_Order} in a two-photon process, will not be included in the laser phase, since it is directly proportional to the respective photon flight times. 
	
	\item The spatial part of the phase $\Phi_L(z, \dot{z})$ in Eq.~\eqref{eq: Effective_Laser_Phase_Atomic_Rest_Frame} needs to be evaluated at $\Phi_L(z(\NewCoordsTime_\text{int}), \dot{z}(\NewCoordsTime_\text{int}))$ and not at $\Phi_L(z(\NewCoordsTime_\text{emit}), \dot{z}(\NewCoordsTime_\text{emit}))$.
	
	\item The integration limits in the propagation phase integrals need to be shifted $\NewCoordsTime_\text{emit} \longmapsto \NewCoordsTime_\text{int}$.
	
	\item The velocity $v$ in the Doppler effect, i.e., the atomic velocity needs to be evaluated as $v = \dot{z}(\NewCoordsTime_\text{int})$ and not as $v = \dot{z}(\NewCoordsTime_\text{emit})$.
\end{enumerate}

Note that the first of those four effects is by far the biggest, since the phase of an EM wave is oscillating fast, even on time scales of $\Delta \NewCoordsTime$. The other three contributions directly depend on the atomic position/velocity difference between emission and interaction time, which is rather small due to the slow atomic velocity. We will comment on those approximations further in App.~\ref{Appendix: Mach-Zehnder-Interferometer} in the analysis of an explicit Mach-Zehnder type IF of~\cite{dimopoulos2008general}.

\section{Mach-Zehnder IF -- Comparison to Dimopoulos et al.}\label{Appendix: Mach-Zehnder-Interferometer}

The papers~\cite{dimopoulos2008general, dimopoulos2007PRL} described a Raman matter wave IF of a Mach-Zehnder type in the PPN spacetime. In order to compare our results to their findings we need to adapt their notation to ours. We list all needed differences in sign conventions and notation in Table~\ref{table: Comparison_Notation_Dimopoulos}. Note that due to a different sign convention in the phases of the EM waves one needs to redefine $\bm{k}_i \longmapsto - \bm{k}_i$ and $\omega_i \longmapsto - \omega_i$ for $i = 1,2$, which gives an overall sign change in $\omega_R$ but not in $k_R$.
\begin{table}
    \centering
    \begin{tabular}{|c|c|}
        \hline
        Dimopoulos et al.   &  Our notation                                     \\
        \hline
        $g$                 & $\quad -\qty(1 + 2 \gamma \frac{\phi_0}{c^2})\, g \quad $ \\
        \hline
        $\partial_r g$      & $\qty(1 + 3 \gamma \frac{\phi_0}{c^2}) \, \Gamma$ \\
        \hline
        $\partial_r^2 g$    & $- \frac{1}{2} \Lambda  $                         \\
        \hline
        $T$                 & $T_R$                                             \\
        \hline
        $\omega_a  $        & 0                                                 \\
        \hline
        $\kappa_\text{eff}$ & $k_R$                                             \\
        \hline
        $\omega_\text{eff}$ & $- \omega_R$                                      \\
        \hline
        $v_L$               & $v_0$                                             \\
        \hline
    \end{tabular}
    \caption{Comparison of notation between the analysis by Dimopoulos et al.~\cite{dimopoulos2007PRL, dimopoulos2008general} and this work.}
    \label{table: Comparison_Notation_Dimopoulos}
\end{table}
However, some differences remain, even after aligning the notation. Dimopoulos et al. chose an initial height of $z_0 = 0$ and they included various FSL effects. Note that we assumed that the gravitational parameters $g, \Gamma$ and $\Lambda$ are obtained by evaluating the gravitational potential and their derivatives in the original coordinate system, cf. Eq.~\eqref{eq: Potential_in_original_coords}. It could, however, also be the case that Dimopoulos et al. chose to evaluate those parameters using `metric lengths' and not `coordinate lengths', which would shift those definitions by some factors of $\phi_0/c^2$.

We therefore expect deviations of our results in comparison to~\cite{dimopoulos2008general} at order $\mathcal{O}(4)$. As mentioned before does our provided Python code not include FSL effects for arbitrary IF geometries. We will now, however, show how to calculate the dominant parts of the FSL phases for the explicit setup proposed by Dimopoulos et al. and will elaborate on the orders of magnitude of other FSL effects. For this we will assume that the light fields will be emitted at heights $z_\text{low}$ and $z_\text{up}$ respectively. Note that for 10m baselines the maximal flight time for photons is therefore $\Delta T < 10^{-7} \mathrm{s}$. 

\paragraph{FSL terms of first order:}

The biggest FSL contribution is the neglected temporal part of the phase of each EM field, i.e., the terms given by Eq.~\eqref{eq: FSL_Phase_1_Order}. To evaluate this, we need to know the photonic flight time between emission and interaction. In order to keep the formulas short we will explain the derivation using non-relativistic trajectories. 

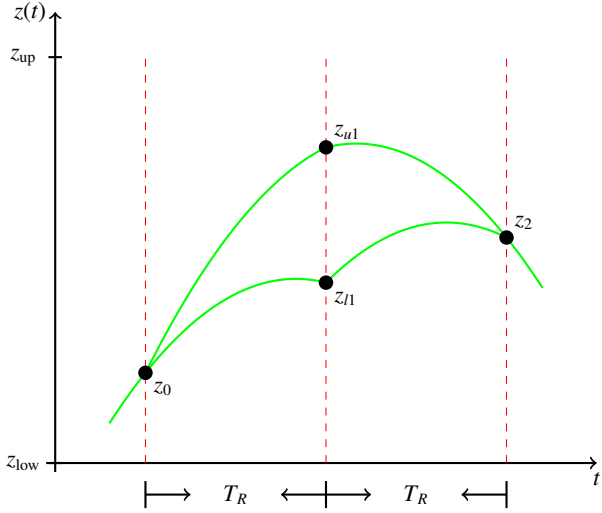
\begin{figure}
	\centering
	\begin{tikzpicture}[scale=1.2]
		\draw[->,thick] (-0.1,0) -- (6,0) node[right, below] {$\NewCoordsTime$}; 		
		\draw[->,thick] (0,-0.1) -- (0,5) node[above, left] {$z(\NewCoordsTime)$}; 		
		\draw[color=red,dashed] (1,0) -- (1,4.5);							    	    
		\draw[color=red,dashed] (3,0) -- (3,4.5);
		\draw[color=red,dashed] (5,0) -- (5,4.5);
		
		\draw[domain=0:1, smooth, variable=\y, green, thick]  plot (2*\y + 1, 4*\y - 1.5*\y*\y + 1);	
		\draw[domain=-0.2:1, smooth, variable=\y, green, thick]  plot (2*\y + 1, 2.5*\y - 1.5*\y*\y + 1);	
		
		\draw[domain=0:1.2, smooth, variable=\y, green, thick]  plot (2*\y + 3, 0.5*\y - 1.5*\y*\y + 3.5);	
		\draw[domain=0:1, smooth, variable=\y, green, thick]  plot (2*\y + 3, 2*\y - 1.5*\y*\y + 2);	

		\draw[color=black,thick] (-0.075,0) -- (0.075,0);
		\draw (-0.35,0) node {$z_\text{low}$};	
		\draw[color=black,thick] (-0.075,4.5) -- (0.075,4.5);
		\draw (-0.35,4.5) node {$z_\text{up}$};	
		
		\draw (2,-0.35) node {$T_R$};					            	
		\draw (4,-0.35) node {$T_R$};
		\draw[->,thick] (1,-0.35) -- (1.5,-0.35);		                
		\draw[<-,thick] (2.5,-0.35) -- (3,-0.35);
		\draw[->,thick] (3,-0.35) -- (3.5,-0.35);
		\draw[<-,thick] (4.5,-0.35) -- (5,-0.35);
		\draw[color=black,thick] (1,-0.2) -- (1,-0.5);		            
		\draw[color=black,thick] (3,-0.2) -- (3,-0.5);
		\draw[color=black,thick] (5,-0.2) -- (5,-0.5);
		
		\fill[black] (1,1) circle (0.08) node[below right] {$z_0$};	    
		\fill[black] (3,2) circle (0.08) node[below right] {$z_{l1}$};
		\fill[black] (3,3.5) circle (0.08) node[above right] {$z_{u1}$};
		\fill[black] (5,2.5) circle (0.08) node[above right] {$z_2$};
	\end{tikzpicture}
	\caption{Graphical description of interaction points (black dots) of photon paths (red dashed lines) and atoms (green solid lines) in the laboratory frame. The photon flight times are then defined by the emission height of $z_\text{up}$ or $z_\text{low}$ and the interaction points.}
	\label{Fig: Depiction_of_MZI_Interferometer}
\end{figure}

Let us denote the interaction heights of the upper path by $z_0$ and $z_{u1} = z_0 + v_0 T_R - \frac{1}{2} g T_R^2 + \frac{\hbar k_R}{m} T_R$ and those of the lower path by $z_{l1} = z_0 + v_0 T_R - \frac{1}{2} g T_R^2$ and $z_2 = z_0 + 2 v_0 T_R - 2 g T_R^2 + \frac{\hbar k_R}{m} T_R$. We also depicted these trajectories abstractly in Fig.~\ref{Fig: Depiction_of_MZI_Interferometer}.
We can then write down the propagation times for each photon abstractly \footnote{Note that here we assume that light rays are given by $z=c \NewCoordsTime$, which is only valid to leading order. To be correct to higher order, one would need to solve the geodesic equation. Relativistic corrections of order $\mathcal{O}(c^{-2})$ are, however, not of interest when considering effects that are naturally of order $\mathcal{O}(c^{-1})$.} as $\overline{\Delta \NewCoordsTime}^{(t_j)}_i$ for the upper atomic path and $\underline{\Delta \NewCoordsTime}^{(t_j)}_i$ for the lower atomic path, where $i = 1, 2$ indicates the first or second laser and $t_j$ is the time of laser emission, i.e.,
\begin{align*}
	\overline{\Delta \NewCoordsTime}^{(0)}_{1} &= \frac{z_0 - z_\text{low}}{c}, 
	&\quad \overline{\Delta \NewCoordsTime}^{(0)}_{2} &= \frac{z_\text{up} - z_0}{c}, \\[8pt]
	\overline{\Delta \NewCoordsTime}^{(T_R)}_{1} &= \frac{z_{u1} - z_\text{low}}{c}, 
	&\quad \overline{\Delta \NewCoordsTime}^{(T_R)}_{2} &= \frac{z_\text{up} - z_{u1}}{c}, \\[8pt]
	\underline{\Delta \NewCoordsTime}^{(T_R)}_{1} &= \frac{z_{l1} - z_\text{low}}{c}, 
	&\quad \underline{\Delta \NewCoordsTime}^{(T_R)}_{2} &= \frac{z_\text{up} - z_{l1}}{c}, \\[8pt]
	\underline{\Delta \NewCoordsTime}^{(2 T_R)}_{1} &= \frac{z_2 - z_\text{low}}{c}, 
	&\quad \underline{\Delta \NewCoordsTime}^{(2 T_R)}_{2} &= \frac{z_\text{up} - z_2}{c}.
\end{align*}

The FSL phase of first order is then the sum over all previously neglected terms $\Phi_\text{FSL}$ at each interaction point, i.e.,
\begin{align*}
	&\Phi_\text{FSL,1} = 
	 \Phi_\text{FSL} \qty(v_\text{up}(0), \overline{\Delta \NewCoordsTime}^{(0)}_{1}, \overline{\Delta \NewCoordsTime}^{(0)}_{2}) \\
	 & \hspace{3cm} - \Phi_\text{FSL} \qty(v_\text{up}(T_R), \overline{\Delta \NewCoordsTime}^{(T_R)}_{1}, \overline{\Delta \NewCoordsTime}^{(T_R)}_{2}) \\ 
	 &- \bigg[ \Phi_\text{FSL} \qty(v_\text{low}(T_R), \underline{\Delta \NewCoordsTime}^{(T_R)}_{1}, \underline{\Delta \NewCoordsTime}^{(T_R)}_{2}) \\
	 &\hspace{3cm} - \Phi_\text{FSL} \qty(v_\text{low}(2 T_R), \underline{\Delta \NewCoordsTime}^{(2 T_R)}_{1}, \underline{\Delta \NewCoordsTime}^{(2 T_R)}_{2})
	\bigg].
\end{align*}
Here $v_\text{up}$ and $v_\text{low}$ denote the atomic (coordinate) velocities of each the respective path as a function of time. Again, this contribution needs to be added or subtracted to the final phase shift, depending on whether momentum was gained or lost in the interaction.This contribution will be manually added in the Python code in order to compare our results with~\cite{dimopoulos2008general}.

\paragraph{FSL terms of second order:}

The next-order FSL correction one needs to keep in mind is that the spatial part of the laser phase also needs to be evaluated at the correct interaction time and not at the time of laser emission. One will therefore make an error that corresponds to the additional movement of the atoms in the time $\Delta T$ and their resulting change of height, i.e., to correct for this neglection, we need to make the replacement $k_R z(T_R) \longmapsto k_R z(T_R + \Delta T)$. Since $v_0 \approx g T_R \approx 10 \, \frac{\mathrm{m}}{\mathrm{s}}$ is roughly the maximal velocity of the atoms, the possible error will be bounded by $v_0 \Delta T < 10^{-6} \, \mathrm{m}$ and needs to be accounted for in every laser interaction. Note that this length variation needs to be compared to the measurement uncertainty of the atomic position, which is analyzed in \cite{loriani2020colocation} for a satellite mission, and assumed to also be at the order of $10^{-6} \mathrm{m}$. 

\paragraph{FSL terms of third order:}

The third-order FSL correction arises in the propagation phase. It stems from the fact that the integration limits of the action integral need to be adjusted for the correct flight times of the photons, i.e., it needs to be accounted for according to $T_R + \Delta T$. For example, for the kinetic energy part of the Lagrangian this correction would manifest as
\begin{align}
    \int\limits_{T_R}^{\Delta T_\text{up}} \frac{m}{2} v_\text{up}(\NewCoordsTime)^2 \dd \NewCoordsTime - \int\limits_{T_R}^{\Delta T_\text{low}} \frac{m}{2} v_\text{low}(\NewCoordsTime)^2 \dd \NewCoordsTime.
\end{align}
This effect would then manifest for differences between the photon flight times of the upper and lower IF path, since common photon flight times would effect both IF arms identically. At a path separation of 10cm, this would correspond to a time delay between the IF arms of at most $3 \times 10^{-10}\, \mathrm{s}$.

\paragraph{FSL terms of fourth order:}

Lastly, the velocity used in the Doppler shift formulas need to be adjusted. If we assume an additional atomic velocity of $g \Delta T  < 10^{-6} \, \frac{\textrm{m}}{\textrm{s}}$ than this would manifest in the first-order Doppler shift, due to an additional factor of $c^{-1}$, at around $10^{-14}$.

\subsection*{Phase shift comparison:}

The comparison of our results to~\cite{dimopoulos2008general} can be found in Fig.~\ref{Table: MZI_Phase}. To orders $\mathcal{O}(2)$ and $\mathcal{O}(3)$ all terms except $\#$8, $\#$9 are reproduced, whereas term $\#$8 may result from a different definition of $g$ by Dimopoulos et al., as discussed before. Term $\#$9 appears to arise due to a discrepancy in the computation of FSL effects in~\cite{dimopoulos2008general}, wherein atomic positions, and therefore the photonic propagation times, were calculated using the atomic trajectory without any momentum kicks. Notably, when the number of imprinted photon momenta is set to zero in our calculation of the atomic position in the FSL phase, the outcome aligns with the prefactor of 3/2 as reported by Dimopoulos et al. We have included a comment in the Python algorithm~\cite{PythonCode} that highlights this particular aspect. At order $\mathcal{O}(4)$, differences were expected, for example, due to our neglection of FSL effects. Note that we also reproduce all terms of~\cite[Table 1]{hogan2008light} which are non-zero for our system, if we leave out $\Lambda$-corrections to the atomic trajectory.

\newcommand{\Green}{\cellcolor{green!15}}

\begin{table*}
	\footnotesize
	\centering
	\begin{tabularx}{\linewidth}{|c|X|c|c|c|c|c|}
		\hline
		\multicolumn{7}{|c|}{\textbf{Comparison to the Mach-Zehnder IF from~\cite{dimopoulos2008general}}} \\
		\hline
		$\#$ in Dim. & 
		Original term in \cite{dimopoulos2008general} & 
		\cite{dimopoulos2008general} in our notation & 
		\cite{dimopoulos2008general} in dim. param. mod $\Comp$ & 
		Our result & $\mathcal{O}(n)$ & 
		Comment \\ 
		\hline
		1 & 
		$- \kappa_\text{eff} g  T^{2}$ & 
		$\kappa_R g T_R^{2}$ & 
		$\RR \GOneR T_R$ & 
		\Green $\RR \GOneR T_R$ & 
		2 & \\ 
		\hline
		2 & 
		$- \kappa_\text{eff} (\partial_r g) v_{L} T^{3}$ & 
		$-\kappa_R \Gamma v_0 T_R^{3}$ & 
		$-\RR \GTwoR \VZero T_R$ & 
		\Green $-\RR \GTwoR \VZero T_R$ & 
		3 & \\ 
		\hline
		3 & 
		$-\frac{7}{12} \kappa_\text{eff} (\partial_r g) g T^{4}$ & 
		$\frac{7}{12} \Gamma g \kappa_R T_R^{4}$ & 
		$\frac{7}{12} \RR \GOneR \GTwoR T_R$ & 
		\Green $\frac{7}{12} \RR \GOneR \GTwoR T_R$ & 
		3 & \\ 
		\hline
		4 & 
		$-3 \frac{\kappa_\text{eff} g^2}{{\color{orange} c}} T^3$ & 
		$-3 \frac{\kappa_R g^2}{c} T_R^3$ & 
		$-3 \RR \GOneR^2 T_R$ & 
		\Green $-3 \RR \GOneR^2 T_R$ & 
		3 & FSL \\ 
		\hline
		5 & 
		$-3 \frac{ \kappa_\text{eff} g v_L}{{\color{orange} c}} T^2$ & 
		$3 \frac{ \kappa_R g v_0}{c} T_R^2$ & 
		$3 \GOneR \VZero \RR T_R$ & 
		\Green $3 \GOneR \VZero \RR T_R$ & 
		3 & 
		FSL \\ 
		\hline
		6 & 
		$- \frac{{\color{orange} \hbar} \kappa_\text{eff}^2}{2 m} (\partial_r g) T^3$ & 
		$- \frac{\hbar \kappa_R^2}{2 m} \Gamma T_R^3$ & 
		$-\frac{1}{2} \RR^2 \GTwoR T_R$ & 
		\Green $-\frac{1}{2} \RR^2 \GTwoR T_R$ & 
		3 & \\ 
		\hline
		7 & 
		$(\omega_\text{eff} - \omega_a) \frac{g T^2}{{\color{orange} c}}$ & 
		$\omega_R \frac{g T_R^2}{c}$ & 
		$\FreqR \GOneR T_R$ & 
		\Green $\FreqR \GOneR T_R$ & 
		2 & 
		FSL \\ 
		\hline
		8 & 
		$(2 - 2 \beta - \gamma) \frac{ \kappa_\text{eff} g \phi T^2 }{{\color{orange} c^2}}$ & 
		$(2 - 2 \beta + \gamma) \frac{\kappa_R g \phi_0 T_R^2}{c^2}$ & 
		$(2 - 2 \beta + \gamma) \GZero \GOneR \RR T_R$ & 
		$(2 - 2 \beta + 2 \gamma) \GZero \GOneR \RR T_R$ & 
		3 & 
		PPN \\ 
		\hline
		9 & 
		$-\frac{3 {\color{orange} \hbar} \kappa_\text{eff}^2}{2 m {\color{orange} c} } g T^2$ & 
		$\frac{3 \hbar \kappa_R^2}{2 m c } g T_R^2$ & $\frac{3}{2} \RR^2 \GOneR T_R $ & 
		$\frac{5}{2} \RR^2 \GOneR T_R $ & 
		3 & 
		FSL \\ 
		\hline
		10 & 
		$-\frac{7}{12} \kappa_\text{eff} v_L^2 \left(\partial_r^2 g \right) T^4 $ & 
		$\frac{7}{6} \kappa_R v_0^2 \Lambda T_R^4$ & $\frac{7}{6} \RR \VZero^2 \GThreeR$ & 
		\Green $\left( \frac{7}{6} - \frac{1}{36} \right) \RR \VZero^2 \GThreeR$ & 
		4 &  \\ 
		\hline
		11 & 
		$-\frac{35}{4 {\color{orange} c}} \kappa_\text{eff} (\partial_r g) g v_L T^4 $ & 
		$\frac{35}{4} \frac{\kappa_R \Gamma g v_0}{c} T_R^4 $ & 
		$\frac{35}{4} \RR \VZero \GOneR \GTwoR T_R$ & 
		\Green $\frac{35}{4} \RR \VZero \GOneR \GTwoR T_R$ & 
		4 & 
		FSL \\ 
		\hline
		12 & 
		$-\frac{4}{{\color{orange} c}} \kappa_\text{eff} (\partial_r g) v_L^2 T^3 $ & 
		$-4 \frac{\kappa_R \Gamma v_0^2}{c} T_R^3 $ & 
		$-4 \RR \VZero^2 \GTwoR T_R$ & 
		\Green $-4 \RR \VZero^2 \GTwoR T_R$ & 
		4 & 
		FSL \\ 
		\hline
		13 & 
		$2 \frac{\omega_a g^2 T^3}{{\color{orange} c^2}}$ & 
		0 & 
		0 & 
		0 & 
		3 & 
		Inel. scat. \\ 
		\hline
		14 & 
		$2 \frac{\omega_a g v_L T^2}{{\color{orange} c^2}}$ & 
		0 & 
		0 & 
		0 & 
		3 & 
		Inel. scat. \\ 
		\hline
		15 & 
		$-\frac{7 {\color{orange} \hbar}\kappa_\text{eff}^2}{12m} v_L \left(\partial_r^2 g \right) T^4 $ & 
		$\frac{7}{6} \frac{\hbar \kappa_R^2}{m} v_0 \Lambda T_R^4 $ & 
		$\frac{7}{6} \RR^2 \VZero \GThreeR$ & 
		\Green $\left( \frac{7}{6} - \frac{1}{36} \right) \RR^2 \VZero \GThreeR$ & 
		4 &  \\ 
		\hline
		16 + 19$^\dagger$ & 
		$ \left[- 12 g^2  + (2 - 2\beta - \gamma) \partial_r (g \phi) \right] \frac{\kappa_\text{eff} v_L T^3}{{\color{orange} c^2}}$ 
		& $-(14 - 2\beta - \gamma) \frac{\kappa_R g^2 v_0 T_R^3}{c^2}$ & 
		$-(14 - 2\beta - \gamma) \RR \GOneR^2 \VZero T_R$ & 
		$(19 + 2 \beta + 20 \gamma) \RR \GOneR^2 \VZero T_R$ & 
		4 & 
		PPN \\ 
		\hline
		17 + 23$^\ast$ & 
		$\left[- 7 - \frac{7}{12} (2 - 2 \beta - \gamma) \right] \frac{\kappa_\text{eff} g^3 T^4}{{\color{orange} c^2}}$ & 
		$\frac{7}{12} (14 - 2 \beta -  \gamma) \frac{\kappa_R g^3 T_R^4}{c^2}$ &
		$\frac{7}{12} (14 - 2 \beta -  \gamma) \RR \GOneR^3 T_R$ &
		$-\frac{151 + 4 \beta + 117 \gamma}{12} \RR \GOneR^3 T_R$ & 
		4 & 
		PPN \\ 
		\hline
		18 & 
		$-5 \frac{ \kappa_\text{eff} g v_L^2 T^2}{{\color{orange} c^2}}$ & 
		$ 5 \frac{\kappa_R g v_0^2}{c^2}  T_R^2 $ & 
		$5 \RR \GOneR \VZero^2 T_R$ & 
		$-\frac{11 + 12\gamma}{2} \RR \GOneR \VZero^2 T_R$ & 
		4 & 
		PPN \\ 
		\hline
		20 & 
		$- \frac{7}{12} (4 - 4 \beta - 3 \gamma) \frac{\kappa_\text{eff} \phi (\partial_r g) g}{{\color{orange} c^2}} T^4$ & 
		$\frac{7}{12} (4 - 4 \beta - 3 \gamma) \frac{\kappa_R \phi_0 \Gamma g}{c^2} T_R^4$ & $\frac{7}{12} (4 - 4 \beta - 3 \gamma) \GZero \GOneR \GTwoR \RR$ & 
		$ \frac{100 \beta - 100 - 14 \gamma}{12} \GZero \GOneR \GTwoR \RR$ & 
		4 & 
		PPN \\ 
		\hline
		21 & 
		$(\omega_\text{eff} - \omega_a) \frac{ (\partial_r g)  v_L T^3}{{\color{orange} c}}$ & 
		$-\omega_R \frac{ \Gamma  v_0 T_R^3}{ c}$ & 
		$-\FreqR \GTwoR \VZero T_R $ & 
		\Green $- \FreqR \GTwoR \VZero T_R$ & 
		3 & 
		FSL \\ 
		\hline
		22 & 
		$\frac{7}{12} (\omega_\text{eff} - \omega_a) \frac{(\partial_r g) g T^4}{{\color{orange} c}} $ & 
		$\frac{7}{12} \omega_R \frac{\Gamma g T_R^4}{c} $ & 
		$\frac{7}{12} \FreqR \GOneR \GTwoR T_R$ & 
		\Green $\frac{7}{12} \FreqR \GOneR \GTwoR T_R$ & 
		3 & 
		FSL \\ 
		\hline
		24 & 
		$\frac{7 {\color{orange} \hbar} \kappa_\text{eff}^2}{2 m {\color{orange} c}} (\partial_r g) v_L T^3 $ & 
		$\frac{7 \hbar \kappa_R^2}{2 m c} \Gamma v_0 T_R^3 $ & 
		$\frac{7}{2} \RR^2 \GTwoR \VZero T_R$ & 
		$-\frac{19}{6} \RR^2 \GTwoR \VZero T_R$ & 
		4 & 
		FSL \\ 
		\hline
		25 & 
		$\frac{27 {\color{orange} \hbar} \kappa_\text{eff}^2}{8 m {\color{orange} c}} (\partial_r g) g T^4 $ & 
		$-\frac{27 \hbar \kappa_R^2}{8 m c} \Gamma g T_R^4 $ & 
		$-\frac{27}{8} \RR^2 \GOneR \GTwoR T_R$ & 
		$\frac{65}{24} \RR^2 \GOneR \GTwoR T_R$ & 
		4 & 
		FSL \\ 
		\hline
		26 & 
		$\frac{{\color{orange} \hbar} \kappa_\text{eff} \omega_a}{m {\color{orange} c^2}} g T^2 $ & 
		0 & 
		0 & 
		0 & 
		3 & 
		Inel. scat. \\ 
		\hline
		27 & 
		$6 (2 - 2 \beta - \gamma) \frac{ \kappa_\text{eff} \phi g^2 T^3}{{\color{orange} c^3}}$ & 
		$6 (2 - 2 \beta - \gamma) \frac{ \kappa_R \phi_0 g^2 T_R^3}{c^3}$ & 
		$6 (2 - 2 \beta - \gamma) \RR \GZero \GOneR^2$ & 
		$6 (4 -  4\beta + \gamma) \RR \GZero \GOneR^2$ & 
		4 & 
		FSL \\ 
		\hline
		28 & 
		$3 (\omega_\text{eff} - \omega_a) \frac{g^2 T^3 }{{\color{orange} c^2}}$ & 
		$-3 \omega_R \frac{g^2 T_R^3 }{c^2}$ & 
		$-3 \FreqR \GOneR^2 T_R$ & 
		\Green $- 3 \FreqR \GOneR^2 T_R$ & 
		3 & \\ 
		\hline
		29 & 
		$3 (\omega_\text{eff} - \omega_a) \frac{ g v_L T^2 }{{\color{orange} c^2}}$ & 
		$3 \omega_R \frac{ g v_0 T_R^2 }{c^2}$ & 
		$3 \FreqR \GOneR \VZero T_R$ & 
		\Green $3 \FreqR \GOneR \VZero T_R$ & 
		3 & \\ 
		\hline
		30 & 
		$6 (1 - \beta) \frac{\phi g v_L T^2}{{\color{orange} c^3}} \kappa_\text{eff}$ & 
		$-6 (1 - \beta) \frac{\phi_0 g v_0 T_R^2}{c^3} \kappa_\text{eff}$ & 
		$-6 (1 - \beta) \GZero \GOneR \VZero \RR$ & 
		$-6 (2 - 2 \beta + \gamma) \GZero \GOneR \VZero \RR$ & 
		4 & 
		FSL \\
		\hline
	\end{tabularx}
	\caption{Comparison of phase shifts for a Mach-Zehnder IF of our results and Dimopoulos et al. Factors of $\hbar$ and $c$ are restored and highlighted in orange. Cells which are highlighted green coincide between our results and those of Dimopoulos et al. One can see that all terms of orders $\mathcal{O}(2)$ and $\mathcal{O}(3)$ are reproduced apart from terms $\#$8 and $\#$9. The prefactor in terms $\#$10 and $\#$15 are written in green form since the contribution of $\frac{7}{6}$ results from adding the $\Lambda$-dependent part of the gravitational potential to the Lagrangian, which manifests in the propagation phase. The additional $-\frac{1}{36}$ results from taking into account this part of the potential in the Euler-Lagrange equation, resulting in altered atomic trajectories. The latter effect seems to be left out in~\cite{dimopoulos2008general}. $\dagger, \ast$: Terms $\#$16 + $\#$19 and $\#$17 + $\#$23 can be combined and therefore share a row. The comment 'FSL' marks the first-order FSL effect described before.}
	\label{Table: MZI_Phase}
\end{table*}

\newpage


\bibliographystyle{apsrev4-1_costum}
\bibliography{bibliography}

\end{document}